\documentclass[12pt]{article}

\usepackage{CJKutf8}
\usepackage{cite}
\usepackage{graphicx}
\usepackage{amsmath}
\usepackage{amsthm}

\usepackage{geometry}
\usepackage{framed}
\usepackage{multirow}
\usepackage{rotating}
\usepackage{amsfonts}
\usepackage{longtable}
\usepackage{array} 

\usepackage{fancyhdr}
\begin{document}

\title{State Estimation}
\date{}

\author{Hao Li 
\thanks{Namely \begin{CJK}{UTF8}{gbsn}李颢\end{CJK}, the same author of the works \cite{Li2026ACTPA_SJTU_2, Li2026ACTPA_SJTU_1}.} }

\maketitle

\begin{abstract}
\textbf{Control science} is a core representative of the third industrial revolution and is so important to modern civilization. \textbf{Control systems} are the main subject of control science and may involve many aspects of consideration, such as hardware consideration, software consideration, operation consideration, maintenance consideration, economy consideration, society consideration. However, besides all such aspects of consideration, one aspect that is most essential to the control system is methodology consideration in mathematical sense, knowledge on which is what we refer to as \textbf{control theory}. Besides its importance from the mathematical perspective, control theory is even more charming as it is deeply rooted in practical applications. Charms of control theory consist in both \textit{know-why} and \textit{know-how} and it is the fusion of control theory and practical applications that highlights such charms. Control theory for practical applications, especially when somewhat with so-called ``advanced'' flavour, involves several fundamental aspects. This article introduces the \textit{State Estimation} aspect of \textit{Advanced Control Theory for Practical Applications} \cite{Li2026ACTPA_SJTU_2, Li2026ACTPA_SJTU_1}.
\end{abstract}

\section{Introduction}

Sometimes in study of control theory, it is assumed by default that the control system's state is known. In practical applications, a question arises naturally: \textit{how do we know the state}? Readers may think this is a trivial issue, because it seems that we can simply install suitable sensors to measure the complete state directly namely to measure all elements of the state directly. However, above question is not a trivial issue at all in practical applications.

The process of using certain method to know or infer the state is formally coined as \textbf{state estimation} or \textbf{estimation} for short. Estimation is a ubiquitous process in our daily life and exists even without our attention in every aspect of human activities. No matter in which activity, we normally do not act blindly but act reasonably according to our conscious estimation of things and situations involved in the activity. Take some routine and ordinary behaviours as examples. Dressing is indispensable to humans, and during dressing, we dynamically estimate the spatial relationship between our body and the clothes and adjust our body movement accordingly, until the clothes are properly dressed on our body. Similarly, eating is also an indispensable activity, during which we dynamically estimate the spatial relationship between our mouth and the food and adjust our arm and hand movement accordingly, until the food is desirably put into our mouth. For another daily-life example, walking is a common activity, during which we unceasingly observe environment objects around us and estimate the spatial relationship between our body and them so that we can dynamically determine a navigable (and often somehow optimal) path for our potential walking movements. 

By analogy to humans, any autonomous or partially-autonomous system that interacts with the environment should possess estimation ability  according to its operation requirements. Intuitively speaking, \textbf{estimation} aims at \textit{revealing} or \textit{deriving} the truth of something essential to a system's operation.

\subsection{State}  \label{sec:state_def}

States can potentially be of unlimited kinds. A state can be the temperature, heart rate, pulse strength, and even the overall healthy status of a patient. A state can be the trading volume of the capital market. A state can be economical performance of an area, a town, a city, a country, or the world. A state can be statistics of traffic flow in daily life. A state can be the structure of relationships among people in an enterprise or a society. A state can be some aspect of human feeling towards an object or an event. A state can be purchasing habits of people in different seasons or under different circumstances. In the context of mobile robots (including outdoor intelligent vehicles) which are typical kinds of automatic control systems, a state can be the system's pose in a two-dimensional or three-dimensional global reference \cite{Li2010, Li2011a}. A state can be the spatial representation of both stationary and dynamic objects in the system's local environment \cite{Li2014, Li2024IV, Li2024ITS}. A state can be certain special condition of the system's surroundings \cite{Li2021cvpr}.

In Chapter 1 and Chapter 2,
\footnote{Namely Chapters 1 and 2 of the author's works \cite{Li2026ACTPA_SJTU_2, Li2026ACTPA_SJTU_1}. Note that this article is Chapter 3 of the works.}
the author has been using the concept \textbf{state} all along, without providing a strict definition of state beforehand. Readers may have intuitive imagination or daily-life understanding of what the state is. In fact, the author still refrains from providing a strict or so-called reasonable definition of state, and believes such intuitive imagination or daily-life understanding would already be sufficient for study of control theory presented by so far and to be presented throughout this book
\footnote{It is indeed difficult to provide a strict definition of state. The difficulty here is somehow like that in answering the famous philosophical question: what is the human? Until today, people still cannot provide a strict definition of human. On the other hand, the seemingly embarrassing fact that there is no strict definition of human does not influence humans' recognition of themselves, humans' interaction among themselves, or human-related activities at all. The situation is similarly so for the concept state as well.}.
However, readers may still wonder: what is the state?

In the author's previous book \cite{Li2024CTPA_Springer, Li2024CTPA_SJTU_1}, the author mentions that for a control system with certain target process, the set of properties that characterizes the target process is called its \textit{process state} (from the perspective of the target process itself), or its \textit{system state} (from the perspective of the control system holistically), or simply its \textit{state}. Besides, the intuitive phrase ``something essential to a system's operation'' just mentioned above may be called \textit{state}. Both seem to be definitions, but they are away from being so in strict sense. Further clarifications are needed to alleviate ambiguity of the statements. 

Readers had better first bear in mind that no state would characterize all aspects of details of a control system. For example, consider an intelligent vehicle, its hardware must be made by certain materials (usually a large variety of materials). Can we have any state describing positions and moving conditions of all atoms in the materials that make up the intelligent vehicle? Even if information of positions and moving conditions of all the atoms is theoretically obtainable, yet the answer is definitely negative for practical applications, because computational and memory resources normally available to an intelligent vehicle can neither store nor process such huge amount of information. Not to say that such information tends to be theoretically unobtainable. On the other hand, even if we had certain state to describe positions and moving conditions of all the atoms, it would be totally unnecessary to do so, because such state is not related to normal operation of an intelligent vehicle at all. In other words, such state is not something essential to the intelligent vehicle's operation or does not play an essential role in characterizing the target process of the intelligent vehicle.

Following above reflection with this intelligent vehicle example, readers had better then bear in mind that the state must play an essential role in characterizing the target process, or simply speaking, the state must be something essential or the properties that form the state must be essential. Yet a question arises naturally: how can the term essential be interpreted exactly in the discussed context? The term essential may be interpreted from two perspectives, namely the objective perspective and the subjective perspective.

Both the objective perspective and the subjective perspective are somehow related to system modelling that will be discussed in more details later in the context of state estimation. From the objective perspective, the state must be \textit{deterministic} in the sense that the state is able to determine its own dynamics namely to determine how the state itself evolves. Still consider the intelligent vehicle example. For an intelligent vehicle, an important target process is its motion, dynamics of which can be described by the bicycle kinematics model
\begin{align}  \label{eq:bicycle_kinematics_model}  
\left\{
\begin{array}{l l}
\frac{\mathrm{d}}{\mathrm{d} t} x &= v \cos \theta \\
\frac{\mathrm{d}}{\mathrm{d} t} y &= v \sin \theta \\
\frac{\mathrm{d}}{\mathrm{d} t} \theta &= \frac{v}{L} \tan \beta \\
\frac{\mathrm{d}}{\mathrm{d} t} \beta &= \frac{1}{\tau_{\beta}} (\beta_I - \beta) 
\end{array}
\right.
\end{align}
In (\ref{eq:bicycle_kinematics_model}), the state 
\begin{align*}
\mathbf{x} \equiv \begin{bmatrix} x & y & \theta & \beta \end{bmatrix}^\mathrm{T}
\end{align*}
is able to determine how $\mathbf{x}$ itself evolves, given that the control input $\beta_I$ is treated as a known. In contrast, if $\theta$ is removed from the state $\mathbf{x}$ to form a reduced state 
\begin{align*}
\mathbf{x}_r \equiv \begin{bmatrix} x & y & \beta \end{bmatrix}^\mathrm{T}, 
\end{align*}
then the reduced state $\mathbf{x}_r$ is unable to determine how $\mathbf{x}_r$ itself evolves and hence cannot be regarded as a state.

From the subjective perspective, the state must be \textit{complete} in the sense that the state contains all the properties that we care during the system's operation. For vehicle navigation on a generic plane, the state 
\begin{align*}
\mathbf{x} \equiv \begin{bmatrix} x & y & \theta & \beta \end{bmatrix}^\mathrm{T}
\end{align*}
namely the set of the vehicle longitudinal position, the vehicle lateral position, the vehicle orientation, and the vehicle steering angle contain all the properties that we care. Sometimes, when vehicle steering dynamics is neglected, the steering angle $\beta$ can be removed from the set of properties that we care and instead serve directly as control input to the vehicle, i.e. 
\begin{align*}
\beta \equiv \beta_I. 
\end{align*}
Then the state is reduced to a new state 
\begin{align*}
\mathbf{x} \equiv \begin{bmatrix} x & y & \theta \end{bmatrix}^\mathrm{T}
\end{align*}
and the bicycle kinematics model (\ref{eq:bicycle_kinematics_model}) is reduced to a new one
\begin{align*}
\left\{
\begin{array}{l l}
\frac{\mathrm{d}}{\mathrm{d} t} x &= v \cos \theta \\
\frac{\mathrm{d}}{\mathrm{d} t} y &= v \sin \theta \\
\frac{\mathrm{d}}{\mathrm{d} t} \theta &= \frac{v}{L} \tan \beta \equiv \frac{v}{L} \tan \beta_I 
\end{array}
\right.
\end{align*}
By the way, we can easily see that if above reduced bicycle kinematics model is adopted, the reduced state 
\begin{align*}
\mathbf{x} \equiv \begin{bmatrix} x & y & \theta \end{bmatrix}^\mathrm{T}
\end{align*}
is deterministic as well.

Similarly, when vehicle lateral dynamics is considered only
\footnote{Instead of analysing vehicle navigation on a generic plane, analyse vehicle navigation in certain (dynamically sliding) local road reference whose horizontal axis namely $x$-axis is always aligned with the local road direction.},
the vehicle longitudinal position $x$ can be removed from the set of properties that we care. Then the state is reduced to the vehicle lateral state 
\begin{align*}
\mathbf{x} \equiv \begin{bmatrix} y & \theta & \beta \end{bmatrix}^\mathrm{T}
\end{align*}
and the bicycle kinematics model (\ref{eq:bicycle_kinematics_model}) is reduced to the bicycle lateral kinematics model
\begin{align}  \label{eq:vehicle_lateral_control}
\left\{
\begin{array}{l l}
\frac{\mathrm{d}}{\mathrm{d} t} y &= v \sin \theta \\
\frac{\mathrm{d}}{\mathrm{d} t} \theta &= \frac{v}{L} \tan \beta \\
\frac{\mathrm{d}}{\mathrm{d} t} \beta &= \frac{1}{\tau_{\beta}} (\beta_I - \beta) 
\end{array}
\right.
\end{align}
We can also see that if the bicycle lateral kinematics model (\ref{eq:vehicle_lateral_control}) is adopted, the vehicle lateral state 
\begin{align*}
\mathbf{x} \equiv \begin{bmatrix} y & \theta & \beta \end{bmatrix}^\mathrm{T}
\end{align*}
is deterministic as well.

On one hand, roughly speaking, if a state is both \textit{deterministic} and \textit{complete}, then it plays an essential role in characterizing the target process. On the other hand, as clarified above, both the objective perspective and the subjective perspective are somehow related to system modelling. In other words, we cannot purely discuss the essential role of the state by itself without involving the system model. For one adopted system model, a state can be both deterministic and complete and hence is indeed a state, whereas it may be neither deterministic nor complete if another system model is adopted. So we discuss system modelling next.

\subsection{System model}  \label{sec:system_model}

Besides \textbf{state}, another concept essential to estimation is \textbf{system model}. The system model of a state is a mathematical model that describes how the state evolves namely that describes dynamics of the state. The state evolution is governed by certain objective laws which may be established according to our systematic knowledge of physics, chemistry, biology, and other disciplines and may also be established according to our \textit{ad hoc} empirical knowledge. The system model can be of continuous-variable form or of discrete-variable form. It can be of equation formula form or of look-up table form. The key point is that \textit{the system model conveys all mathematical essence that describes clearly how the state evolves}.

A system model intended for state estimation should satisfy two basic requirements: \textit{first, it is reasonable enough for valid state estimation; second, it is tractable enough for feasible state estimation}. Before continuing, we had better distinguish between the world that exists objectively beyond human spirit and the world that exists in human knowledge. The former is called \textit{objective world} whereas the latter is called \textit{image world}. Throughout history, philosophers have all along been reflecting on the relationship between the objective world and the image world. Some philosophers hold a view of materialism and believe that an accurate knowledge of the objective world can be achieved so that the image world can be identical to the objective world. Some philosophers hold a view of idealism and believe that human knowledge of the objective world can never touch the objective world but can only serve as empirical or virtual reflection of the objective world. Some philosophers hold a dialectic view that harmonizes idealism with materialism or harmonizes materialism with idealism. 

The author has no intention to deviate into the philosophy kingdom which is much more colourful than above simple categorization. From the diverse kinds of philosophical meditation that coexist and even argue with each other, we can see that it is difficult to obtain an absolutely correct knowledge of the objective world. System modelling in absolutely correct way is also difficult and even impossible.  System modelling may only be \textit{approximation} of the objective world, yet it is still valuable and even indispensable in practical applications. How to properly establish a system model largely depends on specific field knowledge.

Like the fact that a system model is not absolutely necessary for automatic control (e.g., the probably most famous and popular control method namely the \textit{proportional-integral-derivative control method} or for short the \textit{PID control method} is completely model-free), a system model is not absolutely necessary for state estimation either. On the other hand, a reasonable system model can help enhance the performance and quality of automatic control as well as state estimation. For state estimation, the better the system model is, the better it can be. To facilitate understanding, a thought experiment can be performed for the following hypothetical scenario: suppose certain entity was in Shanghai one hour ago, and we need to estimate its current position. Without any \textit{a priori} knowledge of the entity, we can only have a weak system model no better than random guessing. Imagine that the entity is a hyper-developed superman-like alien who masters mysterious method of instant movement in the universe, then its current position may be potentially anywhere in the universe. 

In contrast, if certain \textit{a priori} knowledge is available, we may have a better system model taking into account the \textit{a priori} knowledge. For example, if we know that the entity is a hypersonic missile, then we can establish a system model considering the missile's operation limitations and conclude that the missile is somewhere on the earth instead of being arbitrarily elsewhere in outer space; the estimate is obviously better than a random guess without help of any system model. If we know that the entity is an aeroplane, then we can adopt a system model suitable for describing aeroplane movement and conclude that the aeroplane is somewhere in east Asia. If we further know that the aeroplane has been heading west during last one hour, then we can adopt an even more precise system model and conclude that the aeroplane is currently somewhere around Wuhan; the estimate becomes even better. 

\subsection{Measurement model}  \label{sec:measurement_model}

The next concept essential to estimation is \textbf{measurement} also known as \textbf{observation}. A measurement is a measured value of the state or related to the state and is normally provided by a sensor or sensing module. There are a variety of sensors such as positioning sensors, ranging sensors, force sensors, motion sensors, electricity sensors, and magnetic sensors \cite{Fraden2010}.

A question arises naturally: why do we need to \textit{estimate} the state if direct state measurements are available? Reasons for such necessity are two-folds. First, direct measurements inevitably suffer from errors and noises in practical applications and may not serve as state estimates of quality. In other words, direct measurements may not reveal the true state value well. Therefore, we need some proper estimation method to obtain state estimates better than direct measurements. 

Second, sometimes the system only has \textit{partial measurements} (or \textit{partial observations}) which reflects the state partially. A typical application example is GPS (global positioning system such as Bei Dou) based vehicle localization. The measurement device GPS provides measurements of the vehicle position which reflects the vehicle state (i.e. position and orientation) partially. The orientation component is not directly \textit{measurable} (or \textit{observable}). Another typical application example is landmark ranging based vehicle localization, where the landmark ranging module provides measurements of the distance between the vehicle and certain registered landmark at each instant. Such ranging measurements also reflects the vehicle state partially. In these kinds of applications, we need some proper estimation method to reveal the whole state that cannot be revealed directly by measurements.

The concept measurement (or observation) leads to one more concept essential to estimation, namely \textbf{measurement model} (or \textbf{observation model}). The measurement model is a mathematical model that describes the causal relationship between the state and the measurement, as if the measurement model ``predicts'' what the measurement would be given a known state. 

\begin{figure}[h!]
\begin{center}
\includegraphics[width=0.4\columnwidth]{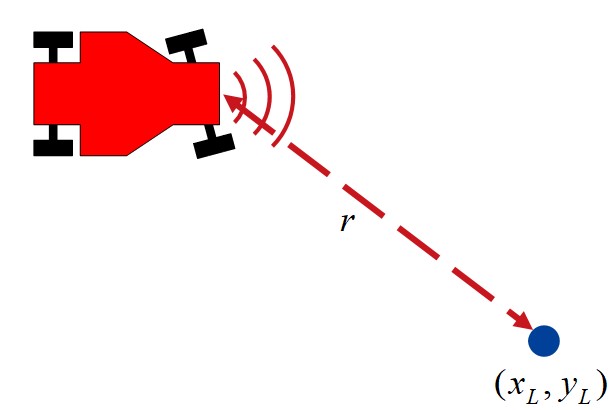}
\end{center}
\caption{Landmark distance measurement model}
\label{fig:landmark_distance_measurement_model}
\end{figure}

A measurement model example is the landmark distance measurement model illustrated in Figure \ref{fig:landmark_distance_measurement_model}. The landmark distance measurement model is formalized as
\begin{equation}  \label{eq:landmark_distance_msr_model}
r = \sqrt{(x - x_\mathrm{L})^2 + (y - y_\mathrm{L})^2},
\end{equation}
where $r$ denotes the distance measurement obtained by an intelligent system such as a robot or an intelligent vehicle, $(x, y)$ denotes the position of the intelligent system, and $(x_\mathrm{L}, y_\mathrm{L})$ denotes the registered position of the landmark corresponding to the distance measurement. The landmark distance measurement model can be involved in practical applications such as special tag based visual localization (especially when singularity of homography matrix computation is encountered) \cite{Li2025GFCV_SJTU}.

In the given example, the landmark distance measurement model described in (\ref{eq:landmark_distance_msr_model}) conveys the causal relationship between the intelligent system state 
\begin{align*}
\mathbf{x} \equiv \begin{bmatrix} x & y & \theta \end{bmatrix}^\mathrm{T}
\end{align*}
and the distance measurement $r$, as if the landmark distance measurement model predicts what the distance measurement would be given a known intelligent system state.

This logic seems unnatural and even confusing to our intuition. If the state is known, then why do we bother to use some so-called measurement model and some so-called estimation method? To facilitate understanding of measurement model, we may resort to an analogue with a famous statement of René DESCARTES (one of the greatest thinkers and scientists in history): ``Je pense donc je suis (Cogito ergo sum)'' i.e. ``I think therefore I am''. Here, ``I am (exist there)'' can be regarded as a state, whereas ``I think'' can be regarded as a measurement. The logic is not that ``I think'' causes the fact that ``I am''. The logic is exactly in the inverse direction: It is the state ``I am'' that causes the measurement ``I think''; otherwise, there will be no possibility at all for my thinking behaviour. The relationship between the state of our existence and our thinking behaviour is what the measurement model conveys here. From the measurement ``I think'' we can fairly infer the state ``I am''. It is such kind of \textit{inverse inference} that makes the measurement model useful to estimation.

\begin{figure}[h!]
\begin{center}
\includegraphics[width=0.6\columnwidth]{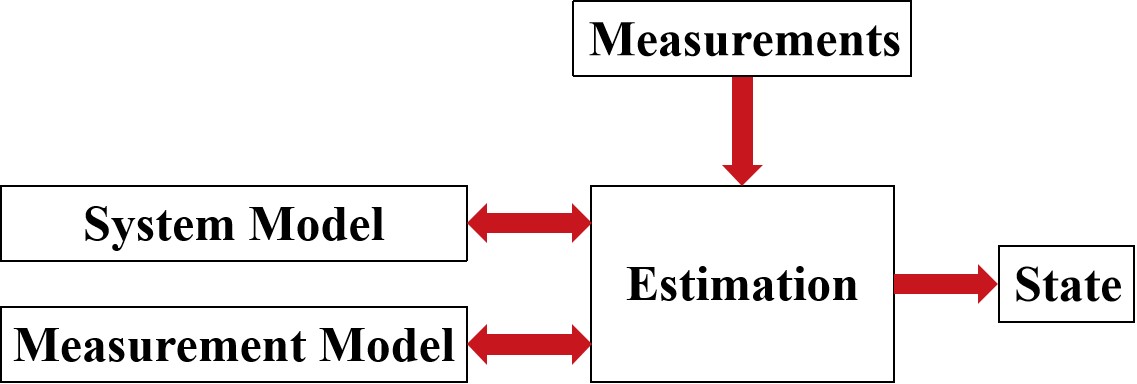}
\end{center}
\caption{Methodology of estimation}
\label{fig:estimation_methodology}
\end{figure}

After introduction of the important concepts \textbf{state}, \textbf{system model}, \textbf{measurement} and \textbf{measurement model}, we can clarify the concept estimation: For a system, \textbf{estimation} is the process of inferring the state from measurements, with the help of a given system model that describes state evolution and a given measurement model that describes state-measurement causal relationship, as demonstrated in Figure \ref{fig:estimation_methodology}.

\subsection{Observability}  \label{sec:observability}

As clarified in Section 1.5 in Chapter 1, 
\footnote{Namely Chapter 1 of the author's works \cite{Li2026ACTPA_SJTU_2, Li2026ACTPA_SJTU_1}. Note that this article is Chapter 3 of the works.}
to design a control system, it is worth making a preliminary theoretical judgement (if possible) of controllability of the target process. Similarly, it is also worth making a preliminary theoretical judgement (if possible) of observability of the target process.

A control system's target process is \textbf{observable} if its state $\mathbf{x}$ can be inferred with its available measurements, be the state inferred directly or indirectly. To better understand \textbf{observability}, it is worth clarifying the difference between \textit{unobservability} and \textit{stochastic uncertainty}. In practical applications, the estimate of any entity suffers from stochastic uncertainty that exists universally, and in this sense we cannot have absolute knowledge of any entity. However, this does not mean that any entity is unobservable.

Readers may resort to an example to understand the difference between unobservability and stochastic uncertainty. Imagine two vehicles that are both located at the origin initially: the first vehicle is equipped with an speedometer, whereas the second vehicle is not equipped with any motion sensor. For the first vehicle, although the equipped speedometer can never be absolutely accurate, we at least have a way to infer the vehicle speed within certain error tolerance associated with the odometer. Despite existence of stochastic uncertainty, we still treat the speed of the first vehicle as observable. Besides, we can also treat the position of the first vehicle as observable because we can infer the position indirectly by integral of the speed, though the estimated position also suffers from certain stochastic uncertainty inevitably. In contrast, for the second vehicle, both its speed and position are unobservable, because we do not have any way to infer its speed and position, neither directly nor indirectly. 

From this example we can see that whether a control system's target process is observable depends on whether there is a way to infer its state, regardless of the fact that the estimated state suffers from stochastic uncertainty. Therefore, to analyse observability of the target process, we can neglect stochastic uncertainty (though concrete state estimation methods normally need to take stochastic uncertainty into account).

Like for controllability analysis, it is difficult as well to have a general method to theoretically determine observability of any arbitrary target process especially severely nonlinear target process. But fortunately, many target processes encountered in practical applications can fairly adopt linear state-space modelling. For a target process that can fairly adopt linear state-space modelling, we have a systematic method to theoretically determine its observability.

Given a control system that adopts linear state-space modelling described by (\ref{eq:state_differential_equation_linear}), with its state denoted as $\mathbf{x}$ and its control input to the target process denoted as $\mathbf{u}$, i.e.
\begin{align}  \label{eq:state_differential_equation_linear}
\frac{\mathrm{d}}{\mathrm{d} t} \mathbf{x} = \mathbf{A} \mathbf{x} + \mathbf{B} \mathbf{u}.
\end{align}
Suppose its \textbf{linearized measurement model} is
\begin{align}  \label{eq:SDE_linear_measurement}
\mathbf{z} = \mathbf{H} \mathbf{x},
\end{align}
where $\mathbf{z}$ denotes raw sensor measurement of the state $\mathbf{x}$ and $\mathbf{H}$ denotes the \textbf{measurement matrix}. State estimation aims at inferring or revealing the complete state with measurements available in finite time.

When the measurement matrix $\mathbf{H}$ is of full rank in terms of row vectors or in other words the rank of $\mathbf{H}$ equals the state dimension $n$, then we can take measurement transform as
\begin{align*}
\bar{\mathbf{z}} = \mathbf{H}^{-1} \mathbf{z} 
\end{align*}
or
\begin{align*}
\bar{\mathbf{z}} = (\mathbf{H}^\mathrm{T} \mathbf{H})^{-1} \mathbf{H}^\mathrm{T} \mathbf{z}
\end{align*}
and transform the measurement model into
\begin{align*}
\bar{\mathbf{z}} = \mathbf{x},
\end{align*}
which implies that the sensors can measure the complete state directly and hence the target process is apparently observable. 

On the other hand, when the measurement matrix $\mathbf{H}$ is rank-deficient, we need some systematic method to theoretically determine observability of the target process. Consider the solution of (\ref{eq:state_differential_equation_linear}), which is described in
\begin{align}  \label{eq:state_differential_equation_linear_solution}
\mathbf{x} = \mathrm{e}^{\mathbf{A} t} \mathbf{x}_0 + \int_0^t \mathrm{e}^{\mathbf{A} (t - \tau)} \mathbf{B} \mathbf{u}(\tau) \mathrm{d} \tau.
\end{align}
The solution described in (\ref{eq:state_differential_equation_linear_solution}) implies that once the initial state $\mathbf{x}_0$ is known, we can infer the state at any time via (\ref{eq:state_differential_equation_linear_solution}). Transform (\ref{eq:state_differential_equation_linear_solution}) into
\begin{align*}
\mathbf{x}_0 = \mathrm{e}^{-\mathbf{A} t} (\mathbf{x} - \int_0^t \mathrm{e}^{\mathbf{A} (t - \tau)} \mathbf{B} \mathbf{u}(\tau) \mathrm{d} \tau),
\end{align*}
which further implies that we can also recover the initial state from the state at any time. Therefore, observability of the target process is equivalent to observability of the initial state $\mathbf{x}_0$. In other words, whether a control system's target process is observable depends on whether we can infer its initial state $\mathbf{x}_0$ with measurements available in finite time.

Substitute (\ref{eq:state_differential_equation_linear_solution}) into (\ref{eq:SDE_linear_measurement}) and obtain 
\begin{align}  \label{eq:SDE_linear_initial_state_equation}
&\mathbf{z} = \mathbf{H} \mathbf{x} = \mathbf{H} \mathrm{e}^{\mathbf{A} t} \mathbf{x}_0 + \mathbf{H} \int_0^t \mathrm{e}^{\mathbf{A} (t - \tau)} \mathbf{B} \mathbf{u}(\tau) \mathrm{d} \tau  \nonumber \\
\iff & \mathbf{H} \mathrm{e}^{\mathbf{A} t} \mathbf{x}_0 = \mathbf{z} - \mathbf{H} \int_0^t \mathrm{e}^{\mathbf{A} (t - \tau)} \mathbf{B} \mathbf{u}(\tau) \mathrm{d} \tau \equiv \bar{\mathbf{z}}.
\end{align}
For each time $t$, (\ref{eq:SDE_linear_initial_state_equation}) provides a linear equation in terms of the initial state $\mathbf{x}_0$. Given $t$ in a time interval $[0, T]$, an infinite number of linear equations in terms of the initial state $\mathbf{x}_0$ can be established. Whether we can infer the initial state $\mathbf{x}_0$ with measurements available in finite time depends on whether we can always solve the initial state $\mathbf{x}_0$ according to the linear equations established by (\ref{eq:SDE_linear_initial_state_equation}) for certain finite time interval $[0, T]$, which further depends on whether $\langle \mathbf{H} \mathrm{e}^{\mathbf{A} t} \rangle_\mathrm{R}$ namely the span of $\mathbf{H} \mathrm{e}^{\mathbf{A} t}$ in terms of row vectors for certain finite time interval $[0, T]$ is of full rank or in other words whether the rank of $\langle \mathbf{H} \mathrm{e}^{\mathbf{A} t} \rangle_\mathrm{R}$ for certain finite time interval $[0, T]$ equals the state dimension $n$.
\begin{align*}
\mathbf{H} \mathrm{e}^{\mathbf{A} t} = \mathbf{H} \sum_{k=0}^{\infty} \mathbf{A}^k \frac{t^k}{k !} = \mathbf{U}^\mathrm{T} \mathbf{O}_{\mathbf{H}, \mathbf{A}},
\end{align*}
where
\begin{align*}
\mathbf{U} = \begin{bmatrix} \mathbf{I} \\ t \mathbf{I} \\ \frac{t^2}{2 !} \mathbf{I} \\ \frac{t^3}{3 !} \mathbf{I} \\ \vdots \end{bmatrix}, \quad \mathbf{O}_{\mathbf{H}, \mathbf{A}} = \begin{bmatrix} \mathbf{H} \\ \mathbf{H} \mathbf{A} \\ \mathbf{H} \mathbf{A}^2 \\ \mathbf{H} \mathbf{A}^3 \\ \vdots \end{bmatrix}.
\end{align*}
We have
\begin{align*}
\langle \mathbf{H} \mathrm{e}^{\mathbf{A} t} \rangle_\mathrm{R} = \langle \mathbf{U}^\mathrm{T} \rangle_\mathrm{R} \mathbf{O}_{\mathbf{H}, \mathbf{A}}.
\end{align*}

Since 
\begin{align*}
1, \quad t, \quad \frac{t^2}{2 !}, \quad \frac{t^3}{3 !}, \quad \cdots 
\end{align*}
are linearly independent, $\langle \mathbf{U}^\mathrm{T} \rangle_\mathrm{R}$ for any real-value interval is of full rank 
\footnote{This rank is an abstract ``number'' that equals the total ``number'' of natural numbers.}
and
\begin{align*}
\mbox{rank of } \langle \mathbf{H} \mathrm{e}^{\mathbf{A} t} \rangle_\mathrm{R}  = \mbox{rank of } \mathbf{O}_{\mathbf{H}, \mathbf{A}}.
\end{align*}
Therefore, the necessary and sufficient condition for the target process to be observable is that $\mathbf{O}_{\mathbf{H}, \mathbf{A}}$ is of full rank in terms of row vectors or in other words the rank of $\mathbf{O}_{\mathbf{H}, \mathbf{A}}$ equals the state dimension $n$. The infinite matrix $\mathbf{O}_{\mathbf{H}, \mathbf{A}}$ is called the \textbf{observability matrix} of the control system.

The \textit{Hamilton-Cayley theorem} implies that each element after $\mathbf{H} \mathbf{A}^{n-1}$ in the observability matrix $\mathbf{O}_{\mathbf{H}, \mathbf{A}}$ namely each $\mathbf{H} \mathbf{A}^k$ for $k \geq n$ can be transformed into a linear combination of 
\begin{align*}
\mathbf{H}, \quad \mathbf{H} \mathbf{A}, \quad \mathbf{H} \mathbf{A}^2, \quad \cdots \quad, \quad \mathbf{H} \mathbf{A}^{n-1}. 
\end{align*}
So we have
\begin{align*}
\mbox{rank of } \begin{bmatrix} \mathbf{H} \\ \mathbf{H} \mathbf{A} \\ \mathbf{H} \mathbf{A}^2 \\ \mathbf{H} \mathbf{A}^3 \\ \vdots \end{bmatrix} = \mbox{rank of } \begin{bmatrix} \mathbf{H} \\ \mathbf{H} \mathbf{A} \\ \mathbf{H} \mathbf{A}^2 \\ \vdots \\ \mathbf{H} \mathbf{A}^{n-1} \end{bmatrix}
\end{align*}
and we can define the \textbf{observability matrix} simply as
\begin{equation}  \label{eq:observability_matrix}
\mathbf{O}_{\mathbf{H}, \mathbf{A}} = \begin{bmatrix} \mathbf{H} \\ \mathbf{H} \mathbf{A} \\ \mathbf{H} \mathbf{A}^2 \\ \vdots \\ \mathbf{H} \mathbf{A}^{n-1} \end{bmatrix}.
\end{equation}
If the observability matrix $\mathbf{O}_{\mathbf{H}, \mathbf{A}}$ defined in (\ref{eq:observability_matrix}) is rank-deficient in terms of row vectors, then the target process is unobservable. If the observability matrix $\mathbf{O}_{\mathbf{H}, \mathbf{A}}$ is of full rank in terms of row vectors, then the target process is observable.

\begin{framed} 
\noindent \textbf{Control system observability criterion}: \textit{For a linear control system, if its observability matrix is rank-deficient in terms of row vectors, then its target process is unobservable. If its observability matrix is of full rank in terms of row vectors, then its target process is observable}.
\end{framed}

\subsubsection*{Application: double inverted pendulum observability analysis}

Apply the observability criterion to determine observability of the double inverted pendulum that adopts linear state-space modelling described by (1.13) or (1.14). 
\footnote{Namely (1.13) in Chapter 1 of the author's works \cite{Li2026ACTPA_SJTU_2, Li2026ACTPA_SJTU_1}. Note that this article is Chapter 3 of the works.}
The state transition matrix $\mathbf{A}$ is
\begin{align*}
\mathbf{A} = \begin{bmatrix} 0 & 1 & 0 & 0 & 0 & 0 \\
(1 + \frac{m_2}{m_1}) \frac{g}{L_1} & 0 & -\frac{m_2}{m_1} \frac{g}{L_1} & 0 & 0 & 0 \\
0 & 0 & 0 & 1 & 0 & 0 \\
-(1 + \frac{m_2}{m_1}) \frac{g}{L_2} & 0 & (1 + \frac{m_2}{m_1}) \frac{g}{L_2} & 0 & 0 & 0 \\
0 & 0 & 0 & 0 & 0 & 1  \\ 0 & 0 & 0 & 0 & 0 & 0 \end{bmatrix}.
\end{align*}
Suppose we can measure the first inverted pendulum angle $\theta_1$ and the cart position $x$ directly. The measurement model is
\begin{align*}
\mathbf{z} = \begin{bmatrix} 1 & 0 & 0 & 0 & 0 & 0 \\ 0 & 0 & 0 & 0 & 1 & 0 \end{bmatrix} \mathbf{x} \equiv \mathbf{H} \mathbf{x}.
\end{align*}
Compute the observability matrix via (\ref{eq:observability_matrix}) as
\begin{align*}
\mathbf{O}_{\mathbf{H}, \mathbf{A}} = \begin{bmatrix} \mathbf{H} \\ \mathbf{H} \mathbf{A} \\ \mathbf{H} \mathbf{A}^2 \\ \mathbf{H} \mathbf{A}^3 \\ \mathbf{H} \mathbf{A}^4 \\ \mathbf{H} \mathbf{A}^5 \end{bmatrix} = \begin{bmatrix} 1 & 0 & 0 & 0 & 0 & 0 \\ 
0 & 0 & 0 & 0 & 1 & 0 \\ 0 & 1 & 0 & 0 & 0 & 0 \\ 0 & 0 & 0 & 0 & 0 & 1 \\ 
o_1 & 0 & -o_2 & 0 & 0 & 0 \\ 0 & 0 & 0 & 0 & 0 & 0 \\ 
0 & o_1 & 0 & -o_2 & 0 & 0 \\ 0 & 0 & 0 & 0 & 0 & 0 \\
o_1 (o_1 + o_3) & 0 & - o_1 (o_2 + o_3) & 0 & 0 & 0 \\ 0 & 0 & 0 & 0 & 0 & 0 \\
0 & o_1 (o_1 + o_3) & 0 & - o_1 (o_2 + o_3) & 0 & 0 \\ 
0 & 0 & 0 & 0 & 0 & 0 \end{bmatrix},
\end{align*}
where
\begin{align*}
o_1 &\equiv (1 + \frac{m_2}{m_1}) \frac{g}{L_1},  \\ 
o_2 &\equiv \frac{m_2 g}{m_1 L_1},  \\ 
o_3 &\equiv \frac{m_2 g}{m_1 L_2}.
\end{align*}
Matlab code for computing the observability matrix via symbolic operation is given as follows.

\begin{framed} 
\noindent \textbf{ObservabilityMatrixSymDIP.m} \\
\noindent \%\% Double inverted pendulum parameters \\
syms m1  m2  L1  L2  g \\
A = [0, 1, 0, 0, 0, 0; ... \\
$~~~~$ (m1+m2)*g/(m1*L1), 0, -m2*g/(m1*L1), 0, 0, 0; ... \\
$~~~~$ 0, 0, 0, 1, 0, 0; ... \\
$~~~~$ -(m1+m2)*g/(m1*L2), 0, (m1+m2)*g/(m1*L2), 0, 0, 0; ... \\
$~~~~$ 0, 0, 0, 0, 0, 1; ... \\
$~~~~$ 0, 0, 0, 0, 0, 0]; \\
H = sym([1, 0, 0, 0, 0, 0; 0, 0, 0, 0, 1, 0]); \\
n = size(A,1); \% State dimension \\
nz = size(H,1); \% Measurement dimension \\
 \\
OM = sym(zeros(n*nz, n)); OM(1:nz,:) = H; \\
for k = 2:n \\
$~~~~$ OM((k*nz-nz+1):(k*nz),:) = OM((k*nz-2*nz+1):(k*nz-nz),:)*A; \\
end \% OM = [H; H*A; H*A\^{}2; H*A\^{}3; H*A\^{}4; H*A\^{}5]; \\
fprintf('Observability matrix: '); OM \\
 \\
\%\% Check if the observability matrix is of full rank \\
if (rank(OM) == n) \\
$~~~~$ fprintf('The double inverted pendulum is observable$\backslash$n'); \\
else \\
$~~~~$ fprintf('The double inverted pendulum is unobservable$\backslash$n'); \\
end
\end{framed}

After trying above Matlab code, readers will find that the double inverted pendulum is indeed observable given measurements of first inverted pendulum angle $\theta_1$ and cart position $x$. In other words, the complete double inverted pendulum state $\mathbf{x}$ can be inferred or revealed with measurements only of first inverted pendulum angle $\theta_1$ and cart position $x$.

\subsubsection*{Application: low-speed autonomous vehicle observability analysis}

Apply the observability criterion to determine observability of the low-speed autonomous vehicle lateral control system that adopts linear state-space modelling described by (1.15). 
\footnote{Namely (1.15) in Chapter 1 of the author's works \cite{Li2026ACTPA_SJTU_2, Li2026ACTPA_SJTU_1}. Note that this article is Chapter 3 of the works.}
The state transition matrix $\mathbf{A}$ is
\begin{align*}
\mathbf{A} = \begin{bmatrix} 0 & v & 0 \\ 0 & 0 & \frac{v}{L} \\ 0 & 0 & -\frac{1}{\tau_{\beta}} \end{bmatrix}.
\end{align*}
Suppose we can measure the vehicle lateral position $y$ directly. The measurement model is
\begin{align*}
\mathbf{z} = \begin{bmatrix} 1 & 0 & 0 \end{bmatrix} \mathbf{x} \equiv \mathbf{H} \mathbf{x}.
\end{align*}
Compute the observability matrix via (\ref{eq:observability_matrix}) as
\begin{align*}
\mathbf{O}_{\mathbf{H}, \mathbf{A}} = \begin{bmatrix} \mathbf{H} \\ \mathbf{H} \mathbf{A} \\ \mathbf{H} \mathbf{A}^2 \end{bmatrix} = \begin{bmatrix} 1 & 0 & 0 \\ 0 & v & 0 \\ 0 & 0 & \frac{v^2}{L} \end{bmatrix},
\end{align*}
which is of full rank in terms of row vectors. So the vehicle lateral control system is indeed observable given measurements of vehicle lateral position $y$. In other words, the complete vehicle lateral state $\mathbf{x}$ can be inferred or revealed with measurements only of vehicle lateral position $y$.

\section{Kalman filter}  \label{sec:Kalman_filter}

Given a control system that adopts linear state-space modelling described by (\ref{eq:state_differential_equation_linear}), with its state denoted as $\mathbf{x}$ and its control input to the target process denoted as $\mathbf{u}$, i.e.
\begin{align*}
\frac{\mathrm{d}}{\mathrm{d} t} \mathbf{x} = \mathbf{A} \mathbf{x} + \mathbf{B} \mathbf{u}.
\end{align*}
Besides, the control system adopts linear measurement modelling described by (\ref{eq:SDE_linear_measurement})
\begin{align*}
\mathbf{z} = \mathbf{H} \mathbf{x}.
\end{align*}
We hardly rely on solving (\ref{eq:SDE_linear_initial_state_equation}) directly to estimate the state. In fact, much more realistic and effective state estimation methods are available, among which the Kalman filter \cite{Kalman1960} is probably the most representative one.

\subsection{Recursive estimation}  \label{sec:recursive_estimation}

\begin{figure}[h!]
\begin{center}
\includegraphics[width=0.9\columnwidth]{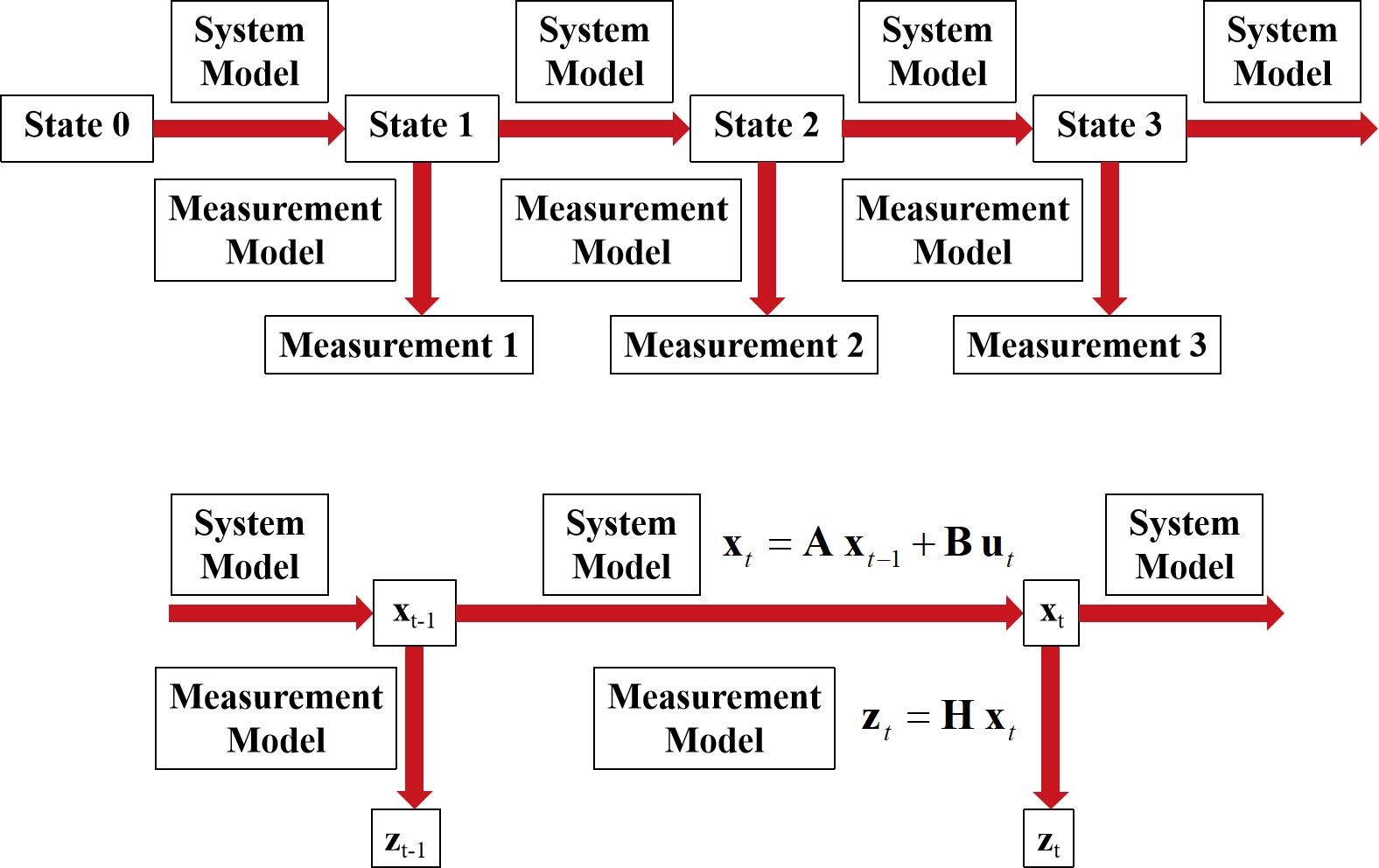}
\end{center}
\caption{Recursive estimation: (top) dynamic Bayesian network (DBN) perspective; (bottom) instantiation as the Kalman filter}
\label{fig:recursive_estimation_KF}
\end{figure}

In many applications especially real-time applications, the state evolves dynamically over time and its corresponding measurements are also obtained dynamically. Whenever a new measurement is available, instead of using all historical measurements to estimate current state from scratch, we can fairly take advantage of last state estimate which contains historical information implicitly and fuse it only with the new measurement to obtain current state estimate. This is a general methodology called \textbf{recursive estimation} \cite{Li2022FARET_2, Li2022FARET_1}, which can render the estimation process much more efficient but essentially no less effective than estimating the state from scratch with all historical measurements. The basic spirit of recursive estimation can be illustrated by a \textit{dynamic Bayesian network} (DBN) \cite{Murphy2002} as in the top sub-figure of Figure \ref{fig:recursive_estimation_KF}. 

\subsubsection*{Bayesian inference}

Conventionally, a state is denoted as $\mathbf{x}$ whereas a measurement is denoted as $\mathbf{z}$ --- A lower-case letter in bold-font style usually represents a vector but may compatibly represent a scalar value as well --- Nowadays, since digital computational devices such as computers are used ubiquitously in engineering activities, a system normally adopts digital processing and operates at discrete time instants called \textbf{control periods}. So recursive estimation is also performed discretely at control periods. We use integer subscripts to denote indices of control periods. For example, $\mathbf{x}_t$ denotes the state at the $t$-th control period or \textit{at time $t$} for simplicity; $\mathbf{x}_{t_1:t_2}$ denotes the states from the $t_1$-th control period to the $t_2$-th control period or \textit{from time $t_1$ to $t_2$}. Similarly, $\mathbf{z}_t$ denotes the measurement \textit{at time $t$} and $\mathbf{z}_{t_1:t_2}$ denotes the measurements \textit{from time $t_1$ to $t_2$}.

The \textbf{Markov assumption} is an important and axiom-alike assumption in probability theory: it states that \textit{the future is independent of the past given the present}. Based on the Markov assumption, we have Bayesian inference as (\ref{eq:predictionDBN}) and (\ref{eq:updateDBN})
\begin{align} \label{eq:predictionDBN}
p(\mathbf{x}_t | \mathbf{z}_{1:t-1}) &= \int_{\mathbf{x}_{t-1}} p(\mathbf{x}_t | \mathbf{x}_{t-1}, \mathbf{z}_{1:t-1}) p(\mathbf{x}_{t-1} | \mathbf{z}_{1:t-1})\, \mathrm{d} \mathbf{x}_{t-1}  \nonumber \\
& \quad (\mbox{total probability law}) \nonumber \\
&= \int_{\mathbf{x}_{t-1}} p(\mathbf{x}_t | \mathbf{x}_{t-1}) p(\mathbf{x}_{t-1} | \mathbf{z}_{1:t-1})\, \mathrm{d} \mathbf{x}_{t-1},
\end{align}
\begin{align} \label{eq:updateDBN}
p(\mathbf{x}_t | \mathbf{z}_{1:t}) &= \frac{p(\mathbf{z}_t | \mathbf{x}_t, \mathbf{z}_{1:t-1}) p(\mathbf{x}_t | \mathbf{z}_{1:t-1})} {p(\mathbf{z}_t | \mathbf{z}_{1:t-1})} = \frac{p(\mathbf{z}_t | \mathbf{x}_t) p(\mathbf{x}_t | \mathbf{z}_{1:t-1})} {p(\mathbf{z}_t | \mathbf{z}_{1:t-1})},
\end{align}
where $p(\mathbf{z}_t | \mathbf{z}_{1:t-1})$ is a normalization constant computed as
\begin{equation*}
p(\mathbf{z}_t | \mathbf{z}_{1:t-1}) = \int_{\mathbf{x}_t} p(\mathbf{z}_t | \mathbf{x}_t) p(\mathbf{x}_t | \mathbf{z}_{1:t-1})\, \mathrm{d}\mathbf{x}_t.
\end{equation*} 
The formalism (\ref{eq:predictionDBN}) and (\ref{eq:updateDBN}) describes the mechanism of inferring the new state distribution $p(\mathbf{x}_t | \mathbf{z}_{1:t})$ from the old state distribution $p(\mathbf{x}_{t-1} | \mathbf{z}_{1:t-1})$ and the new measurement $\mathbf{z}_t$ recursively. (\ref{eq:predictionDBN}) serves as the \textbf{prediction} step that \textit{predicts} the \textit{a prior} distribution of the state $\mathbf{x}_t$. (\ref{eq:updateDBN}) serves as the \textbf{update} step that \textit{updates} the state distribution to form the \textit{a posterior} state distribution by taking the newly available measurement $\mathbf{z}_t$ into account. The prediction step (\ref{eq:predictionDBN}) and the update step (\ref{eq:updateDBN}) form a generic two-steps methodology of recursive estimation using Bayesian inference. 

The generic Bayesian inference methodology described by (\ref{eq:predictionDBN}) and (\ref{eq:updateDBN}) is mainly of theoretical value, but is hardly put into practice directly due to difficulty in computing the involved integrals generally. Certain assumption is normally needed to instantiate this generic methodology. If the \textit{linear-Gaussian assumption} is adopted, then the Bayesian inference methodology can be instantiated as a famous recursive estimation method namely the \textbf{Kalman filter} that will be presented immediately. If the \textit{sampling assumption} namely the assumption that the state distribution can be fairly represented by a given number of state samples is adopted, then the Bayesian inference methodology can be instantiated as another representative recursive estimation method that will be presented in Section \ref{sec:particle_filter}.

\subsubsection*{Linear-Gaussian assumption}

The ``linear'' aspect of the linear-Gaussian assumption consists in \textit{linear state-space modelling}, namely (\ref{eq:state_differential_equation_linear}) for the system model and (\ref{eq:SDE_linear_measurement}) for the measurement model. The ``Gaussian'' aspect of the linear-Gaussian assumption consists in \textit{approximating all involved distributions in the Bayesian inference formalism by Gaussian distributions}. 

More specifically concerning the ``Gaussian'' aspect, use $\hat{\mathbf{x}}$ namely a \textit{hat} on the state $\mathbf{x}$ to denote the state estimate namely the estimated expectation of the state distribution $p(\mathbf{x}_t | \mathbf{z}_{1:t})$. 
A covariance associated with the state estimate is adopted to characterize estimate uncertainty
\footnote{In practical applications, any estimate itself is of no practical use if there is no specification of its uncertainty. For example, if a responsible economist estimates that the product price will decrease, then we are likely to trust the economist's estimate. In contrast, if a stranger estimates that the product price will decrease, then we are unlikely to treat the estimate seriously. The economist and the stranger provide the same estimate but our attitude towards their estimates are different. What makes a difference here is estimate uncertainty: we tend to believe that the economist's estimate has small uncertainty whereas the stranger's estimate has large uncertainty. From this example we can see that an estimate itself does not convey meaningful information; it is the \textbf{estimate} together with its \textbf{uncertainty} that conveys meaningful information --- It is true that people may not always specify estimate uncertainty explicitly, but instead may bear some implicit specification of estimate uncertainty in mind.}. 
For the state $\mathbf{x}$ and its estimate $\hat{\mathbf{x}}$, the covariance $\hat{\mathbf{\Sigma}}_{\mathbf{x}}$ or simply $\hat{\mathbf{\Sigma}}$ denotes the uncertainty of the state estimate $\hat{\mathbf{x}}$. In some matrix norm sense \cite{Horn2012, Golub1996}, the larger the state estimate covariance $\hat{\mathbf{\Sigma}}_{\mathbf{x}}$ or $\hat{\mathbf{\Sigma}}$ is, the larger the uncertainty of the state estimate $\hat{\mathbf{x}}$ is. Then the state distribution $p(\mathbf{x}_t | \mathbf{z}_{1:t})$ is approximated by a Gaussian distribution
\begin{equation}  \label{eq:state_distribution_Gaussian_approx}
p(\mathbf{x}_t | \mathbf{z}_{1:t}) = N(\hat{\mathbf{x}}_t, \hat{\mathbf{\Sigma}}_t) \quad \mbox{i.e.} \quad \mathbf{x}_t \sim N(\hat{\mathbf{x}}_t, \hat{\mathbf{\Sigma}}_t).
\end{equation}
In the Gaussian distribution notation $N(\cdot, \cdot)$, the random variable is saved for simplicity, if there is not any ambiguity of the random variable. On the other hand, the random variable may also be added explicitly into the Gaussian distribution notation as
\begin{equation}  \label{eq:Gaussian_distribution_notation}
N(\mathbf{x} \mbox{ } | \mu, \mathbf{\Sigma}) \equiv N(\mu, \mathbf{\Sigma}) \equiv \frac{1}{\sqrt{(2 \pi)^n | \mathbf{\Sigma} |}} \mathrm{e}^{-\frac{1}{2} (\mathbf{x} - \mu)^\mathrm{T} \mathbf{\Sigma}^{-1} (\mathbf{x} - \mu)}.
\end{equation}
So the Gaussian distribution $N(\hat{\mathbf{x}}_t, \hat{\mathbf{\Sigma}}_t)$ in (\ref{eq:state_distribution_Gaussian_approx}) can also have the random variable $\mathbf{x}_t$ explicitly expressed as
\begin{align*}
N(\hat{\mathbf{x}}_t, \hat{\mathbf{\Sigma}}_t) = N(\mathbf{x}_t \mbox{ } | \hat{\mathbf{x}}_t, \hat{\mathbf{\Sigma}}_t) = \frac{1}{\sqrt{(2 \pi)^n | \hat{\mathbf{\Sigma}}_t |}} \mathrm{e}^{-\frac{1}{2} (\mathbf{x}_t - \hat{\mathbf{x}}_t)^\mathrm{T} \hat{\mathbf{\Sigma}}_t^{-1} (\mathbf{x}_t - \hat{\mathbf{x}}_t)}.
\end{align*}
If there is only one random variable under consideration, then the simpler notation $N(\cdot, \cdot)$ may be adopted. In contrast, if there are different random variables under consideration, then the more detailed notation $N(\cdot \mbox{ } | \cdot, \cdot)$ had better be adopted for clarity.

Similarly, use the covariance $\mathbf{\Sigma_z}$ to denote the uncertainty of the measurement $\mathbf{z}$, and use the covariance $\mathbf{\Sigma}_{\mathbf{u}}$ to denote the uncertainty of the control input $\mathbf{u}$. By default, a variable ``wearing'' the \textit{hat} denotes certain estimate, whereas a variable without \textit{hat} denotes certain measurement or certain value that can be known without estimation.

\subsubsection*{Discrete-time version of linear state-space modelling}

Since recursive estimation is performed discretely at control periods, discrete-time versions of linear system modelling described by (\ref{eq:state_differential_equation_linear}) and linear measurement modelling described by (\ref{eq:SDE_linear_measurement}) are needed. 
The discrete-time version of (\ref{eq:state_differential_equation_linear}) can be derived according to (\ref{eq:state_differential_equation_linear_solution})
\begin{equation}  \label{eq:SDE_linear_discrete}
\mathbf{x}_t = \mathrm{e}^{\mathbf{A} \Delta t} \mathbf{x}_{t-1} + [\int_0^{\Delta t} \mathrm{e}^{\mathbf{A} (\Delta t - \tau)} \mathbf{B} \mathrm{d} \tau] \mathbf{u}_t \equiv \mathbf{A}^* \mathbf{x}_{t-1} + \mathbf{B}^* \mathbf{u}_t,
\end{equation}
where $\Delta t$ denotes the control period and
\begin{align*}
\mathbf{A}^* &\equiv \mathrm{e}^{\mathbf{A} \Delta t} = \sum_{k=0}^{\infty} \frac{\mathbf{A}^k \Delta t^k}{k!} = \mathbf{I} + \mathbf{A} \Delta t + \cdots, \\
\mathbf{B}^* &\equiv \int_0^{\Delta t} \mathrm{e}^{\mathbf{A} (\Delta t - \tau)} \mathbf{B} \mathrm{d} \tau = (\int_0^{\Delta t} \mathrm{e}^{\mathbf{A} \tau} \mathrm{d} \tau) \mathbf{B} = (\int_0^{\Delta t} \sum_{k=0}^{\infty} \frac{\mathbf{A}^k \tau^k}{k!} \mathrm{d} \tau) \mathbf{B}  \\
  &= [\sum_{k=0}^{\infty} \frac{\mathbf{A}^k \Delta t^{k+1}}{(k+1)!}] \mathbf{B} = [\sum_{k=0}^{\infty} \frac{\mathbf{A}^k \Delta t^k}{(k+1)!}] \mathbf{B} \Delta t = (\mathbf{I} + \frac{\mathbf{A} \Delta t}{2} + \cdots) \mathbf{B} \Delta t.
\end{align*}
Derivation of (\ref{eq:SDE_linear_discrete}) follows the natural assumption that the control input during current control period $[(t-1) \Delta t, t \Delta t]$ is constantly $\mathbf{u}_t$. For formalism simplicity yet without causing confusion, we abuse notations $\mathbf{A}$ and $\mathbf{B}$ to replace respectively $\mathbf{A}^*$ and $\mathbf{B}^*$ in (\ref{eq:SDE_linear_discrete}) and the discrete-time system model is formalized as
\begin{equation}  \label{eq:SDE_linear_discrete2}
\mathbf{x}_t = \mathbf{A} \mathbf{x}_{t-1} + \mathbf{B} \mathbf{u}_t.
\end{equation}
The discrete-time version of (\ref{eq:SDE_linear_measurement}) is obtained simply by adding time indices to relevant variables
\begin{equation}  \label{eq:SDE_linear_measurement_discrete}
\mathbf{z}_t = \mathbf{H} \mathbf{x}_t.
\end{equation}

\subsection{Prediction-update formalism}

The Kalman filter is a milestone instantiation of the recursive estimation methodology for applications where both the system model and the measurement model can be fairly linearized, as illustrated in the bottom sub-figure of Figure \ref{fig:recursive_estimation_KF}.
The Kalman filter consists of two essential steps: \textbf{prediction} and \textbf{update}. The prediction step propagates the estimate from time $t-1$ to time $t$, namely, to predict \{$\mathbf{x}_t$, $\mathbf{\Sigma}_t$\} from \{$\hat{\mathbf{x}}_{t-1}$, $\hat{\mathbf{\Sigma}}_{t-1}$\} according to the system model. To avoid notation confusion, we denote the predicted estimate with a \textit{bar} on variable, as \{$\bar{\mathbf{x}}_t$, $\bar{\mathbf{\Sigma}}_t$\}. The predicted estimate is also called the \textit{a priori} estimate. The update step updates the \textit{a priori} estimate \{$\bar{\mathbf{x}}_t$, $\bar{\mathbf{\Sigma}}_t$\} with the new measurement $\mathbf{z}_t$ to obtain the \textit{a posteriori} estimate \{$\hat{\mathbf{x}}_t$, $\hat{\mathbf{\Sigma}}_t$\} which is the final estimate at current time $t$. 

The original Kalman filter relies on the linear-Gaussian assumption, namely the system model and the measurement model are formalized as linear relationships (\ref{eq:SDE_linear_discrete2}) and (\ref{eq:SDE_linear_measurement_discrete}) respectively, where all involved random variables follow the \textit{Gaussian} or \textit{normal} distribution assumption. In other words, we have
\begin{align*}
\mathbf{x}_t \sim N(\hat{\mathbf{x}}_t, \hat{\mathbf{\Sigma}}_t), \quad  \mathbf{x}_{t-1} \sim N(\hat{\mathbf{x}}_{t-1}, \hat{\mathbf{\Sigma}}_{t-1}), \quad \mathbf{u}_t \sim N(\hat{\mathbf{u}}_t, \mathbf{\Sigma_u}), \quad \mathbf{z}_t \sim N(\mathbf{H} \mathbf{x}_t, \mathbf{\Sigma_z}).
\end{align*}
Here, $\hat{\mathbf{u}}$ with \textit{hat} denotes monitored control input, whereas $\mathbf{u}$ without \textit{hat} denotes real control input which is used only in theoretical sense in system modelling. It is always the monitored control input $\hat{\mathbf{u}}$ that is actually used in concrete procedures of recursive estimation and is used somehow as time-variant parameters of the system model. Without causing confusion, we abuse the notation $\mathbf{u}$ to denote both real control input that is used theoretically in system modelling and monitored control input that is used somehow as system model parameters in concrete procedures of recursive estimation. 

The prediction and update steps of the Kalman filter are realized as (\ref{eq:KFprediction}) and (\ref{eq:KFupdate}) respectively.

\noindent \textbf{Prediction}: $\mathbf{x}_t$ (\textit{a priori}) $\sim N(\bar{\mathbf{x}}_t,\bar{\mathbf{\Sigma}}_t)$ 
\begin{subequations}  \label{eq:KFprediction}
\begin{align} 
\bar{\mathbf{x}}_t &= \mathbf{A} \hat{\mathbf{x}}_{t-1} + \mathbf{B} \mathbf{u}_t,  \\
\bar{\mathbf{\Sigma}}_t &= \mathbf{A} \hat{\mathbf{\Sigma}}_{t-1} \mathbf{A}^{\mathrm{T}} + \mathbf{B} \mathbf{\Sigma_u} \mathbf{B}^{\mathrm{T}}.
\end{align}
\end{subequations}

\noindent \textbf{Update}: $\mathbf{x}_t$ (\textit{a posteriori}) $\sim N(\hat{\mathbf{x}}_t,\hat{\mathbf{\Sigma}}_t)$ 
\begin{subequations}  \label{eq:KFupdate}
\begin{align} 
\mathbf{K} &= \bar{\mathbf{\Sigma}}_t \mathbf{H}^{\mathrm{T}} (\mathbf{H} \bar{\mathbf{\Sigma}}_t \mathbf{H}^{\mathrm{T}} + \mathbf{\Sigma_z})^{-1},  \\
\hat{\mathbf{x}}_t &= \bar{\mathbf{x}}_t + \mathbf{K} (\mathbf{z}_t - \mathbf{H} \bar{\mathbf{x}}_t),  \\
\hat{\mathbf{\Sigma}}_t &= (\mathbf{I} - \mathbf{K H}) \bar{\mathbf{\Sigma}}_t.
\end{align}
\end{subequations}

The formalism (\ref{eq:KFprediction}) and (\ref{eq:KFupdate}) is a commonly-adopted version of the original Kalman filter, whereas other formalism versions also exist. Reasoning of (\ref{eq:KFprediction}) is straightforward, whereas a complete understanding of (\ref{eq:KFupdate}) necessitates a bit more complicated derivation. 

\subsubsection*{$\alpha$-$\beta$ filter and $\alpha$-$\beta$-$\gamma$ filter}

In some practical applications, the Kalman gain $\mathbf{K}$ tends to converge roughly to certain constant after convergence of the state estimation process. In such kind of application context, the Kalman gain $\mathbf{K}$ may be empirically set to the constant, instead of being re-computed dynamically at each control period. This practice leads to a simplified version of the Kalman filter that is computationally much more efficient than the original one but may still perform somewhat as desirably as the original one.

A typical example of such kind of application context is \textit{object tracking} which will be mentioned in Section \ref{sec:interacting_multiple_model}. The Kalman filter using the constant velocity model has a simplified version that uses a constant Kalman gain
\begin{equation}  \label{eq:alpha_beta_filter_K}
\mathbf{K} \equiv \begin{bmatrix} K_p \\ K_v \end{bmatrix} = \begin{bmatrix} \alpha \\ \beta / \Delta T \end{bmatrix}.
\end{equation}
The Kalman filter using the constant acceleration model has a simplified version that uses a constant Kalman gain
\begin{equation}  \label{eq:alpha_beta_gamma_filter_K}
\mathbf{K} \equiv \begin{bmatrix} K_p \\ K_v \\ K_a \end{bmatrix} = \begin{bmatrix} \alpha \\ \beta / \Delta T \\ \gamma / 2 \Delta T^2 \end{bmatrix}.
\end{equation}
The Kalman filter versions that use constant Kalman gains specified in (\ref{eq:alpha_beta_filter_K}) and (\ref{eq:alpha_beta_gamma_filter_K}) are also called the \textit{$\alpha$-$\beta$ filter} and the \textit{$\alpha$-$\beta$-$\gamma$ filter} \cite{Kalata1984} respectively.

\subsection{Data fusion perspective}

The update step of the Kalman filter embodies an important spirit of data fusion
\footnote{Based on the Gaussian distribution assumption, (\ref{eq:KFupdate}) can be derived strictly via Bayesian inference. However, we would rather explain (\ref{eq:KFupdate}) from data fusion perspective to intuitively highlight the essence of this famous recursive estimation method.}.
On data fusion, we would like to begin with a daily-life example. Suppose we have two thermometers of the same quality to measure our room temperature. One thermometer indicates a value of 23 degrees centigrade, whereas the other indicates 27 degrees centigrade. Since the two indicated temperature values are different, a question arises naturally: what would be the most-likely room temperature?

Apparently, we should neither trust the first thermometer only nor trust the second thermometer only; instead, we may form our answer to the question by incorporating information conveyed by both thermometers. Since the two thermometers are of the same quality, a natural intuition is to take an average of the two indicated temperature values and we have our answer 
\begin{align*}
\frac{23 + 27}{2} = 25
\end{align*}
degrees centigrade. In other words, we \textit{fuse} the two indicated temperature values by averaging them.

Now suppose the thermometers are of different qualities; the first thermometer is better and its error level is only a third of the second thermometer's error level. In this case, what would be the most-likely room temperature? Since the first thermometer's error level is only a third of the second thermometer's error level; in other words, the first thermometer's quality level is three times the second thermometer's quality level. By intuition, we may give a confidence weight of three to the first thermometer and a confidence weight of one to the second thermometer. Thereafter we take a weighted average of the two indicated temperature values and our answer to the question is 
\begin{align*}
\frac{23 \times 3 + 27 \times 1}{3 + 1} = 24
\end{align*}
degrees centigrade.

Above practice of taking a weighted average has its probability theory foundation. We can treat the two thermometers' indicated temperature values as two independent random variables \{$x_1$, $\sigma_1^2$\} and \{$x_2$, $\sigma_2^2$\} with $\sigma_1^2 = \sigma_2^2/3$. Suppose the fusion weights assigned to the two thermometers are $k_1$ and $k_2$ respectively and 
\begin{align*}
k_1 + k_2 = 1. 
\end{align*}
We follow the Gaussian distribution assumption and aims at finding the optimal weights in the maximum likelihood sense \cite{Durrett2019} as
\begin{align*}
&p(x_1, x_2 | k_1 x_1 + k_2 x_2) \\
&= p(x_1 | k_1 x_1 + k_2 x_2) p(x_2 | k_1 x_1 + k_2 x_2) \\
&\propto exp(-\frac{(k_1 x_1 + k_2 x_2 - x_1)^2}{\sigma_1^2}) exp(-\frac{(k_1 x_1 + k_2 x_2 - x_2)^2}{\sigma_2^2}) \\
&= exp(- \frac{[k_2 (x_2 - x_1)]^2}{\sigma_1^2} - \frac{[k_1 (x_1 - x_2)]^2}{\sigma_2^2}) \\
&= exp(- (\frac{k_2^2}{\sigma_1^2} + \frac{k_1^2}{\sigma_2^2}) (x_1 - x_2)^2).
\end{align*}
Note that
\begin{align*}
(\sigma_1^2 + \sigma_2^2) (\frac{k_2^2}{\sigma_1^2} + \frac{k_1^2}{\sigma_2^2}) \geq (\frac{k_2}{\sigma_1} \sigma_1 + \frac{k_1}{\sigma_2} \sigma_2)^2 = (k_2 + k_1)^2 = 1,
\end{align*}
derivation of which resorts to the \textit{Cauchy inequality} \cite{Mitrinovic1970}. so
\begin{align*}
\frac{k_2^2}{\sigma_1^2} + \frac{k_1^2}{\sigma_2^2} \geq \frac{1}{\sigma_1^2 + \sigma_2^2},
\end{align*}
where the equality condition holds if and only if 
\begin{align*}
\frac{k_2}{\sigma_1}/\sigma_1 = \frac{k_1}{\sigma_2}/\sigma_2 \Rightarrow \frac{k_1}{k_2} = \frac{\sigma_2^2}{\sigma_1^2} = 3.
\end{align*}
Therefore, when the weight assigned to the first thermometer is three times the weight assigned to the second thermometer, the weighted average of the two thermometers' indicated temperature values is of the maximum likelihood.

\subsection{Optimal weighted average}  \label{sec:opt_wgt_average}

Consider fusion of two generic source estimates \{$\mathbf{x}_1$, $\mathbf{\Sigma}_1$\} and \{$\mathbf{x}_2$, $\mathbf{\Sigma}_2$\} of a state $\mathbf{x}$. Their covariance matrices $\mathbf{\Sigma}_1$ and $\mathbf{\Sigma}_2$ reflect their uncertainty respectively, and accordingly their covariance inverses $\mathbf{\Sigma}_1^{-1}$ and $\mathbf{\Sigma}_2^{-1}$ can be used to reflect their \textit{quality} respectively. Such \textit{quality} is also known as \textit{information} and such covariance inverses are called \textit{information matrices}.

Compute the \textit{weighted average} of the two estimates \{$\mathbf{x}_1$, $\mathbf{\Sigma}_1$\} and \{$\mathbf{x}_2$, $\mathbf{\Sigma}_2$\} as
\begin{subequations}  \label{eq:WAfullDim}
\begin{align} 
\hat{\mathbf{\Sigma}}^{-1} &= \mathbf{\Sigma}_1^{-1} + \mathbf{\Sigma}_2^{-1}  \\
\hat{\mathbf{x}} &= \hat{\mathbf{\Sigma}} (\mathbf{\Sigma}_1^{-1} \mathbf{x}_1 + \mathbf{\Sigma}_2^{-1} \mathbf{x}_2)
\end{align}
\end{subequations}
The fusion formula (\ref{eq:WAfullDim}) is intuitively reasonable. In fact, its optimality among all possible weighted averages can also be proved --- The weighted averages mentioned here actually mean linear weighted averages, whereas the concept \textit{weighted average} itself can mean a nonlinear weighted average as well from pure mathematical perspective. However, in practical applications, a nonlinear weighted average of two source estimates is usually of no practical meaning. For example, what would be the meaning of the square or root of a generic state estimate vector? So we neglect consideration of nonlinear weighted averages here.

\begin{proof}
Suppose the fused estimate $\hat{\mathbf{x}}$ is a weighted average of $\mathbf{x}_1$ and $\mathbf{x}_2$ as 
\begin{align*}
\hat{\mathbf{x}} = \mathbf{C} \mathbf{x}_1 + (\mathbf{I} - \mathbf{C}) \mathbf{x}_2
\end{align*}
and hence the covariance of $\hat{\mathbf{x}}$ is
\begin{align*}
\hat{\mathbf{\Sigma}} &= \mathbf{C} \mathbf{\Sigma}_1 \mathbf{C}^{\mathrm{T}} + (\mathbf{I} - \mathbf{C}) \mathbf{\Sigma}_2 (\mathbf{I} - \mathbf{C})^{\mathrm{T}}.
\end{align*}
The covariance $\hat{\mathbf{\Sigma}}$ had better be as less as possible, which implies the minimality of estimate uncertainty. So the optimal weight $\mathbf{C}_{opt}$ can be determined by solving the optimization problem
\begin{equation}  \label{eq:opt_wgt_average}
\mathbf{C}_{opt} = \arg \min_{\mathbf{C}} \hat{\mathbf{\Sigma}}(\mathbf{C}) = \arg \min_{\mathbf{C}} (\mathbf{C} \mathbf{\Sigma}_1 \mathbf{C}^{\mathrm{T}} + (\mathbf{I} - \mathbf{C}) \mathbf{\Sigma}_2 (\mathbf{I} - \mathbf{C})^{\mathrm{T}}).
\end{equation}

Consider differentiation of $\hat{\mathbf{\Sigma}}(\mathbf{C})$ with respect to $\mathbf{C}$ as
\begin{align*}
\Delta \hat{\mathbf{\Sigma}}(\mathbf{C}) &= 2 \mathbf{C} \mathbf{\Sigma}_1 \Delta \mathbf{C}^{\mathrm{T}} + 2 (\mathbf{I} - \mathbf{C}) \mathbf{\Sigma}_2 (- \Delta \mathbf{C})^{\mathrm{T}} \\
 &= 2 [ \mathbf{C} (\mathbf{\Sigma}_1 + \mathbf{\Sigma}_2) - \mathbf{\Sigma}_2 ] \Delta \mathbf{C}^{\mathrm{T}}.
\end{align*}
The variation $\Delta \mathbf{C}$ can be arbitrary; by the first optimality condition, we have
\begin{align*}
\mathbf{C}_{opt} (\mathbf{\Sigma}_1 + \mathbf{\Sigma}_2) - \mathbf{\Sigma}_2 = 0.
\end{align*}
We may neglect mathematical ambiguity in above definition of the objective function, because it has no essential influence on deriving the optimal weight --- In fact, if 
\begin{align*}
\mathbf{C}_{opt} (\mathbf{\Sigma}_1 + \mathbf{\Sigma}_2) - \mathbf{\Sigma}_2 \not= 0, 
\end{align*}
then set 
\begin{align*}
\Delta \mathbf{C} = \mathbf{C}_{opt} (\mathbf{\Sigma}_1 + \mathbf{\Sigma}_2) - \mathbf{\Sigma}_2. 
\end{align*}
So 
\begin{align*}
\Delta \hat{\mathbf{\Sigma}}(\mathbf{C}) = 2 \Delta \mathbf{C} \Delta \mathbf{C}^{\mathrm{T}} \geq 0,
\end{align*}
i.e. positive semi-definite, which contradicts optimality of $\mathbf{C}_{opt}$.

The first optimality condition implies
\begin{align*}
\mathbf{C}_{opt} &= \mathbf{\Sigma}_2 (\mathbf{\Sigma}_1 + \mathbf{\Sigma}_2)^{-1} = (\mathbf{\Sigma}_1^{-1} + \mathbf{\Sigma}_2^{-1})^{-1} \mathbf{\Sigma}_1^{-1}, \\
\mathbf{I} - \mathbf{C}_{opt} &= \mathbf{\Sigma}_1 (\mathbf{\Sigma}_1 + \mathbf{\Sigma}_2)^{-1} = (\mathbf{\Sigma}_1^{-1} + \mathbf{\Sigma}_2^{-1})^{-1} \mathbf{\Sigma}_2^{-1}, \\
\hat{\mathbf{\Sigma}} &= \mathbf{C}_{opt} \mathbf{\Sigma}_1 \mathbf{C}_{opt}^{\mathrm{T}} + (\mathbf{I} - \mathbf{C}_{opt}) \mathbf{\Sigma}_2 (\mathbf{I} - \mathbf{C}_{opt})^{\mathrm{T}} = (\mathbf{\Sigma}_1^{-1} + \mathbf{\Sigma}_2^{-1})^{-1},  \\
\hat{\mathbf{x}} &= \mathbf{C}_{opt} \mathbf{x}_1 + (\mathbf{I} - \mathbf{C}_{opt}) \mathbf{x}_2  \\
  &= (\mathbf{\Sigma}_1^{-1} + \mathbf{\Sigma}_2^{-1})^{-1} \mathbf{\Sigma}_1^{-1} \mathbf{x}_1 + (\mathbf{\Sigma}_1^{-1} + \mathbf{\Sigma}_2^{-1})^{-1} \mathbf{\Sigma}_2^{-1} \mathbf{x}_2 \\
  &= \hat{\mathbf{\Sigma}} (\mathbf{\Sigma}_1^{-1} \mathbf{x}_1 + \mathbf{\Sigma}_2^{-1} \mathbf{x}_2).
\end{align*}
which is exactly the fusion formula (\ref{eq:WAfullDim}).
\end{proof}

\begin{figure}[h!]
\begin{center}
\includegraphics[width=0.8\columnwidth]{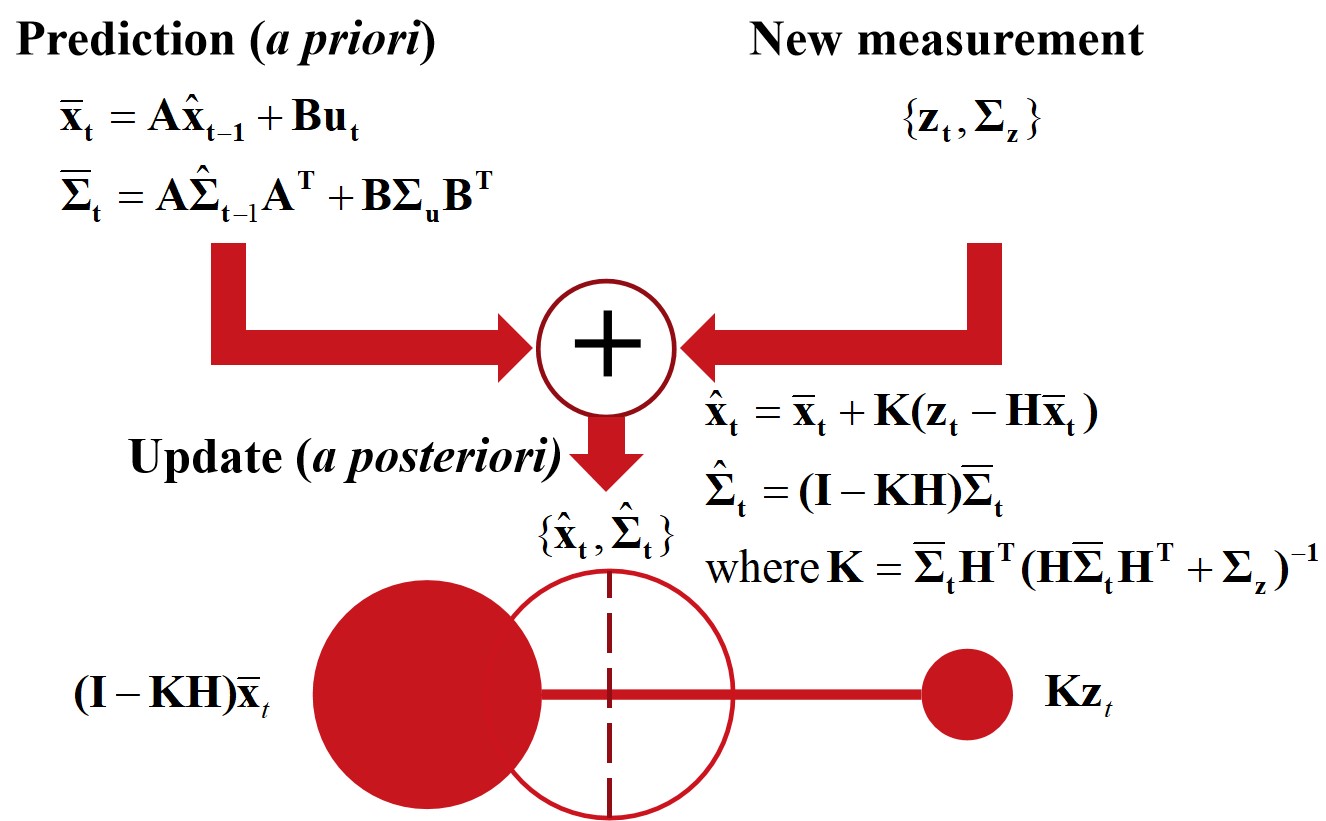}
\end{center}
\caption{Kalman filter update: weighted average of the predicted estimate \{$\bar{\mathbf{x}}_t$, $\bar{\mathbf{\Sigma}}_t$\} and the measurement \{$\mathbf{z}_t$, $\mathbf{\Sigma_z}$\}}
\label{fig:weighted_average_KF}
\end{figure}

In the update step of the Kalman filter, we can treat the predicted state estimate \{$\bar{\mathbf{x}}_t$, $\bar{\mathbf{\Sigma}}_t$\} as one source estimate and the new measurement \{$\mathbf{z}_t$, $\mathbf{\Sigma_z}$\} as another source estimate. What the update step of the Kalman filter does essentially is to fuse these two source estimates according to above weighted averaging strategy, as illustrated in Figure \ref{fig:weighted_average_KF}.

In practical applications, the measurement \{$\mathbf{z}_t$, $\mathbf{\Sigma_z}$\} may be a partial measurement, whereas the predicted state estimate \{$\bar{\mathbf{x}}_t$, $\bar{\mathbf{\Sigma}}_t$\} is always a complete state estimate. Consequently, we cannot always fuse \{$\bar{\mathbf{x}}_t$, $\bar{\mathbf{\Sigma}}_t$\} and \{$\mathbf{z}_t$, $\mathbf{\Sigma_z}$\} via (\ref{eq:WAfullDim}) directly; instead, we can resort to (\ref{eq:KFupdate}) to fuse \{$\bar{\mathbf{x}}_t$, $\bar{\mathbf{\Sigma}}_t$\} and \{$\mathbf{z}_t$, $\mathbf{\Sigma_z}$\}. It seems that (\ref{eq:KFupdate}) is more general than (\ref{eq:WAfullDim}); however, (\ref{eq:KFupdate}) can actually be derived from (\ref{eq:WAfullDim}). This is why we treat the update step of the Kalman filter essentially as weighted averaging of the predicted state estimate \{$\bar{\mathbf{x}}_t$, $\bar{\mathbf{\Sigma}}_t$\} and the new measurement \{$\mathbf{z}_t$, $\mathbf{\Sigma_z}$\}. 

Suppose there exist a complete source estimate \{$\mathbf{x}_1$, $\mathbf{\Sigma}_1$\} and a partial source estimate 
\begin{align*}
\{\mathbf{z} = \mathbf{H} \mathbf{x}_2, \mathbf{\Sigma_z}\}. 
\end{align*}
Augment $\mathbf{z}$ to a complete source estimate 
\begin{align*}
\begin{bmatrix} \mathbf{z} \\ \mathbf{z}_0 \end{bmatrix}
= \begin{bmatrix} \mathbf{H} \\ \mathbf{H}_0  \end{bmatrix} \mathbf{x}_2 
\end{align*}
where $\mathbf{z}_0$ is set arbitrary and its covariance is set as 
\begin{align*}
\mathbf{\Sigma}_{\mathbf{z}_0} = \infty.
\end{align*}
Besides, the augmented measurement matrix
\begin{align*}
\begin{bmatrix} \mathbf{H} \\ \mathbf{H}_0 \end{bmatrix}
\end{align*}
is assumed to be an invertible matrix, so we have
\begin{align*}
\mathbf{x}_2 &= \begin{bmatrix} \mathbf{H} \\ \mathbf{H}_0  \end{bmatrix}^{-1} \begin{bmatrix} \mathbf{z} \\ \mathbf{z}_0  \end{bmatrix},  \\
\mathbf{\Sigma}_2 &= \begin{bmatrix} \mathbf{H} \\  \mathbf{H}_0 \end{bmatrix}^{-1} \begin{bmatrix} \mathbf{\Sigma_z} & \mathbf{0} \\  \mathbf{0} & \infty \end{bmatrix} \begin{bmatrix} \mathbf{H} \\  \mathbf{H}_0 \end{bmatrix}^{-\mathrm{T}},
\end{align*}
and
\begin{align*}
\mathbf{\Sigma}_2^{-1} &= \begin{bmatrix} \mathbf{H} \\  \mathbf{H}_0 \end{bmatrix}^{\mathrm{T}} \begin{bmatrix} \mathbf{\Sigma}_\mathbf{z}^{-1} & \mathbf{0} \\  \mathbf{0} & \mathbf{0} \end{bmatrix} \begin{bmatrix} \mathbf{H} \\  \mathbf{H}_0 \end{bmatrix} = \mathbf{H}^{\mathrm{T}} \mathbf{\Sigma}_\mathbf{z}^{-1} \mathbf{H}.
\end{align*}

Fuse \{$\mathbf{x}_1$, $\mathbf{\Sigma}_1$\} and \{$\mathbf{x}_2$, $\mathbf{\Sigma}_2$\} via (\ref{eq:WAfullDim}) and we have
\begin{align*}
\mathbf{\Sigma} &= (\mathbf{\Sigma}_1^{-1} + \mathbf{\Sigma}_2^{-1})^{-1} = (\mathbf{\Sigma}_1^{-1} + \mathbf{H}^{\mathrm{T}} \mathbf{\Sigma}_\mathbf{z}^{-1} \mathbf{H})^{-1} \\
&= (\mathbf{I} + \mathbf{\Sigma}_1 \mathbf{H}^{\mathrm{T}} \mathbf{\Sigma}_\mathbf{z}^{-1} \mathbf{H})^{-1} \mathbf{\Sigma}_1 = [\sum\limits_{i=0}^\infty (-\mathbf{\Sigma}_1 \mathbf{H}^{\mathrm{T}} \mathbf{\Sigma}_\mathbf{z}^{-1} \mathbf{H})^i] \mathbf{\Sigma}_1 \\
&= \{\mathbf{I} - \mathbf{\Sigma}_1 \mathbf{H}^{\mathrm{T}} \mathbf{\Sigma}_\mathbf{z}^{-1} [\sum\limits_{i=0}^\infty (-\mathbf{H} \mathbf{\Sigma}_1 \mathbf{H}^{\mathrm{T}} \mathbf{\Sigma}_\mathbf{z}^{-1})^i] \mathbf{H}\} \mathbf{\Sigma}_1 \\
&= \{\mathbf{I} - \mathbf{\Sigma}_1 \mathbf{H}^{\mathrm{T}} \mathbf{\Sigma}_\mathbf{z}^{-1} (\mathbf{I} + \mathbf{H} \mathbf{\Sigma}_1 \mathbf{H}^{\mathrm{T}} \mathbf{\Sigma}_\mathbf{z}^{-1})^{-1} \mathbf{H}\} \mathbf{\Sigma}_1 \\
&= \{\mathbf{I} - \mathbf{\Sigma}_1 \mathbf{H}^{\mathrm{T}} (\mathbf{\Sigma}_\mathbf{z} + \mathbf{H} \mathbf{\Sigma}_1 \mathbf{H}^{\mathrm{T}})^{-1} \mathbf{H}\} \mathbf{\Sigma}_1 \\
&= (\mathbf{I} - \mathbf{K} \mathbf{H}) \mathbf{\Sigma}_1
\end{align*}
where
\footnote{It is worth noting that the infinite matrix series expansion in above derivation holds true only when the eigenvalues of relevant matrices are within the unit circle in the complex plane. But this does not influence the equality $\mathbf{\Sigma} = (\mathbf{I} - \mathbf{K} \mathbf{H}) \mathbf{\Sigma}_1$ which is equivalent to the equality of two finite-order polynomials. Since the equality holds true for infinite choices of matrix elements involved, the equality must always hold true.}
\begin{align*}
\mathbf{K} = \mathbf{\Sigma}_1 \mathbf{H}^{\mathrm{T}} (\mathbf{\Sigma}_\mathbf{z} + \mathbf{H} \mathbf{\Sigma}_1 \mathbf{H}^{\mathrm{T}})^{-1}.
\end{align*}
By (\ref{eq:WAfullDim}) we also have
\begin{align*}
\mathbf{x} &= \mathbf{\Sigma} (\mathbf{\Sigma}_1^{-1} \mathbf{x}_1 + \mathbf{\Sigma}_2^{-1} \mathbf{x}_2) = \mathbf{\Sigma} \mathbf{\Sigma}_1^{-1} \mathbf{x}_1 + \mathbf{\Sigma} \mathbf{\Sigma}_2^{-1} \mathbf{x}_2 \\
&= \mathbf{\Sigma} \mathbf{\Sigma}_1^{-1} \mathbf{x}_1 + (\mathbf{I} - \mathbf{\Sigma} \mathbf{\Sigma}_1^{-1}) \mathbf{x}_2 \\
&= (\mathbf{I} - \mathbf{K} \mathbf{H}) \mathbf{x}_1 + \mathbf{K} \mathbf{H} \mathbf{x}_2 \\
&= (\mathbf{I} - \mathbf{K} \mathbf{H}) \mathbf{x}_1 + \mathbf{K} \mathbf{H} \begin{bmatrix} \mathbf{H} \\ \mathbf{H}_0 \end{bmatrix}^{-1} \begin{bmatrix} \mathbf{z} \\ \mathbf{z}_0 \end{bmatrix} \\
&= (\mathbf{I} - \mathbf{K} \mathbf{H}) \mathbf{x}_1 + \mathbf{K} \begin{bmatrix} \mathbf{I} & \mathbf{0} \end{bmatrix} \begin{bmatrix} \mathbf{z} \\ \mathbf{z}_0 \end{bmatrix} \\
&= (\mathbf{I} - \mathbf{K} \mathbf{H}) \mathbf{x}_1 + \mathbf{K} \mathbf{z} \\
&= \mathbf{x}_1 + \mathbf{K} (\mathbf{z} - \mathbf{H} \mathbf{x}_1).
\end{align*}
If we substitute \{$\bar{\mathbf{x}}_t$, $\bar{\mathbf{\Sigma}}_t$\} for \{$\mathbf{x}_1$, $\mathbf{\Sigma}_1$\} and \{$\mathbf{z}_t$, $\mathbf{\Sigma_z}$\} for \{$\mathbf{z}$, $\mathbf{\Sigma_z}$\} in above derivation, we will have the update formalism (\ref{eq:KFupdate}).

Matlab code for implementing the Kalman filter is given as follows.

\begin{framed} 
\noindent \textbf{KF.m} \\
\noindent \% Suppose X1 is a complete estimate i.e. X1 = X\_true \\
\%$~~~~$ $~~~~$  X2 can be a partial estimate i.e. X2 = H * X\_true \\
function [X, P] = KF(X1, P1, X2, P2, H) \\
$~~~~$ I = eye(length(X1)); \\
$~~~~$ K = P1*H'*inv(H*P1*H'+P2); \\
$~~~~$ X = X1 + K*(X2-H*X1); \\
$~~~~$ P = (I-K*H)*P1; \\
end
\end{framed}

\section{Handle systems of special aspects}

\subsection{Handle non-determinism: interacting multiple model method}  \label{sec:interacting_multiple_model}

Consider the generic state differential equation
\begin{align}  \label{eq:state_differential_equation}
\frac{\mathrm{d}}{\mathrm{d} t} \mathbf{x} = f(\mathbf{x}, \mathbf{u})
\end{align}
and suppose its solution $\mathbf{x}$ parametrized in terms of the function $\mathbf{u}$ is implicitly denoted as
\begin{align*}
\mathbf{x}_t = \psi (\mathbf{x}_0, \mathbf{u}_{0:t}).
\end{align*}
Replace $t$ by $t-1$ and obtain
\begin{align*}
\mathbf{x}_{t-1} = \psi (\mathbf{x}_0, \mathbf{u}_{0:t-1}),
\end{align*}
from which the initial value $\mathbf{x}_0$ can be inversely recovered as
\begin{align*}
\mathbf{x}_0 = \psi^{-1} (\mathbf{x}_{t-1}, \mathbf{u}_{0:t-1}).
\end{align*}
So we have
\begin{align*}
\mathbf{x}_t = \psi (\mathbf{x}_0, \mathbf{u}_{0:t}) = \psi (\psi^{-1} (\mathbf{x}_{t-1}, \mathbf{u}_{0:t-1}), \mathbf{u}_{0:t}),
\end{align*}
which conveys
\begin{equation} \label{eq:generic_sys_model_discrete}
\mathbf{x}_t = g(\mathbf{x}_{t-1}, \mathbf{u}_t)
\end{equation}
where
\footnote{Note that state dynamics described by the generic state differential equation (\ref{eq:state_differential_equation}) apparently follows the \textit{Markov assumption}, namely that control input before current state has no influence on future states once current state is given.}
\begin{align*}
g(\mathbf{x}_{t-1}, \mathbf{u}_t) \equiv \psi (\psi^{-1} (\mathbf{x}_{t-1}, \mathbf{u}_{0:t-1}), \mathbf{u}_{0:t}).
\end{align*}
In fact, like (\ref{eq:SDE_linear_discrete}) or (\ref{eq:SDE_linear_discrete2}) is the discrete-time version of (\ref{eq:state_differential_equation_linear}) for linear system modelling, (\ref{eq:generic_sys_model_discrete}) is the discrete-time version of (\ref{eq:state_differential_equation}) for generic system modelling.

\subsubsection*{Non-deterministic system modelling and system model mismatch}

System modelling for a control system's target process is \textbf{deterministic} when there exists a system model which performs state prediction purely using known information. More specifically, given a generic system model described in (\ref{eq:generic_sys_model_discrete})
\begin{align*}
\mathbf{x}_t = g(\mathbf{x}_{t-1}, \mathbf{u}_t).
\end{align*}
When the control input $\mathbf{u}$ can be completely known in practice, then the system model (\ref{eq:generic_sys_model_discrete}) can predict $\mathbf{x}_t$ from $\mathbf{x}_{t-1}$ which can be known iteratively from previous states. In such case, the system model is deterministic. It is true that the prediction usually suffers from unknown \textit{ad hoc} random error in practices, yet the prediction's probabilistic characteristics are deterministic.

In contrast, system modelling for a control system's target process is \textbf{non-deterministic} when there is no system model which can perform state prediction purely using known information. More specifically, when the control input $\mathbf{u}$ can never be completely known in practice, then the system model (\ref{eq:generic_sys_model_discrete}) will always have some inherent uncertainty in its prediction of $\mathbf{x}_t$ from $\mathbf{x}_{t-1}$. In such case, the system model is non-deterministic --- Sometimes certain set of \textit{meta-parameters} (or \textit{meta-data}) $\mathbf{m}$ may be incorporated explicitly into the system model (\ref{eq:generic_sys_model_discrete}) as
\begin{align*}
\mathbf{x}_t = g(\mathbf{x}_{t-1}, \mathbf{u}_t, \mathbf{m})
\end{align*}
to characterize the model structure. However, we do not distinguish between the control input $\mathbf{u}$ and the meta-parameter set $\mathbf{m}$. We treat the meta-parameter set $\mathbf{m}$ as a special kind of control input that implicitly forms part of the generalized control input ``$\mathbf{u}$'' in the system model (\ref{eq:generic_sys_model_discrete}). So sometimes when the control input $\mathbf{u}$ can never be completely known, perhaps it is actually such meta-parameter part of control input that can never be completely known and hence incurs non-deterministic system modelling.

A typical example of non-deterministic modelling is the \textit{object tracking} system \cite{LiX2003part1, LiX2010part2, LiX2001part3, LiX2002part4, LiX2005part5}, the non-deterministic nature of which consists in two aspects: First, the object state $\mathbf{x}$ can theoretically have a dimensionality of infinity, i.e.
\begin{align*}
\mathbf{x} \equiv \begin{bmatrix} p & v & a & \cdots \end{bmatrix}^\mathrm{T},
\end{align*}
where $p$ denotes the object position, 
\begin{align*}
v = \frac{\mathrm{d} p}{\mathrm{d} t}
\end{align*}
denotes the object velocity, 
\begin{align*}
a = \frac{\mathrm{d}^2 p}{\mathrm{d} t^2}
\end{align*}
denotes the object acceleration, and 
\begin{align*}
\frac{\mathrm{d}^3 p}{\mathrm{d} t^3}, \quad \frac{\mathrm{d}^4 p}{\mathrm{d} t^4}, \quad \cdots
\end{align*}
denote the object position $p$'s even higher-order derivative terms until infinity.

Second, the motion pattern of the tracked object is \textit{a priori} unknown, and hence we cannot determine which dimensionality of the state is consistent with the actual motion pattern. For example, if the object is stationary, then the \textit{constant position (CP) model}
\begin{equation} \label{eq:modelCP_simple}
p_t = p_{t-1}
\end{equation}
should be adopted, where the one-dimensional state 
\begin{align*}
\mathbf{x} \equiv p
\end{align*}
is consistent with the actual motion pattern. 

If the object is moving at stable velocity, then the \textit{constant velocity (CV) model}
\begin{equation} \label{eq:modelCV_simple}
\begin{bmatrix} p_t \\ v_t \end{bmatrix} = \begin{bmatrix} 1 & \Delta T \\ 0 & 1 \end{bmatrix} \begin{bmatrix} p_{t-1} \\ v_{t-1} \end{bmatrix}
\end{equation}
should be adopted, where the two-dimensional state 
\begin{align*}
\mathbf{x} \equiv \begin{bmatrix} p & v \end{bmatrix}^\mathrm{T}
\end{align*}
is consistent with the actual motion pattern. 

If the object is at a stable acceleration stage, then the \textit{constant acceleration (CA) model}
\begin{equation} \label{eq:modelCA_simple}
\begin{bmatrix} p_t \\ v_t \\ a_t \end{bmatrix} = \begin{bmatrix} 1 & \Delta T & \Delta T^2/2 \\ 0 & 1 & \Delta T \\ 0 & 0 & 1 \end{bmatrix} \begin{bmatrix} p_{t-1} \\ v_{t-1} \\ a_{t-1} \end{bmatrix}
\end{equation}
should be adopted, where the three-dimensional state 
\begin{align*}
\mathbf{x} \equiv \begin{bmatrix} p & v & a \end{bmatrix}^\mathrm{T}
\end{align*}
is consistent with the actual motion pattern. However, we cannot know \textit{a priori} whether the tracked object is stationary, is moving at stable velocity, or is at a stable acceleration stage, and hence cannot know either which state and motion model should be adopted.

It is worth noting that an object does not tend to be ideally stationary, or moving at ideally stable velocity, or at an ideally stable acceleration stage in practical applications. Instead, a roughly stationary object may demonstrate certain position variation, an object moving at roughly stable velocity may demonstrate certain velocity variation, and an object with roughly stable acceleration may demonstrate certain acceleration variation. Therefore, the more practical version of the constant position (CP) model is
\begin{equation} \label{eq:modelCP}
p_t = p_{t-1} + \Delta p_t
\end{equation}
where $\Delta p_t$ represents object position variation which is treated as a random error following the Gaussian distribution $N(0,\Sigma_p)$. 

The more practical version of the constant velocity (CV) model is
\begin{equation} \label{eq:modelCV}
\begin{bmatrix} p_t \\ v_t \end{bmatrix} = \begin{bmatrix} 1 & \Delta T \\ 0 & 1 \end{bmatrix} \begin{bmatrix} p_{t-1} \\ v_{t-1} \end{bmatrix} + \begin{bmatrix} 0 \\ \Delta v_t \end{bmatrix}
\end{equation}
where $\Delta v_t$ represents object velocity variation which is treated as a random error following the Gaussian distribution $N(0,\Sigma_v)$. 

The more practical version of the constant acceleration (CA) model is
\begin{equation} \label{eq:modelCA}
\begin{bmatrix} p_t \\ v_t \\ a_t \end{bmatrix} = \begin{bmatrix} 1 & \Delta T & \Delta T^2/2 \\ 0 & 1 & \Delta T \\ 0 & 0 & 1 \end{bmatrix} \begin{bmatrix} p_{t-1} \\ v_{t-1} \\ a_{t-1} \end{bmatrix} + \begin{bmatrix} 0 \\ 0 \\ \Delta a_t \end{bmatrix}
\end{equation}
where $\Delta a_t$ represents object acceleration variation which is treated as a random error following the Gaussian distribution $N(0,\Sigma_a)$.

In fact, the constant position model (\ref{eq:modelCP}), the constant velocity model (\ref{eq:modelCV}), and the constant acceleration model (\ref{eq:modelCA}) can be unified into one formalism
\begin{equation}  \label{eq:modelCP+CV+CA}
\mathbf{x}_t \equiv \begin{bmatrix} p_t \\ v_t \\ a_t \end{bmatrix} = \begin{bmatrix} 1 & \Delta T & \Delta T^2/2 \\ 0 & 1 & \Delta T \\ 0 & 0 & 1 \end{bmatrix} \begin{bmatrix} p_{t-1} \\ v_{t-1} \\ a_{t-1} \end{bmatrix} + \begin{bmatrix} 1 & 0 & 0 \\ 0 & 1 & 0 \\ 0 & 0 & 1 \end{bmatrix} \begin{bmatrix} \Delta p_t \\ \Delta v_t \\ \Delta a_t \end{bmatrix} \equiv \mathbf{A} \mathbf{x}_{t-1} + \mathbf{B} \mathbf{u}_t
\end{equation}
where
\begin{align*}
\mathbf{A} \equiv \begin{bmatrix} 1 & \Delta T & \Delta T^2/2 \\ 0 & 1 & \Delta T \\ 0 & 0 & 1 \end{bmatrix}, \quad \mathbf{B} = \begin{bmatrix} 1 & 0 & 0 \\ 0 & 1 & 0 \\ 0 & 0 & 1 \end{bmatrix}.
\end{align*}
Configuring the control input 
\begin{align*}
\mathbf{u} \equiv \begin{bmatrix} \Delta p & \Delta v & \Delta a \end{bmatrix}^\mathrm{T}
\end{align*}
in different ways will cause the unified model (\ref{eq:modelCP+CV+CA}) to transform into the constant position model, the constant velocity model, and the constant acceleration model respectively. If we configure 
\begin{align*}
\Delta p = 0, \quad \Delta v = 0, 
\end{align*}
then the unified model (\ref{eq:modelCP+CV+CA}) becomes the constant acceleration model (\ref{eq:modelCA}). If we configure 
\begin{align*}
\Delta p = 0, \quad \Delta a = 0 
\end{align*}
together with the initial acceleration 
\begin{align*}
a_0 = 0, 
\end{align*}
then the unified model (\ref{eq:modelCP+CV+CA}) becomes the constant velocity model (\ref{eq:modelCV}). If we configure 
\begin{align*}
\Delta v = 0, \quad \Delta a = 0 
\end{align*}
together with the initial velocity 
\begin{align*}
v_0 = 0 
\end{align*}
and the initial acceleration 
\begin{align*}
a_0 = 0, 
\end{align*}
then the unified model (\ref{eq:modelCP+CV+CA}) becomes the constant position model (\ref{eq:modelCP}). 

However, under the circumstance of non-deterministic system modelling, the problem is that we never know how the control input 
\begin{align*}
\mathbf{u} \equiv \begin{bmatrix} \Delta p & \Delta v & \Delta a \end{bmatrix}^\mathrm{T}
\end{align*}
should be configured. In other words, the control input $\mathbf{u}$ can never be completely known, which implies the non-deterministic nature of object tracking.

Non-deterministic system modelling tends to incur \textbf{system model mismatch}, including \textbf{under-modelling mismatch} and \textbf{over-modelling mismatch}. For example, the constant velocity (CV) model may be used as the system model for object tracking whereas the tracked object is actually accelerating. Such kind of system model mismatch belongs to the former one which means mismatch between a low-order system model and a high-order object motion pattern. For another example, the constant acceleration (CA) model may be used as the system model for object tracking whereas the tracked object is actually stationary. Such kind of system model mismatch belongs to the latter one which means mismatch between a high-order system model and a low-order object motion pattern.

As clarified in \cite{Li2022FARET_2, Li2022FARET_1}, both under-modelling mismatch and over-modelling mismatch are undesirable to state estimation. We had better neither establish a system model that is not flexible enough to characterize state evolution (i.e. suffering from the under-modelling mismatch problem) nor establish a system model that is too generalized and captures useless trends in state evolution (i.e. suffering from the over-modelling mismatch problem).

\subsubsection*{Spirit of multiple hypotheses merging}

Consistent system modelling, neither with under-modelling mismatch nor with over-modelling mismatch, is important to recursive estimation. However, consistent system modelling is sometimes difficult especially in practical applications involving non-deterministic system modelling. Fortunately, in many such practical applications, although the control input can never be completely known and hence the actual state evolution pattern is unknown in advance, it is known that the control input normally belongs to a set of multiple deterministic control input patterns and hence the state evolution pattern belongs to a set of multiple deterministic state evolution patterns. We can resort to a self-adaptive mechanism of system modelling which aims at alleviating and even eliminating potential influence of system model mismatch. This self-adaptive mechanism of system modelling is called the \textbf{interacting multiple model (IMM)} method \cite{Blom1988, Mazor1998}.

The basic spirit of the interacting multiple model method is dynamic merging of multiple representative system models that would cover potential state evolution patterns and dynamic adaptation of their contribution weights for state estimation. More specifically, each representative system model has a corresponding estimation track that performs recursive estimation using this system model. Regulated by a model-to-model transition mechanism, each estimation track shares its own estimate with other estimation tracks and also fuses estimates shared by other estimation tracks with its own estimate. The overall state estimate is obtained by a weighted merging of state estimates provided by all estimation tracks.

Given a generic system model described in (\ref{eq:generic_sys_model_discrete})
\begin{align*}
\mathbf{x}_t = g(\mathbf{x}_{t-1}, \mathbf{u}_t)
\end{align*}
and a generic measurement model
\begin{equation} \label{eq:generic_msr_model_discrete}
\mathbf{z}_t = h(\mathbf{x}_t).
\end{equation}
Sometimes we may add a ``tail'' $\mathbf{\epsilon}$ to the system model as
\begin{align*}
\mathbf{x}_t = g(\mathbf{x}_{t-1}, \mathbf{u}_t) + \mathbf{\epsilon}_t
\end{align*}
to explicitly highlight existence of the system model error. Similarly, we may also add a ``tail'' $\mathbf{\gamma}$ to the measurement model as
\begin{align*}
\mathbf{z}_t = h(\mathbf{x}_t) + \mathbf{\gamma}_t
\end{align*}
to explicitly highlight existence of the measurement model error. Here, the ``tails'' $\mathbf{\epsilon}$ and $\mathbf{\gamma}$ are removed from the generic system model (\ref{eq:generic_sys_model_discrete}) and the generic measurement model (\ref{eq:generic_msr_model_discrete}) for expression simplicity, yet readers had better realize that the ``tails'' which are normally modelled as stochastic variables always exist implicitly in (\ref{eq:generic_sys_model_discrete}) and (\ref{eq:generic_msr_model_discrete}).

For non-deterministic system modelling, suppose the control input belongs to $m$ representative deterministic control input patterns 
\begin{align*}
\mathbf{u}^{[1]}, \quad \mathbf{u}^{[2]}, \quad \cdots \quad, \quad \mathbf{u}^{[m]} 
\end{align*}
and hence we have $m$ derived system models (namely $m$ representative system models) describing potential state evolution patterns. A subscript is added to above notations $g$ to indicate the index of the derived system model, i.e.
\begin{align*}
\mathbf{x}_t &= g(\mathbf{x}_{t-1}, \mathbf{u}_t^{[1]}) \equiv g_1(\mathbf{x}_{t-1}, \mathbf{u}_t)  \\
\mathbf{x}_t &= g(\mathbf{x}_{t-1}, \mathbf{u}_t^{[2]}) \equiv g_2(\mathbf{x}_{t-1}, \mathbf{u}_t)  \\
\vdots & \qquad \vdots  \\
\mathbf{x}_t &= g(\mathbf{x}_{t-1}, \mathbf{u}_t^{[m]}) \equiv g_m(\mathbf{x}_{t-1}, \mathbf{u}_t)
\end{align*}
Their corresponding measurement models are also distinguished from each other by adding a subscript that indicates the measurement model index, i.e.
\begin{align*}
\mathbf{z}_t &= h_1(\mathbf{x}_t)  \\
\mathbf{z}_t &= h_2(\mathbf{x}_t)  \\
\vdots & \qquad \vdots  \\
\mathbf{z}_t &= h_m(\mathbf{x}_t)
\end{align*}

\begin{figure}[h!]
\begin{center}
\includegraphics[width=0.7\columnwidth]{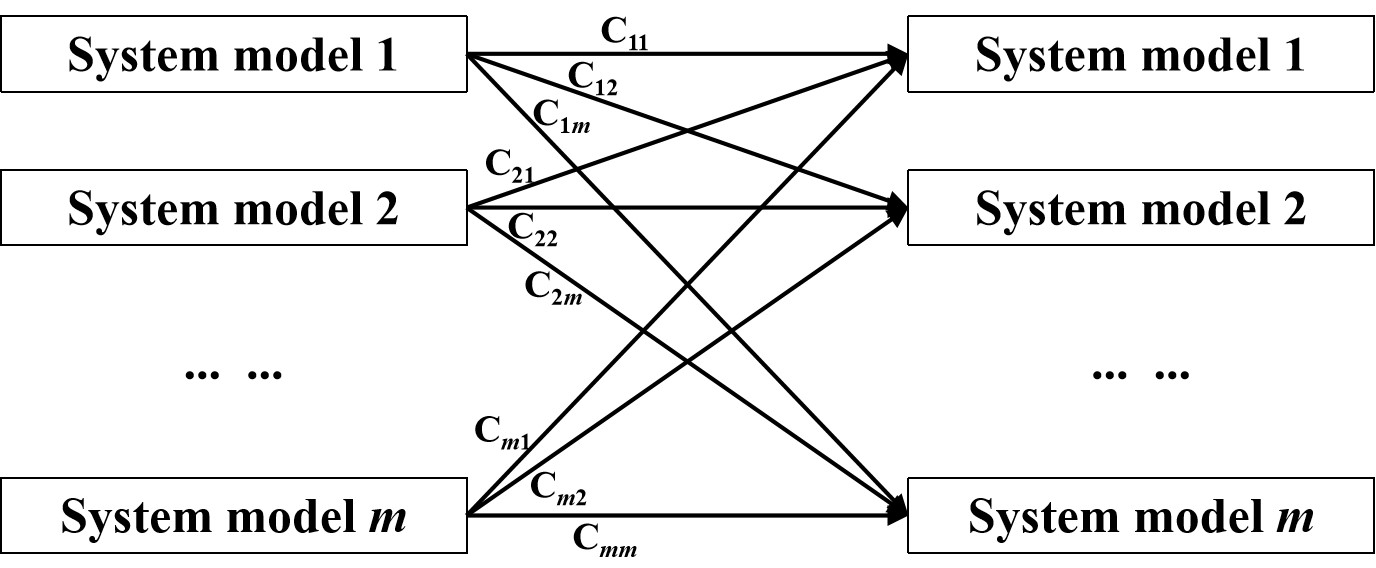}
\end{center}
\caption{Interacting multiple model transition}
\label{fig:imm_transition}
\end{figure}

The representative system models can change to each other and a $m$-by-$m$ transition matrix $\mathbf{C}$ denotes transition probabilities among them. Here, \textit{model change} means a change from the state evolution pattern described by a representative system model to the state evolution pattern described by another representative system model. For example, in the context of object tracking, the tracked object may switch among multiple representative motion patterns. $\mathbf{C}_{ij}$ denotes the probability of the $i$-th representative system model changing to the $j$-th representative system model for 
\begin{align*}
i, j \in \{1, 2, \cdots, m\}, 
\end{align*}
as illustrated in Figure \ref{fig:imm_transition}.

The normalization condition 
\begin{align*}
\sum_{j=1}^m \mathbf{C}_{ij} = 1
\end{align*}
holds for each $i$. It is worth noting that 
\begin{align*}
\sum_{i=1}^m \mathbf{C}_{ij} = 1
\end{align*}
does not necessarily hold, though the transition matrix $\mathbf{C}$ may be set to satisfy this plausible normalization condition by coincidence in practical applications.

\subsubsection*{Algorithm overview}

Each recursive cycle of the interacting multiple model algorithm \cite{Blom1988} consists of four steps: 1) multiple-model merging (or hypotheses merging), 2) distributed estimation, 3) model weight update, and 4) output synthesis. A quick overview of the interacting multiple model algorithm is put forward immediately, followed by explanations and analysis on its key points. 

Each recursive cycle starts with the $m$ model weights 
\begin{align*}
p(\hat{\mathbf{I}}_{t-1} = i) = w_{i,t-1}, 
\end{align*}
the $m$ state means $\hat{\mathbf{x}}_{i,t-1}$ and the $m$ associated covariances $\hat{\mathbf{\Sigma}}_{i,t-1}$, where
\begin{align*}
i \in \{1, 2, \cdots, m\}
\end{align*}
and
\begin{align*}
\mathbf{I} = i
\end{align*}
means that the representative system model with the index $i$ takes effect.

\textbf{(1) Multiple-model merging} 

For the $k$-th estimation track its \textit{initial} estimate $\mathbf{x}^M_{k,t}$ is obtained by a weighted merging of all $m$ state means $\hat{\mathbf{x}}_{i,t-1}$ that contribute to $\mathbf{x}^M_{k,t}$ more or less according to their model weights $w_{i,t-1}$ and transition probabilities $\mathbf{C}_{ik}$, where
\begin{align*}
k, i \in \{1, 2, \cdots, m\}.
\end{align*}
The initially merged model weight is
\begin{align} \label{eq:imm_wgt_merge}
w^M_{k,t} \equiv p(\mathbf{I}^M_t = k) = \sum_{i=1}^m p(\mathbf{I}^M_t = k | \hat{\mathbf{I}}_{t-1} = i) p(\hat{\mathbf{I}}_{t-1} = i) = \sum_{i=1}^m \mathbf{C}_{ik} w_{i,t-1}.
\end{align}
The initially merged state estimate is
\begin{align} \label{eq:imm_stt_merge}
\mathbf{x}^M_{k,t} &= \int \mathbf{x}_{k,t} p(\mathbf{x}_{k,t}) \mathrm{d} \mathbf{x}_{k,t}  = \int \mathbf{x}_{k,t} [\sum_{i=1}^m p(\mathbf{x}_{k,t}|\hat{\mathbf{x}}_{i,t-1}) p(\hat{\mathbf{x}}_{i,t-1})] \mathrm{d} \mathbf{x}_{k,t}   \nonumber \\
  &= \int \mathbf{x}_{k,t} [\sum_{i=1}^m \delta(\mathbf{x}_{k,t}-\hat{\mathbf{x}}_{i,t-1}) \mathbf{C}_{ik} w_{i,t-1} / \sum_{i=1}^m \mathbf{C}_{ik} w_{i,t-1}] \mathrm{d} \mathbf{x}_{k,t}  \nonumber \\
  &= \sum_{i=1}^m \hat{\mathbf{x}}_{i,t-1} \mathbf{C}_{ik} w_{i,t-1} / w^M_{k,t}.
\end{align}
The initially merged state estimate covariance is
\begin{align} \label{eq:imm_cov_merge}
\mathbf{\Sigma}^M_{k,t} = \sum_{i=1}^m \mathbf{C}_{ik} w_{i,t-1} [ \hat{\mathbf{\Sigma}}_{i,t-1} + (\hat{\mathbf{x}}_{i,t-1}-\mathbf{x}^M_{k,t}) (\hat{\mathbf{x}}_{i,t-1}-\mathbf{x}^M_{k,t})^\mathrm{T}] / w^M_{k,t}.
\end{align}
Reasoning for model weight merging (\ref{eq:imm_wgt_merge}) and state mean merging (\ref{eq:imm_stt_merge}) is obvious, whereas reasoning for state covariance merging (\ref{eq:imm_cov_merge}) is not so and will be explained later.

\textbf{(2) Distributed estimation}

For the $k$-th estimation track where
\begin{align*}
k \in \{1, 2, \cdots, m\},
\end{align*}
its initial estimate $\{\mathbf{x}^M_{k,t}, \mathbf{\Sigma}^M_{k,t}\}$ is used to predict its \textit{a priori} estimate $\{\bar{\mathbf{x}}_{k,t}, \bar{\mathbf{\Sigma}}_{k,t}\}$ via the $k$-th representative system model $g_k$. Its \textit{a priori} estimate $\{\bar{\mathbf{x}}_{k,t}, \bar{\mathbf{\Sigma}}_{k,t}\}$ is fused with the measurement $\mathbf{z}_t$ that follows the $k$-th measurement model $h_k$ to obtain its \textit{a posteriori} estimate $\{\hat{\mathbf{x}}_{k,t}, \hat{\mathbf{\Sigma}}_{k,t}\}$. The Kalman filter serves as the recursive estimation method for the $m$ distributed estimation tracks.

\textbf{(3) Model weight update}

For the $k$-th estimation track where
\begin{align*}
k \in \{1, 2, \cdots, m\},
\end{align*}
its weight $w_{k,t}$ is updated from its initial weight $w^M_{k,t}$ according to its measurement innovation as 
\begin{align}
w_{k,t} &\equiv p(\hat{\mathbf{I}}_t = k | \mathbf{z}_t) = p(\mathbf{z}_t | \mathbf{I}^M_t = k) p(\mathbf{I}^M_t = k) / p(\mathbf{z}_t) \nonumber \\
  &= \eta \frac{1}{\sqrt{\det \mathbf{Q}_{k,t}}} \mathrm{e}^{-\frac{1}{2} (\mathbf{z}_t - h_k (\hat{\mathbf{x}}_{k,t}))^\mathrm{T} \mathbf{Q}_{k,t}^{-1} (\mathbf{z}_t - h_k (\hat{\mathbf{x}}_{k,t})) } w^M_{k,t},
\end{align}
where $\eta$ is a normalization constant and $\mathbf{Q}_{k,t}$ is computed as
\begin{align*}
\mathbf{Q}_{k,t} = \nabla h_k (\hat{\mathbf{x}}_{k,t})^\mathrm{T} \hat{\mathbf{\Sigma}}_{k,t} \nabla h_k (\hat{\mathbf{x}}_{k,t}) + \mathbf{\Sigma}_{\mathbf{z},t}.
\end{align*}

\textbf{(4) Output synthesis}

The recursive cycle outputs the synthesized state estimate $\{\hat{\mathbf{x}}_t, \hat{\mathbf{\Sigma}}_t\}$ which is computed via
\begin{subequations}  \label{eq:imm_output_synthesis}
\begin{align}
\hat{\mathbf{x}}_t &= \sum_{k=1}^m w_{k,t} \hat{\mathbf{x}}_{k,t}, \\
\hat{\mathbf{\Sigma}}_t &= \sum_{k=1}^m w_{k,t} [\hat{\mathbf{\Sigma}}_{k,t} + (\hat{\mathbf{x}}_{k,t}-\hat{\mathbf{x}}_t) (\hat{\mathbf{x}}_{k,t}-\hat{\mathbf{x}}_t)^\mathrm{T} ].
\end{align}
\end{subequations}

\subsubsection*{Reasoning for state covariance merging}

For the $k$-th estimation track, denote 
\begin{align*}
\lambda_i \equiv \mathbf{C}_{ik} w_{i,t-1} / (\sum_{i=1}^m \mathbf{C}_{ik} w_{i,t-1}) = \mathbf{C}_{ik} w_{i,t-1} / w^M_{k,t}
\end{align*}
for
\begin{align*}
k, i \in \{1, 2, \cdots, m\}.
\end{align*}
Then (\ref{eq:imm_stt_merge}) becomes
\begin{align} \label{eq:imm_stt_merge_short}
\mathbf{x}^M_{k,t} &= \sum_{i=1}^m \lambda_i \hat{\mathbf{x}}_{i,t-1}.
\end{align} 

Let $\mathbf{cov}(\mathbf{x},\mathbf{y})$ denote the covariance between two random vector variables $\mathbf{x}$ and $\mathbf{y}$, namely
\begin{equation} \label{eq:cov_definition}
\mathbf{cov}(\mathbf{x},\mathbf{y}) = E[(\mathbf{x}-E[\mathbf{x}]) (\mathbf{y}-E[\mathbf{y}])^\mathrm{T}].
\end{equation}
If $\mathbf{x} \equiv \mathbf{y}$, let $\mathbf{var}(\mathbf{x})$ denote the variance of $\mathbf{x}$ as
\begin{equation}
\mathbf{var}(\mathbf{x}) \equiv \mathbf{cov}(\mathbf{x},\mathbf{x}) = E[(\mathbf{x}-E[\mathbf{x}]) (\mathbf{x}-E[\mathbf{x}])^\mathrm{T}].
\end{equation}
Note that 
\begin{align*}
(\mathbf{x}-E[\mathbf{x}]) (\mathbf{x}-E[\mathbf{x}])^\mathrm{T} \geq 0
\end{align*}
always hold, where $\geq 0$ means \textit{positive semi-definite}. So 
\begin{align*}
\mathbf{var}(\mathbf{x}) \equiv \mathbf{cov}(\mathbf{x},\mathbf{x})
\end{align*}
is at least \textit{positive semi-definite} and even \textit{positive definite} namely 
\begin{align*}
\mathbf{var}(\mathbf{x}) > 0. 
\end{align*}
The state variance being at least positive semi-definite can be interpreted as the state estimate having some degree of uncertainty, which apparently makes sense in practical applications.

If all $\mathbf{x}_{i,t-1}$ in (\ref{eq:imm_stt_merge_short}) were independent of each other, then state covariance merging would be realized as
\begin{align*}
\mathbf{var} (\mathbf{x}_{k,t}) &= \mathbf{cov} (\sum_{i=1}^m \lambda_i \mathbf{x}_{i,t-1}, \sum_{i=1}^m \lambda_i \mathbf{x}_{i,t-1}) \\
 &= \sum_{i=1}^m \mathbf{cov} (\lambda_i \mathbf{x}_{i,t-1}, \lambda_i \mathbf{x}_{i,t-1}) \\
 &= \sum_{i=1}^m \lambda_i^2 \mathbf{cov} ( \mathbf{x}_{i,t-1}, \mathbf{x}_{i,t-1}) = \sum_{i=1}^m \lambda_i^2 \hat{\mathbf{\Sigma}}_{i,t-1}.
\end{align*}
However, those $\mathbf{x}_{i,t-1}$ in (\ref{eq:imm_stt_merge_short}) usually are not independent of each other and may even be heavily correlated with each other. Consequently, state covariance merging cannot be simply performed in above way.

It is usually difficult to estimate the variance of $\mathbf{x}_{k,t}$ exactly because it is usually difficult to know \textit{a priori} how those $\mathbf{x}_{i,t-1}$ in (\ref{eq:imm_stt_merge_short}) are exactly correlated with each other. On the other hand, an upper bound for the variance of $\mathbf{x}_{k,t}$ may be estimated to approximate the variance of $\mathbf{x}_{k,t}$.
\begin{align} \label{eq:imm_cov_merge_derive1}
\mathbf{\Sigma}^M_{k,t} &\equiv \mathbf{var} (\mathbf{x}_{k,t}) = \mathbf{cov} (\sum_{i=1}^m \lambda_i \mathbf{x}_{i,t-1}, \sum_{i=1}^m \lambda_i \mathbf{x}_{i,t-1}) \nonumber \\
 &= E[ (\sum_{i=1}^m \lambda_i \mathbf{x}_{i,t-1} - E[\mathbf{x}_{k,t}]) (\sum_{i=1}^m \lambda_i \mathbf{x}_{i,t-1} - E[\mathbf{x}_{k,t}])^\mathrm{T} ] \nonumber \\
 &= E[ (\sum_{i=1}^m \lambda_i \mathbf{x}_{i,t-1} - \mathbf{x}^M_{k,t}) (\sum_{i=1}^m \lambda_i \mathbf{x}_{i,t-1} - \mathbf{x}^M_{k,t})^\mathrm{T} ] \nonumber \\
 &= E[ (\sum_{i=1}^m \lambda_i (\mathbf{x}_{i,t-1} - \mathbf{x}^M_{k,t})) (\sum_{i=1}^m \lambda_i (\mathbf{x}_{i,t-1} - \mathbf{x}^M_{k,t}))^\mathrm{T} ] \quad (\text{as} \sum_{i=1}^m \lambda_i = 1) \nonumber \\
 &\leq \sum_{i=1}^m \lambda_i E[(\mathbf{x}_{i,t-1} - \mathbf{x}^M_{k,t}) (\mathbf{x}_{i,t-1} - \mathbf{x}^M_{k,t})^\mathrm{T}].
\end{align}
The validity of the last inequality in (\ref{eq:imm_cov_merge_derive1}) relies on the following inequality
\begin{equation} \label{eq:imm_cov_merge_inequality1}
(\sum_{i=1}^m \lambda_i \mathbf{z}_i) (\sum_{i=1}^m \lambda_i \mathbf{z}_i)^\mathrm{T} \leq (\sum_{i=1}^m \lambda_i) (\sum_{i=1}^m \lambda_i \mathbf{z}_i \mathbf{z}_i^\mathrm{T}),
\end{equation}
where
\begin{align*}
0 \leq \lambda_i \leq 1, \quad i \in \{1, 2, \cdots, m\}.
\end{align*}

\begin{proof}
In the light of the Lagrange identity \cite{Mitrinovic1970},
\begin{align*}
&(\sum_{i=1}^m \lambda_i) (\sum_{i=1}^m \lambda_i \mathbf{z}_i \mathbf{z}_i^\mathrm{T}) - (\sum_{i=1}^m \lambda_i \mathbf{z}_i) (\sum_{i=1}^m \lambda_i \mathbf{z}_i)^\mathrm{T}  \\
 &= \sum_{1 \leq i < j \leq m} (\sqrt{\lambda_i} \sqrt{\lambda_j} \mathbf{z}_j - \sqrt{\lambda_j} \sqrt{\lambda_i} \mathbf{z}_i) (\sqrt{\lambda_i} \sqrt{\lambda_j} \mathbf{z}_j - \sqrt{\lambda_j} \sqrt{\lambda_i} \mathbf{z}_i)^\mathrm{T} \geq 0,
\end{align*}
because each 
\begin{align*}
(\sqrt{\lambda_i} \sqrt{\lambda_j} \mathbf{z}_j - \sqrt{\lambda_j} \sqrt{\lambda_i} \mathbf{z}_i) (\sqrt{\lambda_i} \sqrt{\lambda_j} \mathbf{z}_j - \sqrt{\lambda_j} \sqrt{\lambda_i} \mathbf{z}_i)^\mathrm{T} \geq 0.
\end{align*}
The proof is done.
\end{proof}

If there is an additional condition
\begin{align*}
\sum_{i=1}^m \lambda_i = 1, 
\end{align*}
then we know from (\ref{eq:imm_cov_merge_inequality1}) that
\begin{equation} \label{eq:imm_cov_merge_inequality2}
(\sum_{i=1}^m \lambda_i \mathbf{z}_i) (\sum_{i=1}^m \lambda_i \mathbf{z}_i)^\mathrm{T} \leq \sum_{i=1}^m \lambda_i \mathbf{z}_i \mathbf{z}_i^\mathrm{T}.
\end{equation}
It is actually (\ref{eq:imm_cov_merge_inequality2}) that supports the last inequality in (\ref{eq:imm_cov_merge_derive1}).

Return to (\ref{eq:imm_cov_merge_derive1}) and derive each $E[(\mathbf{x}_{i,t-1} - \mathbf{x}^M_{k,t}) (\mathbf{x}_{i,t-1} - \mathbf{x}^M_{k,t})^\mathrm{T}]$ as
\begin{align*}
&E[(\mathbf{x}_{i,t-1} - \mathbf{x}^M_{k,t}) (\mathbf{x}_{i,t-1} - \mathbf{x}^M_{k,t})^\mathrm{T}] \\
 &= E[(\mathbf{x}_{i,t-1} - \hat{\mathbf{x}}_{i,t-1} + \hat{\mathbf{x}}_{i,t-1} - \mathbf{x}^M_{k,t}) (\mathbf{x}_{i,t-1} - \hat{\mathbf{x}}_{i,t-1} + \hat{\mathbf{x}}_{i,t-1} - \mathbf{x}^M_{k,t})^\mathrm{T}] \\
 &= E[(\mathbf{x}_{i,t-1} - \hat{\mathbf{x}}_{i,t-1}) (\mathbf{x}_{i,t-1} - \hat{\mathbf{x}}_{i,t-1})^\mathrm{T}] 
    + \mathbf{0} \times (\hat{\mathbf{x}}_{i,t-1} - \mathbf{x}^M_{k,t})^\mathrm{T} \\
 & \quad   + (\hat{\mathbf{x}}_{i,t-1} - \mathbf{x}^M_{k,t}) \times \mathbf{0}^\mathrm{T}
    + (\hat{\mathbf{x}}_{i,t-1} - \mathbf{x}^M_{k,t}) (\hat{\mathbf{x}}_{i,t-1} - \mathbf{x}^M_{k,t})^\mathrm{T}  \\
 &= \hat{\mathbf{\Sigma}}_{i,t-1} + (\hat{\mathbf{x}}_{i,t-1}-\mathbf{x}^M_{k,t}) (\hat{\mathbf{x}}_{i,t-1}-\mathbf{x}^M_{k,t})^\mathrm{T}.
\end{align*}
Therefore (\ref{eq:imm_cov_merge_derive1}) becomes
\begin{equation} \label{eq:imm_cov_merge_derive2}
\mathbf{\Sigma}^M_{k,t} \equiv \mathbf{var} (\mathbf{x}_{k,t}) \leq \sum_{i=1}^m \lambda_i [\hat{\mathbf{\Sigma}}_{i,t-1} + (\hat{\mathbf{x}}_{i,t-1}-\mathbf{x}^M_{k,t}) (\hat{\mathbf{x}}_{i,t-1}-\mathbf{x}^M_{k,t})^\mathrm{T}].
\end{equation}
To guarantee \textit{estimate consistency}, set 
\begin{align*}
\mathbf{\Sigma}^M_{k,t} &= \sum_{i=1}^m \lambda_i [\hat{\mathbf{\Sigma}}_{i,t-1} + (\hat{\mathbf{x}}_{i,t-1}-\mathbf{x}^M_{k,t}) (\hat{\mathbf{x}}_{i,t-1}-\mathbf{x}^M_{k,t})^\mathrm{T}]  \\
 &= \sum_{i=1}^m \mathbf{C}_{ik} w_{i,t-1} [ \hat{\mathbf{\Sigma}}_{i,t-1} + (\hat{\mathbf{x}}_{i,t-1}-\mathbf{x}^M_{k,t}) (\hat{\mathbf{x}}_{i,t-1}-\mathbf{x}^M_{k,t})^\mathrm{T} ] / w^M_{k,t}
\end{align*}
and (\ref{eq:imm_cov_merge}) is obtained. The concept of \textit{estimate consistency} will be explained in more details in Section \ref{sec:split_cif}.

\subsection{Handle nonlinearity: extended Kalman filter and other variants}

Consider the generic system model described in (\ref{eq:generic_sys_model_discrete})
\begin{align*}
\mathbf{x}_t = g(\mathbf{x}_{t-1}, \mathbf{u}_t)
\end{align*}
and the generic measurement model described in (\ref{eq:generic_msr_model_discrete})
\begin{align*}
\mathbf{z}_t = h(\mathbf{x}_t).
\end{align*}
Like for the original Kalman filter presented in Section \ref{sec:Kalman_filter}, the Gaussian distribution assumption is still followed in the system model (\ref{eq:generic_sys_model_discrete}) and the measurement model (\ref{eq:generic_msr_model_discrete}) to characterize involved random variables.

The system model (\ref{eq:generic_sys_model_discrete}) may be nonlinear with respect to the state $\mathbf{x}$ and the control input $\mathbf{u}$. The measurement model (\ref{eq:generic_msr_model_discrete}) may be nonlinear with respect to the state $\mathbf{x}$. The Kalman filter, namely (\ref{eq:KFprediction})
\begin{align*}
\bar{\mathbf{x}}_t &= \mathbf{A} \hat{\mathbf{x}}_{t-1} + \mathbf{B} \mathbf{u}_t,  \\
\bar{\mathbf{\Sigma}}_t &= \mathbf{A} \hat{\mathbf{\Sigma}}_{t-1} \mathbf{A}^{\mathrm{T}} + \mathbf{B} \mathbf{\Sigma_u} \mathbf{B}^{\mathrm{T}}
\end{align*}
for prediction and (\ref{eq:KFupdate})
\begin{align*}
\mathbf{K} &= \bar{\mathbf{\Sigma}}_t \mathbf{H}^{\mathrm{T}} (\mathbf{H} \bar{\mathbf{\Sigma}}_t \mathbf{H}^{\mathrm{T}} + \mathbf{\Sigma_z})^{-1},  \\
\hat{\mathbf{x}}_t &= \bar{\mathbf{x}}_t + \mathbf{K} (\mathbf{z}_t - \mathbf{H} \bar{\mathbf{x}}_t),  \\
\hat{\mathbf{\Sigma}}_t &= (\mathbf{I} - \mathbf{K H}) \bar{\mathbf{\Sigma}}_t
\end{align*}
for update, cannot be applied directly with potentially-nonlinear models (\ref{eq:generic_sys_model_discrete}) and (\ref{eq:generic_msr_model_discrete}).

To generally apply the Kalman filter, we can approximate the system model (\ref{eq:generic_sys_model_discrete}) and the measurement model (\ref{eq:generic_msr_model_discrete}) by linearizing (\ref{eq:generic_sys_model_discrete}) and (\ref{eq:generic_msr_model_discrete}) locally about the state $\mathbf{x}$ and the control input $\mathbf{u}$ to transform (\ref{eq:generic_sys_model_discrete}) and (\ref{eq:generic_msr_model_discrete}) into a linear form like (\ref{eq:SDE_linear_discrete2}) and (\ref{eq:SDE_linear_measurement_discrete}).
\begin{align}  \label{eq:sysModelGenericLinearized}
\Delta \mathbf{x}_t &= g(\mathbf{x}_{t-1} + \Delta \mathbf{x}_{t-1}, \mathbf{u}_t + \Delta \mathbf{u}_t) - g(\mathbf{x}_{t-1}, \mathbf{u}_t)  \nonumber \\  
&\approx g_{\mathbf{x}} (\mathbf{x}_{t-1},\mathbf{u}_t) \Delta \mathbf{x}_{t-1}
+ g_{\mathbf{u}} (\mathbf{x}_{t-1},\mathbf{u}_t) \Delta \mathbf{u}_t,
\end{align}
where
\begin{align*}
g_{\mathbf{x}} (\mathbf{x}_{t-1},\mathbf{u}_t)
&= \frac{\partial g(\mathbf{x}, \mathbf{u})}{\partial \mathbf{x}}|_{\mathbf{x} = \mathbf{x}_{t-1}, \mathbf{u}  = \mathbf{u}_t},  \\
g_{\mathbf{u}} (\mathbf{x}_{t-1},\mathbf{u}_t)
&= \frac{\partial g(\mathbf{x}, \mathbf{u})}{\partial \mathbf{u}}|_{\mathbf{x} = \mathbf{x}_{t-1}, \mathbf{u}  = \mathbf{u}_t}
\end{align*}
denote the Jacobian matrices of the state evolution function $g(\mathbf{x}_{t-1}, \mathbf{u}_t)$ with respect to $\mathbf{x}_{t-1}$ and $\mathbf{u}_t$ respectively. 

\begin{equation} \label{eq:msrModelGenericLinearized}
\Delta \mathbf{z}_t = h(\mathbf{x}_t + \Delta \mathbf{x}_t) - h(\mathbf{x}_t) \approx h_{\mathbf{x}} (\mathbf{x}_t) \Delta \mathbf{x}_t,
\end{equation}
where
\begin{align*}
h_{\mathbf{x}} (\mathbf{x}_t) = \frac{\partial h(\mathbf{x})}{\partial \mathbf{x}} |_{\mathbf{x} = \mathbf{x}_t}
\end{align*}
denotes the Jacobian matrix of the measurement function $h(\mathbf{x}_t)$ with respect to $\mathbf{x}_t$.

(\ref{eq:sysModelGenericLinearized}) reflects how the offsets of last state $\mathbf{x}_{t-1}$ and current control input $\mathbf{u}_t$ would influence the offset of current state $\mathbf{x}_t$. (\ref{eq:msrModelGenericLinearized}) reflects how the offset of current state $\mathbf{x}_t$ would influence the offset of current measurement $\mathbf{z}_t$. As mentioned in Section \ref{sec:recursive_estimation}, when we talk about any estimate such as state estimate, system control input, or measurement, we had better bear in mind that a covariance or other mathematical form characterizing its uncertainty is implicitly associated with the estimate. The system model (\ref{eq:generic_sys_model_discrete}) and the measurement model (\ref{eq:generic_msr_model_discrete}) describe relationships among estimates $\mathbf{x}_{t-1}$, $\mathbf{u}_t$, $\mathbf{x}_t$, and $\mathbf{z}_t$, whereas the locally-linearized system model (\ref{eq:sysModelGenericLinearized}) and the locally-linearized measurement model (\ref{eq:msrModelGenericLinearized}) describe relationships among estimate uncertainties.

In the original Kalman filter, relationships among estimates and relationships among estimate uncertainties are described by the same formalism (\ref{eq:SDE_linear_discrete2}) and (\ref{eq:SDE_linear_measurement_discrete}). In contrast, in the extended Kalman filter, relationships among estimates are described by (\ref{eq:generic_sys_model_discrete}) and (\ref{eq:generic_msr_model_discrete}), whereas relationships among estimate uncertainties are described by (\ref{eq:sysModelGenericLinearized}) and (\ref{eq:msrModelGenericLinearized}).

On the other hand, the extended Kalman filter and the original Kalman filter share the same essence of data fusion: an \textit{a priori} estimate \{$\bar{\mathbf{x}}_t$, $\bar{\mathbf{\Sigma}}_t$\} is predicted according to the system model, and then an \textit{a posteriori} estimate \{$\hat{\mathbf{x}}_t$, $\hat{\mathbf{\Sigma}}_t$\} is obtained by weighted averaging of the two source estimates \{$\bar{\mathbf{x}}_t$, $\bar{\mathbf{\Sigma}}_t$\} and \{$\mathbf{z}_t$, $\mathbf{\Sigma_z}$\}. 

Consider similarity between the linear formalism (\ref{eq:sysModelGenericLinearized}) and (\ref{eq:msrModelGenericLinearized}) and the linear formalism (\ref{eq:SDE_linear_discrete2}) and (\ref{eq:SDE_linear_measurement_discrete}), and adapt the formalism of the original Kalman filter to suit generic models (\ref{eq:generic_sys_model_discrete}) and (\ref{eq:generic_msr_model_discrete}) and their locally-linearized counterparts (\ref{eq:sysModelGenericLinearized}) and (\ref{eq:msrModelGenericLinearized}).

\noindent \textbf{Prediction}: $\mathbf{x}_t$ (\textit{a priori}) $\sim N(\bar{\mathbf{x}}_t,\bar{\mathbf{\Sigma}}_t)$ 
\begin{subequations}  \label{eq:EKFprediction}
\begin{align} 
\bar{\mathbf{x}}_t &= g(\hat{\mathbf{x}}_{t-1}, \mathbf{u}_t),  \\
\bar{\mathbf{\Sigma}}_t &= g_{\mathbf{x}} (\mathbf{x}_{t-1},\mathbf{u}_t) \hat{\mathbf{\Sigma}}_{t-1} g_{\mathbf{x}} (\mathbf{x}_{t-1},\mathbf{u}_t)^\mathrm{T} + g_{\mathbf{u}} (\mathbf{x}_{t-1},\mathbf{u}_t) \mathbf{\Sigma_u} g_{\mathbf{u}} (\mathbf{x}_{t-1},\mathbf{u}_t)^\mathrm{T}.
\end{align}
\end{subequations}

\noindent \textbf{Update}: $\mathbf{x}_t$ (\textit{a posteriori}) $\sim N(\hat{\mathbf{x}}_t,\hat{\mathbf{\Sigma}}_t)$ 
\begin{subequations}  \label{eq:EKFupdate}
\begin{align} 
\mathbf{K} &= \bar{\mathbf{\Sigma}}_t h_{\mathbf{x}} (\bar{\mathbf{x}}_t)^\mathrm{T} (h_{\mathbf{x}} (\bar{\mathbf{x}}_t) \bar{\mathbf{\Sigma}}_t h_{\mathbf{x}} (\bar{\mathbf{x}}_t)^\mathrm{T} + \mathbf{\Sigma_z})^{-1},  \\
\hat{\mathbf{x}}_t &= \bar{\mathbf{x}}_t + \mathbf{K} (\mathbf{z}_t - h (\bar{\mathbf{x}}_t) ),  \\
\hat{\mathbf{\Sigma}}_t &= (\mathbf{I} - \mathbf{K} h_{\mathbf{x}} (\bar{\mathbf{x}}_t)) \bar{\mathbf{\Sigma}}_t.
\end{align}
\end{subequations}

Above \textit{local linearization} based variant of the Kalman filter is the \textbf{extended Kalman filter (EKF)} \cite{Grewal2000}. Readers may easily notice commonality between the extended Kalman filter (\ref{eq:EKFprediction}) and (\ref{eq:EKFupdate}) and the original Kalman filter (\ref{eq:KFprediction}) and (\ref{eq:KFupdate}). If generic models (\ref{eq:generic_sys_model_discrete}) and (\ref{eq:generic_msr_model_discrete}) are replaced by linear models (\ref{eq:SDE_linear_discrete2}) and (\ref{eq:SDE_linear_measurement_discrete}), the extended Kalman filter (\ref{eq:EKFprediction}) and (\ref{eq:EKFupdate}) will be reduced to the original Kalman filter (\ref{eq:KFprediction}) and (\ref{eq:KFupdate}).

\subsubsection*{Unscented Kalman filter}

Another variant of the Kalman filter namely the \textbf{unscented Kalman filter} (UKF) \cite{Julier1997, Julier2004}, which is based on \textbf{unscented transformation} \cite{Uhlmann1995}, also aims at handling nonlinearity in the system model and the measurement model, with an extra objective to handle inconsistent statistics that may be caused by local linearization involved in the extended Kalman filter. Summary of the unscented Kalman filter is as follows.

\noindent \textbf{Prediction}:

1. Generate \textit{sigma points} $\mathbf{\hat{x}}_{t-1,i}^a$ ($i \in \{0, 1, \cdots, 2n\}$) from the \textit{augmented state vector} $\mathbf{\hat{x}}_{t-1}^a$ and the corresponding \textit{augmented state covariance} $\mathbf{\hat{\Sigma}}_{t-1}^a$ 
\begin{align*}
\mathbf{\hat{x}}_{t-1}^a &= \begin{bmatrix} \mathbf{\hat{x}}_{t-1} \\ \mathbf{0} \\ \mathbf{0} \end{bmatrix},  \quad  
\mathbf{\hat{\Sigma}}_{t-1}^a = \begin{bmatrix} \mathbf{\hat{\Sigma}}_{t-1} & \mathbf{0} & \mathbf{0} \\ \mathbf{0} & \mathbf{\Sigma_u} & \mathbf{0} \\ \mathbf{0} & \mathbf{0} & \mathbf{\Sigma}_\epsilon \end{bmatrix}  
\end{align*}
according to 
\begin{subequations}  \label{eq:sigma_points_augmented}
\begin{align}
\mathbf{\hat{x}}_{t-1,0}^a &= \mathbf{\hat{x}}_{t-1}^a & w_0 &= \kappa/(n+\kappa),  \\
\mathbf{\hat{x}}_{t-1,j}^a &= \mathbf{\hat{x}}_{t-1}^a+(\sqrt{(n+\kappa)\mathbf{\hat{\Sigma}}_{t-1}^a}) \mathbf{e}_j & w_j &= 1/2(n+\kappa),  \\
\mathbf{\hat{x}}_{t-1,j+n}^a &= \mathbf{\hat{x}}_{t-1}^a-(\sqrt{(n+\kappa)\mathbf{\hat{\Sigma}}_{t-1}^a}) \mathbf{e}_j & w_{j+n} &= 1/2(n+\kappa),
\end{align}
\end{subequations}
where $j \in \{1, 2, \cdots, n\}$ and $\sqrt{(n+\kappa)\mathbf{\hat{\Sigma}}_{t-1}^a}$ does not really denote the square root of the matrix $(n+\kappa)\mathbf{\hat{\Sigma}}_{t-1}^a$ but denotes a \textit{quasi square root} of $(n+\kappa)\mathbf{\hat{\Sigma}}_{t-1}^a$ satisfying
\begin{align*}
\sqrt{(n+\kappa)\mathbf{\hat{\Sigma}}_{t-1}^a} \sqrt{(n+\kappa)\mathbf{\hat{\Sigma}}_{t-1}^a}^\mathrm{T} = (n+\kappa)\mathbf{\hat{\Sigma}}_{t-1}^a.
\end{align*}
The quasi square root $\sqrt{(n+\kappa)\mathbf{\hat{\Sigma}}_{t-1}^a}$ can be obtained via \textit{Cholesky decomposition} \cite{Golub1996}.

2. Compute the predicted state point of each sigma point $\mathbf{\hat{x}}_{t-1,i}^a$ ($i \in \{0, 1, \cdots, 2n\}$) according to the system model
\begin{equation}  \label{eq:UKF_prediction_sigma}
\mathbf{\bar{x}}_{t,i} = g(\mathbf{\hat{x}}_{t-1,i}^a,\mathbf{u}_t).
\end{equation}

3. Compute the predicted state mean $\mathbf{\bar{x}}_t$ as 
\begin{equation}  \label{eq:UKF_prediction_state_mean} 
\mathbf{\bar{x}}_t = \sum_{i=0}^{2n} w_i \mathbf{\bar{x}}_{t,i}.
\end{equation}

4. Compute the predicted state covariance $\mathbf{\bar{\Sigma}}_t$ as
\begin{equation}  \label{eq:UKF_prediction_state_cov} 
\mathbf{\bar{\Sigma}}_t = \sum_{i=0}^{2n} w_i (\mathbf{\bar{x}}_{t,i}-\mathbf{\bar{x}}_t)(\mathbf{\bar{x}}_{t,i}-\mathbf{\bar{x}}_t)^T.
\end{equation}
It is worth noting that system model process noises may be removed from consideration in the augmented state vector $\mathbf{\hat{x}}_{t-1}^a$ and the augmented state covariance $\mathbf{\hat{\Sigma}}_{t-1}^a$ but can instead be taken into account as an extra term $\mathbf{\Sigma}_\epsilon$ added to (\ref{eq:UKF_prediction_state_cov}) as
\begin{equation}  \label{eq:UKF_prediction_state_cov+proc_noise} 
\mathbf{\bar{\Sigma}}_t = \sum_{i=0}^{2n} w_i (\mathbf{\bar{x}}_{t,i}-\mathbf{\bar{x}}_t)(\mathbf{\bar{x}}_{t,i}-\mathbf{\bar{x}}_t)^T + \mathbf{\Sigma}_\epsilon.
\end{equation}

\noindent \textbf{Update}:

5. Compute the predicted measurement point $\mathbf{\hat{z}}_{t,i}$ of each predicted state point $\mathbf{\bar{x}}_{t,i}$ ($i \in \{0, 1, \cdots, 2n\}$) as
\begin{equation}  \label{eq:UKF_prediction_measurement_point}
\mathbf{\hat{z}}_{t,i}=h(\mathbf{\bar{x}}_{t,i}).
\end{equation}

6. Compute the predicted measurement $\mathbf{\hat{z}}_t$ as
\begin{equation}  \label{eq:UKF_prediction_measurement_mean}
\mathbf{\hat{z}}_t = \sum_{i=0}^{2n} w_i \mathbf{\hat{z}}_{t,i}.
\end{equation}

7. Compute the innovation covariance $\mathbf{\Sigma}_{yy}$ as 
\begin{equation}  \label{eq:UKF_innovation_covariance}
\mathbf{\Sigma}_{yy} = \mathbf{\Sigma}_\gamma+\sum_{i=0}^{2n} w_i (\mathbf{\hat{z}}_{t,i}-\mathbf{\hat{z}}_t)(\mathbf{\hat{z}}_{t,i}-\mathbf{\hat{z}}_t)^T.
\end{equation}

8. Compute the cross covariance $\mathbf{\Sigma}_{xz}$ as
\begin{equation}  \label{eq:UKF_cross_covariance} 
\mathbf{\Sigma}_{xz} = \sum_{i=0}^{2n} w_i (\mathbf{\bar{x}}_{t,i}-\mathbf{\bar{x}}_t)(\mathbf{\hat{z}}_{t,i}-\mathbf{\hat{z}}_t)^T.
\end{equation}

9. Compute the \textit{a posteriori} state estimate and covariance estimate as
\begin{subequations}  \label{eq:UKF_a_posteriori}
\begin{align}
\mathbf{K} &= \mathbf{\Sigma}_{xz} \mathbf{\Sigma}_{yy}^{-1},  \\
\mathbf{\hat{y}}_t &= \mathbf{z}_t-\mathbf{\hat{z}}_t,  \\
\mathbf{\hat{x}}_t &= \mathbf{\bar{x}}_t + \mathbf{K}\mathbf{\hat{y}}_t, \\
\mathbf{\hat{\Sigma}}_t &= \mathbf{\bar{\Sigma}}_t-\mathbf{K}\mathbf{\Sigma}_{yy}\mathbf{K}^T. 
\end{align}
\end{subequations}

\subsubsection*{Cubature Kalman filter}

One more representative variant of the Kalman filter is the \textbf{cubature Kalman filter} (CKF) \cite{Haykin2009}. The cubature Kalman filter and the unscented Kalman filter share the same spirit of taking advantage of sampled points to characterize estimate distributions (including their Bayesian inference) instead of using parametrized model structures to do so. The algorithm procedures of the cubature Kalman filter are similar to those of the unscented Kalman filter. If \textit{sigma points} and \textit{unscented transformation} in the unscented Kalman filter are replaced by \textit{cubature points} and \textit{cubature transformation} respectively, then the unscented Kalman filter becomes the cubature Kalman filter. Like the unscented Kalman filter, the cubature Kalman filter also aims at handling nonlinearity in the system model and the measurement model, with the extra objective to handle inconsistent statistics that may be caused by local linearization involved in the extended Kalman filter. Summary of the cubature Kalman filter is as follows.

\noindent \textbf{Prediction}:

1. Generate \textit{cubature points} $\mathbf{\hat{x}}_{t-1,i}^a$ ($i \in \{1, 2, \cdots, 2n\}$) from the \textit{augmented state vector} $\mathbf{\hat{x}}_{t-1}^a$ and the corresponding \textit{augmented state covariance} $\mathbf{\hat{\Sigma}}_{t-1}^a$ according to 
\begin{subequations}  \label{eq:cubature_points_augmented}
\begin{align}
\mathbf{\hat{x}}_{t-1,j}^a &= \mathbf{\hat{x}}_{t-1}^a+(\sqrt{n \mathbf{\hat{\Sigma}}_{t-1}^a}) \mathbf{e}_j & w_j &= 1/2n,  \\
\mathbf{\hat{x}}_{t-1,j+n}^a &= \mathbf{\hat{x}}_{t-1}^a-(\sqrt{n \mathbf{\hat{\Sigma}}_{t-1}^a}) \mathbf{e}_j & w_{j+n} &= 1/2n,
\end{align}
\end{subequations}
where $j \in \{1, 2, \cdots, n\}$ and $\sqrt{n \mathbf{\hat{\Sigma}}_{t-1}^a}$ denotes a quasi square root of $n \mathbf{\hat{\Sigma}}_{t-1}^a$ satisfying
\begin{align*}
\sqrt{n \mathbf{\hat{\Sigma}}_{t-1}^a} \sqrt{n \mathbf{\hat{\Sigma}}_{t-1}^a}^\mathrm{T} = n \mathbf{\hat{\Sigma}}_{t-1}^a.
\end{align*}

2. Compute the predicted state point of each cubature point $\mathbf{\hat{x}}_{t-1,i}^a$ ($i \in \{1, 2, \cdots, 2n\}$) according to the system model
\begin{equation}  \label{eq:CKF_prediction_sigma}
\mathbf{\bar{x}}_{t,i} = g(\mathbf{\hat{x}}_{t-1,i}^a,\mathbf{u}_t).
\end{equation}

3. Compute the predicted state mean $\mathbf{\bar{x}}_t$ as 
\begin{equation}  \label{eq:CKF_prediction_state_mean} 
\mathbf{\bar{x}}_t = \sum_{i=1}^{2n} w_i \mathbf{\bar{x}}_{t,i}.
\end{equation}

4. Compute the predicted state covariance $\mathbf{\bar{\Sigma}}_t$ as
\begin{equation}  \label{eq:CKF_prediction_state_cov} 
\mathbf{\bar{\Sigma}}_t = \sum_{i=1}^{2n} w_i (\mathbf{\bar{x}}_{t,i}-\mathbf{\bar{x}}_t)(\mathbf{\bar{x}}_{t,i}-\mathbf{\bar{x}}_t)^T.
\end{equation}
or
\begin{equation}  \label{eq:CKF_prediction_state_cov+proc_noise} 
\mathbf{\bar{\Sigma}}_t = \sum_{i=1}^{2n} w_i (\mathbf{\bar{x}}_{t,i}-\mathbf{\bar{x}}_t)(\mathbf{\bar{x}}_{t,i}-\mathbf{\bar{x}}_t)^T + \mathbf{\Sigma}_\epsilon.
\end{equation}

4b. Re-sample predicted cubature points as
\begin{subequations}  \label{eq:resample_cubature_points}
\begin{align}
\mathbf{\bar{x}}_{t,j} &:= \mathbf{\bar{x}}_t + (\sqrt{n \mathbf{\bar{\Sigma}}_t}) \mathbf{e}_j,  \\
\mathbf{\bar{x}}_{t,j+n} &:= \mathbf{\bar{x}}_t - (\sqrt{n \mathbf{\bar{\Sigma}}_t}) \mathbf{e}_j,
\end{align}
\end{subequations}
where $j \in \{1, 2, \cdots, n\}$ and $:=$ denotes \textit{value assignment} rather than equality --- This kind of re-sampling step may also be incorporated into the unscented Kalman filter.

\noindent \textbf{Update}:

5. Compute the predicted measurement point $\mathbf{\hat{z}}_{t,i}$ of each predicted state point $\mathbf{\bar{x}}_{t,i}$ ($i \in \{1, 2, \cdots, 2n\}$) as
\begin{equation}  \label{eq:CKF_prediction_measurement_point}
\mathbf{\hat{z}}_{t,i}=h(\mathbf{\bar{x}}_{t,i}).
\end{equation}

6. Compute the predicted measurement $\mathbf{\hat{z}}_t$ as
\begin{equation}  \label{eq:CKF_prediction_measurement_mean}
\mathbf{\hat{z}}_t = \sum_{i=1}^{2n} w_i \mathbf{\hat{z}}_{t,i}.
\end{equation}

7. Compute the innovation covariance $\mathbf{\Sigma}_{yy}$ as 
\begin{equation}  \label{eq:CKF_innovation_covariance}
\mathbf{\Sigma}_{yy} = \mathbf{\Sigma}_\gamma + \sum_{i=1}^{2n} w_i (\mathbf{\hat{z}}_{t,i}-\mathbf{\hat{z}}_t)(\mathbf{\hat{z}}_{t,i}-\mathbf{\hat{z}}_t)^T.
\end{equation}

8. Compute the cross covariance $\mathbf{\Sigma}_{xz}$ as
\begin{equation}  \label{eq:CKF_cross_covariance} 
\mathbf{\Sigma}_{xz} = \sum_{i=1}^{2n} w_i (\mathbf{\bar{x}}_{t,i}-\mathbf{\bar{x}}_t)(\mathbf{\hat{z}}_{t,i}-\mathbf{\hat{z}}_t)^T.
\end{equation}

9. Compute the \textit{a posteriori} state estimate and covariance estimate as
\begin{subequations}  \label{eq:CKF_a_posteriori}
\begin{align}
\mathbf{K} &= \mathbf{\Sigma}_{xz} \mathbf{\Sigma}_{yy}^{-1},  \\
\mathbf{\hat{y}}_t &= \mathbf{z}_t-\mathbf{\hat{z}}_t,  \\
\mathbf{\hat{x}}_t &= \mathbf{\bar{x}}_t + \mathbf{K}\mathbf{\hat{y}}_t, \\
\mathbf{\hat{\Sigma}}_t &= \mathbf{\bar{\Sigma}}_t-\mathbf{K}\mathbf{\Sigma}_{yy}\mathbf{K}^T. 
\end{align}
\end{subequations}

It is worth noting that the original formalism of the cubature Kalman filter presented in \cite{Haykin2009} relies on the \textit{fully symmetric set}
\begin{equation}  \label{eq:fully_symmetric_set_generator}
\begin{bmatrix} \mathbf{c} \end{bmatrix} \equiv \begin{bmatrix} c_1 & c_2 & \cdots c_r & 0 & \cdots & 0 \end{bmatrix} \in \mathrm{R}^n
\end{equation}
that is obtained by permutating and changing the sign of the generator $\mathbf{c}$ in all possible ways. Although conceptualization of the fully symmetric set is general, the cubature Kalman filter involves only a simple case of the fully symmetric set, namely
\begin{equation}  \label{eq:FSSG_simple_case}
\begin{bmatrix} 1 \end{bmatrix} \equiv \begin{bmatrix} 1 & 0 & \cdots & 0 \end{bmatrix}
\end{equation}
which would be more intuitive for readers to understand if the set is ``paraphrased'' as the \textit{standard basis} vectors
\begin{align*}
\pm \mathbf{e}_1, \quad \pm \mathbf{e}_2, \quad \cdots \quad, \quad \pm \mathbf{e}_n
\end{align*}
of $\mathrm{R}^n$. This is why the author ``paraphrases'' original representation of cubature points as that formalized in (\ref{eq:cubature_points_augmented}) and (\ref{eq:resample_cubature_points}).

Besides, the author ``paraphrases'' entirely the original formalism of the cubature Kalman filter presented in \cite{Haykin2009} as (\ref{eq:cubature_points_augmented}) to (\ref{eq:CKF_a_posteriori}), for sake of facilitating a comparison between the formalisms of the cubature Kalman filter and the unscented Kalman filter. Then readers would better see similarity and differences between the two representative variants of the Kalman filter.

\subsection{Handle multiple-modality: Gaussian mixture and particle filter}  \label{sec:particle_filter}

\subsubsection*{Gaussian mixture modelling}

The linear-Gaussian assumption which underlies the famous Kalman filter includes two aspects: one is the ``linear'' aspect namely the \textit{linear modelling assumption} and the other is the ``Gaussian'' aspect namely the \textit{Gaussian distribution assumption}. As linearity of the system model and the measurement model is only to guarantee that Gaussian distributions after model transforms are still Gaussian distributions, the essence of the linear-Gaussian assumption consists in the ``Gaussian'' aspect, namely \textit{all distributions involved in recursive estimation are approximated by Gaussian distributions}. In other figurative words, all involved distributions are approximated by \textit{ellipsoid-shape} (generic analogue of \textit{ellipses}) distributions. The ``Gaussian'' aspect underlies not only the original Kalman filter but also the entire Kalman filter family.

\begin{figure}[h!]
\begin{center}
\includegraphics[width=0.5\columnwidth]{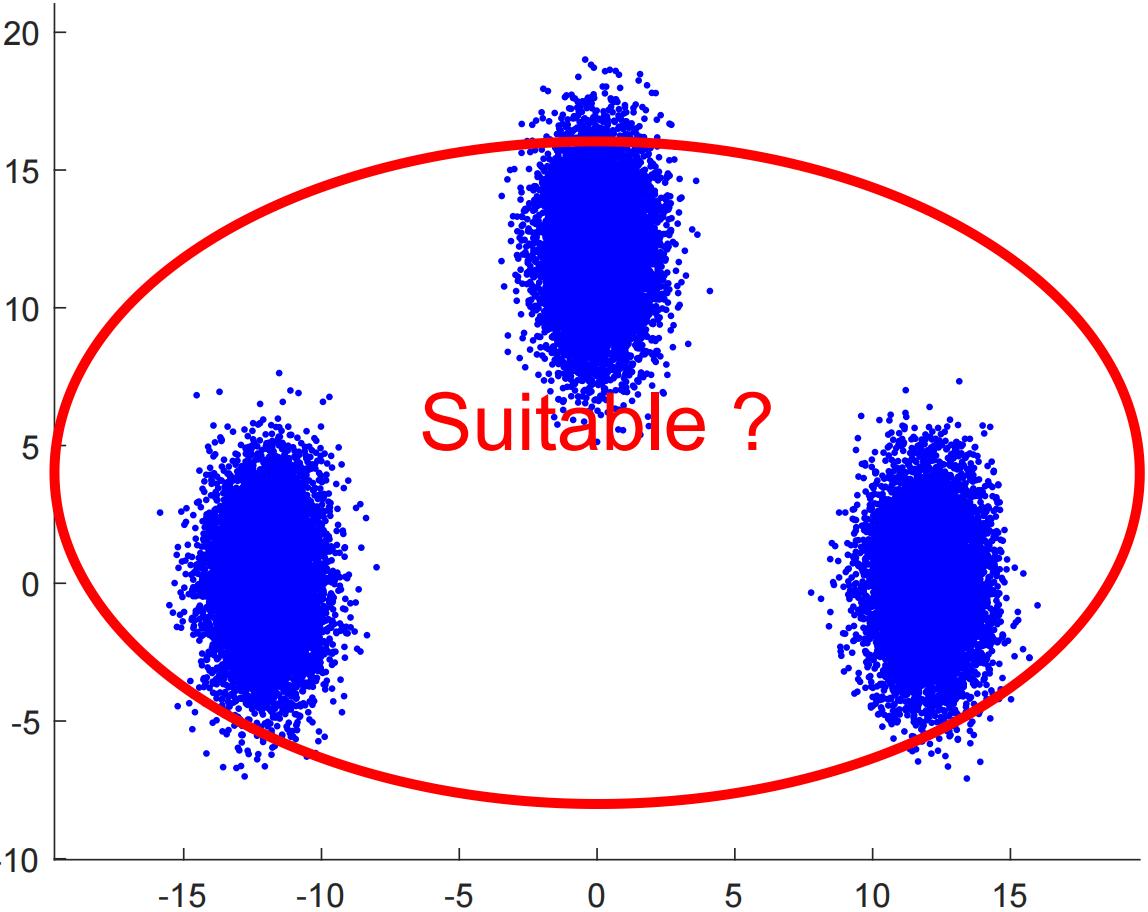}
\end{center}
\caption{Trimodal distribution}
\label{fig:bad_unimodal_eg1}
\end{figure}

\begin{figure}[h!]
\begin{center}
\includegraphics[width=0.5\columnwidth]{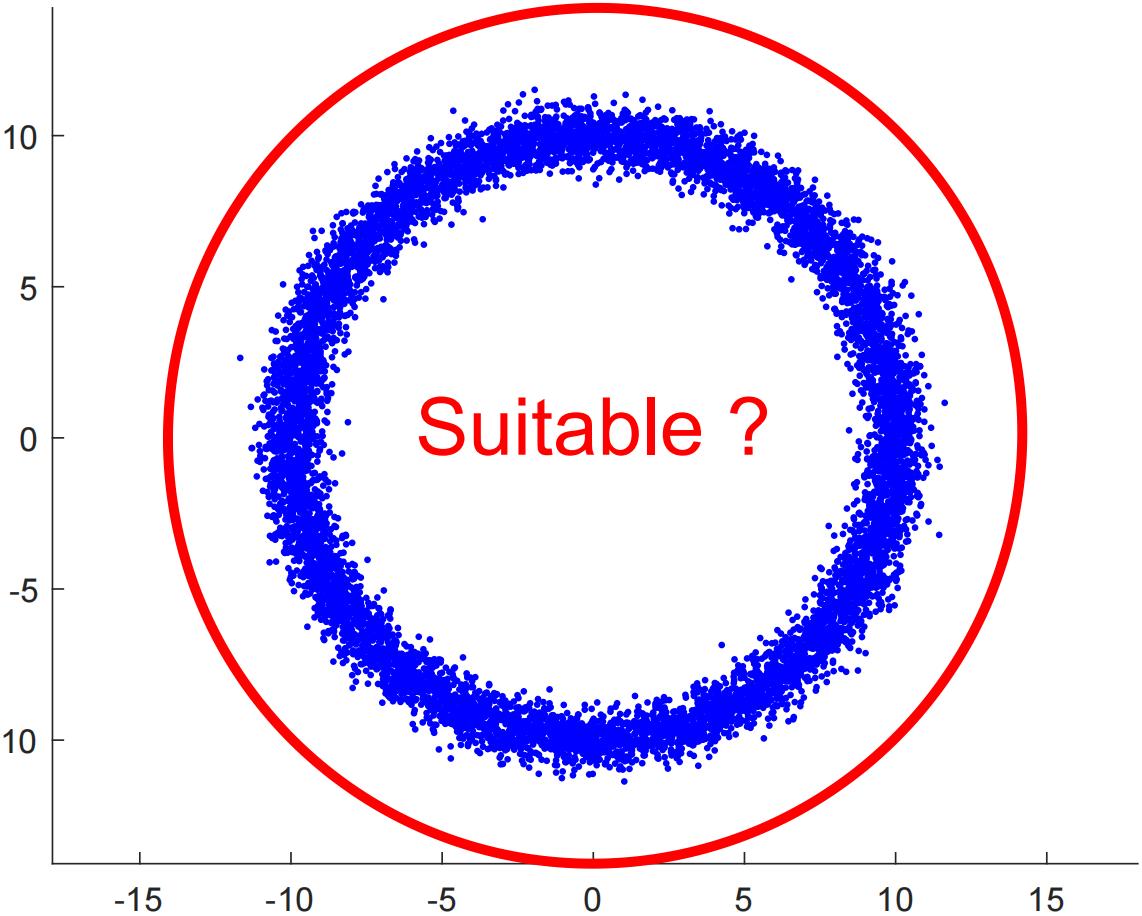}
\end{center}
\caption{Ring-shape distribution}
\label{fig:bad_unimodal_eg2}
\end{figure}

Distributions encountered in reality can hardly be strictly Gaussian, yet the Gaussian distribution assumption is fairly effective for modelling \textit{unique-modal distributions} which are rather common in practical applications. However, if distributions are not unique-modal, then it is inappropriate to approximate them as Gaussian distributions. For example, given a trimodal distribution illustrated in Figure \ref{fig:bad_unimodal_eg1}, would it be suitable to approximate it as a Gaussian distribution? For another example, namely a ring-shape distribution illustrated in Figure \ref{fig:bad_unimodal_eg2}, would it be suitable to approximate it as a Gaussian distribution? The answer is apparently no for both examples.

A representative method to handle multiple-modality is \textbf{Gaussian mixture modelling} \cite{Sorenson1971, Alspach1972}. The idea is to approximate a potentially multiple-modal distribution by a weighted mixture of Gaussian distributions
\begin{subequations}  \label{eq:GMM}
\begin{align}
p(\mathbf{x}) &= \sum_{k=1}^m \alpha_k N(\mathbf{x} \mbox{ } | \hat{\mathbf{x}}_k, \hat{\mathbf{\Sigma}}_k) = \sum_{k=1}^m \alpha_k N(\hat{\mathbf{x}}_k, \hat{\mathbf{\Sigma}}_k)  \\
\mbox{i.e.} \quad \mathbf{x} &\sim \sum_{k=1}^m \alpha_k N(\mathbf{x} \mbox{ } | \hat{\mathbf{x}}_k, \hat{\mathbf{\Sigma}}_k) = \sum_{k=1}^m \alpha_k N(\hat{\mathbf{x}}_k, \hat{\mathbf{\Sigma}}_k)
\end{align}
\end{subequations}
where the Gaussian distribution notations $N(\cdot, \cdot)$ and $N(\cdot \mbox{ } | \cdot, \cdot)$ are defined in (\ref{eq:Gaussian_distribution_notation}). In practice, the total number $m$ of Gaussian distributions is usually set empirically. For recursive estimation, not only the state estimates and covariances $\{\hat{\mathbf{x}}_k, \hat{\mathbf{\Sigma}}_k\}$ associated with the Gaussian distributions but also their mixture weights $\alpha_k$ are revealed dynamically.

\subsubsection*{Sequential Monte Carlo method}

The Gaussian mixture method is already effective and general enough in handling potential multiple-modality involved in a large variety of practical applications. Another representative method that can handle multiple-modality during recursive estimation is the \textbf{sequential Monte Carlo method} \cite{Doucet2001, Arulampalam2002}, which embodies the spirit of \textbf{sampling}
\footnote{\textit{Monte Carlo} refers to a \textit{Côte d'Azur} city famous for its casino, i.e. a place full of \textit{sampling} processes, so the figurative implication of \textit{Monte Carlo} can be easily seen in the statistics domain.}. 
In fact, if given strong enough computational power, the sequential \textit{Monte Carlo} method has the potential to handle arbitrary system and measurement models and arbitrary distributions.

Suppose there is a random variable whose \textit{probability density function} $p(x)$ is not exactly known, but samples $x^i$ ($i \in \{1, 2, 3, \cdots\}$) can be generated according to the distribution of this random variable, i.e. 
\begin{align*}
x^i \sim p(x). 
\end{align*}
We may imagine that a black box outputs random samples consecutively but there is no \textit{a priori} knowledge on the probability density function according to which the black box outputs random samples --- The process of throwing a dice repetitively is an example of such black box processes --- Then this unknown probability density function can be approximated by a large enough number $N$ of generated samples as
\begin{equation} \label{eq:MC}
p(x) \approx \frac{1}{N}\sum_{i=1}^N \delta_{x^i}(x)|_{x^i\sim p(x)}
\end{equation}
and the limit computation
\begin{equation} \label{eq:MC2}
p(x) = \lim_{N \to \infty} \frac{1}{N}\sum_{i=1}^N \delta_{x^i}(x)|_{x^i\sim p(x)},
\end{equation}
where $\delta_{x^i}(x)$ is the \textit{Dirac delta function} with the \textit{Dirac measure} or \textit{Dirac unit mass} at $x^i$.

Now suppose it is difficult to directly generate samples according to the objective probability density function $p(x)$ but easy to generate samples according to another probability density function $q(x)$ as 
\begin{align*}
x^i \sim  q(x). 
\end{align*}
Then this probability density function $p(x)$ can be approximated by a large number of $q(x)$-generated samples as 
\begin{align} \label{eq:importance_smp}
p(x) &= \frac{p(x)}{q(x)}q(x) \approx \frac{p(x)}{q(x)}[\frac{1}{N}\sum_{i=1}^N \delta_{x^i}(x)|_{x^i\sim q(x)}]  \nonumber \\
 &= \frac{1}{N}\sum_{i=1}^N \frac{p(x^i)}{q(x^i)} \delta_{x^i}(x)|_{x^i\sim q(x)} = \sum_{i=1}^N w^i \delta_{x^i}(x)|_{x^i\sim q(x)},
\end{align}
where
\begin{align*}
w^i &= \frac{1}{N} \frac{p(x^i)}{q(x^i)}
\end{align*}
is called \textit{importance weight} or \textit{weight} for short. All weights $w^i$ ($i \in \{1, \cdots, N\}$) are normalized to guarantee that the approximation function described in (\ref{eq:importance_smp}) satisfies the normalization condition. The probability density function $q(x)$, according to which samples are generated, is called \textit{proposal distribution function}, \textit{proposal density}, or \textit{importance density}. The sampling method described by (\ref{eq:importance_smp}) is the \textbf{importance sampling} method.

Given a set of variables 
\begin{align*}
\mathbf{x}_{0:t}=\{x_0, x_1, \cdots, x_t\}, 
\end{align*}
its associated \textit{joint probability density function} $p(\mathbf{x}_{0:t})$ can be approximated via (\ref{eq:MC}) as
\begin{equation*}
p(\mathbf{x}_{0:t}) \approx \frac{1}{N}\sum_{i=1}^N \delta_{\mathbf{x}_{0:t}^i} (\mathbf{x})|_{\mathbf{x}_{0:t}^i \sim p(\mathbf{x}_{0:t})}.
\end{equation*}
However, as $t$ increases, it will become more and more difficult to generate samples directly according to the joint probability density function $p(\mathbf{x}_{0:t})$. To handle the sampling difficulty in the spirit of \textit{divide and conquer}, 
\begin{align*}
x_0^i, \quad x_1^i, \quad \cdots \quad, \quad x_t^i 
\end{align*}
can be generated one by one in a sequential way. Recall the \textit{chain rule}
\begin{equation*}
p(\mathbf{x}_{0:t})=p(x_0) p(x_1 | x_0) p(x_2 | \mathbf{x}_{0:1}) \cdots p(x_t | \mathbf{x}_{0:t-1})
\end{equation*}
and we have a sampling method as
\begin{equation}  \label{eq:sequential_smp}
x_0^i \sim p(x_0), \quad x_1^i \sim p(x_1 | x_0^i), \quad x_2^i \sim p(x_2 | \mathbf{x}_{0:1}^i), \quad \cdots \quad, \quad x_t^i \sim p(x_t | \mathbf{x}_{0:t-1}^i).
\end{equation}
The sampling method described by (\ref{eq:sequential_smp}) is the \textbf{sequential sampling} method. A question arises naturally, can a set of enough samples $\mathbf{x}_{0:t}^i$ generated sequentially via (\ref{eq:sequential_smp}) approximate the joint probability density function $p(\mathbf{x}_{0:t})$? The answer is \textit{yes}. 

Given a set of variables $\mathbf{x}_{0:t}$ and its associated joint probability density function $p(\mathbf{x}_{0:t})$. Now suppose it is even difficult to generate samples of only one element in $\mathbf{x}_{0:t}$, not to mention to generate samples of the whole $\mathbf{x}_{0:t}$ directly according to the joint probability density function $p(\mathbf{x}_{0:t})$. A natural way to handle the sampling difficulty is to combine the importance sampling method and the sequential sampling method, forming the \textbf{sequential importance sampling} method.

More specifically, generate samples $\mathbf{x}_{0:t}^i$ according to a proposal joint probability density function $q(\mathbf{x}_{0:t})$ using the sequential sampling method
\begin{equation}  \label{eq:SIS_1}
x_0^i \sim q(x_0), \quad x_1^i \sim q(x_1 | x_0^i), \quad x_2^i \sim q(x_2 | \mathbf{x}_{0:1}^i), \quad \cdots \quad, \quad x_t^i \sim q(x_t | \mathbf{x}_{0:t-1}^i).
\end{equation}
Represent $p(\mathbf{x}_{0:t})$ by these generated samples via the importance sampling method
\begin{align}  \label{eq:SIS_1b}
p(\mathbf{x}_{0:t}) &= \frac{p(\mathbf{x}_{0:t})}{q(\mathbf{x}_{0:t})} q(\mathbf{x}_{0:t}) \approx \frac{p(\mathbf{x}_{0:t})}{q(\mathbf{x}_{0:t})} \frac{1}{N} \sum_{i=1}^N \delta_{\mathbf{x}_{0:t}^i}(\mathbf{x}_{0:t}) \nonumber \\
 &= \frac{1}{N} \sum_{i=1}^N \frac{p(\mathbf{x}_{0:t}^i)}{q(\mathbf{x}_{0:t}^i)} \delta_{\mathbf{x}_{0:t}^i}(\mathbf{x}_{0:t}) = \frac{1}{N} \sum_{i=1}^N [\frac{p(x_0^i)}{q(x_0^i)} \prod_{k=1}^t \frac{p(x_k^i | \mathbf{x}_{0:k-1}^i)} {q(x_k^i | \mathbf{x}_{0:k-1}^i)}] \delta_{\mathbf{x}_{0:t}^i}(\mathbf{x}_{0:t}),
\end{align}
where samples $\mathbf{x}_{0:t}^i$ ($i \in \{1, \cdots, N\}$) are generated according to $q(\mathbf{x}_{0:t})$ via (\ref{eq:SIS_1}). 

The sequential importance sampling method relies on the assumption that $N$ is large enough to cover the joint distribution of all the $t+1$ single-variable distributions. The needed $N$ increases exponentially as $t$ increases, but in practice only a limited number of samples can be generated. Before long the samples will lose ability to represent the joint distribution of $\mathbf{x}_{0:t}$. This is the \textit{degeneracy problem}.

The \textbf{resampling} method aims at handling the degeneracy problem by dynamically concentrating samples more towards distribution areas of large importance: Given $N$ current samples $\mathbf{x}^i$ with their corresponding weights $w^i$ ($i \in \{1, \cdots, N\}$), generate $N$ new samples $\bar{\mathbf{x}}^i$ from the old samples. Each new sample is generated by randomly selecting an old sample in $\{\mathbf{x}^1, \cdots, \mathbf{x}^N\}$ with the probability of each old sample being selected proportional to its corresponding weight. Set all the weights of the $N$ new samples as 
\begin{align*}
\bar{w}^i = 1/N. 
\end{align*}
Replace old (current) samples by new samples. Resampling may be applied when sample weights are not distributed uniformly enough. An indicator of sample weight uniformness is defined as
\begin{equation} \label{eq:w_uniform}
N_{eff}=\frac{1}{\sum_{i=1}^N (w^i)^2}.
\end{equation}
The more uniform are sample weights, the larger is the indicator $N_{eff}$. Resampling may be applied if $N_{eff}$ is below certain threshold $N_{thr}$. 

Recall (\ref{eq:SIS_1}) and (\ref{eq:SIS_1b}). Define
\begin{equation} \label{eq:SIS_2}
w_t^i \equiv \frac{1}{N} \frac{p(x_0^i)}{q(x_0^i)} \prod_{k=1}^t \frac{p(x_k^i | \mathbf{x}_{0:k-1}^i)} {q(x_k^i | \mathbf{x}_{0:k-1}^i)}.
\end{equation}
Note that
\begin{align*}
w_0^i &= \frac{1}{N} \frac{p(x_0^i)}{q(x_0^i)},  \\
w_k^i &= w_{k-1}^i \frac{p(x_k^i | \mathbf{x}_{0:k-1}^i)} {q(x_k^i | \mathbf{x}_{0:k-1}^i)}  \qquad k \in \{1, \cdots, t\}.
\end{align*}
Formalize (\ref{eq:SIS_1}), (\ref{eq:SIS_1b}), (\ref{eq:SIS_2}) with resampling in a recursive way as follows.
\\

\noindent \rule{\columnwidth}{2pt}
\begin{center}
\textbf{Sequential Importance Sampling with Resampling (SIS/R)} 
\end{center}
\noindent \textbf{Initialization} ($t=0$, $i \in \{1, \cdots, N\}$): \\
$\bullet$ Generate samples $\mathbf{x}_0^i$ as 
\begin{align*} 
\mathbf{x}_0^i \sim q(\mathbf{x}_0).
\end{align*}
$\bullet$ Compute weights $w_0^i$ as
\begin{align*} 
w_0^i &= \frac{p(\mathbf{x}_0^i)}{q(\mathbf{x}_0^i)} 
\end{align*}
\quad and normalize the weights. 

\noindent \textbf{Iteration} ($t\geq 1$, $i \in \{1, \cdots, N\}$): \\
$\bullet$ Generate samples $\mathbf{x}_t^i$ as 
\begin{align*} 
\mathbf{x}_t^i \sim q(\mathbf{x}_t | \mathbf{x}_{0:t-1}^i).
\end{align*}
$\bullet$ Compute weights $w_t^i$ as
\begin{align*} 
w_t^i = w_{t-1}^i \frac{p(\mathbf{x}_t^i | \mathbf{x}_{0:t-1}^i)} {q(\mathbf{x}_t^i | \mathbf{x}_{0:t-1}^i)}
\end{align*}
\quad and normalize the weights. \\
$\bullet$ Compute $N_{eff}$ of $w_t^i$ via (\ref{eq:w_uniform}). If 
\begin{align*} 
N_{eff} < N_{thr}, 
\end{align*}
\quad then resample $\mathbf{x}_{0:t}^i$ and $w_t^i$. \\
\noindent \rule{\columnwidth}{2pt}

\subsubsection*{Particle filter}

Given \textit{a priori} knowledge on $\mathbf{x}_0$, measurements $\mathbf{z}_{1:t}$, the system model $p(\mathbf{x}_t | \mathbf{x}_{t-1})$, and the measurement model $p(\mathbf{z}_t | \mathbf{x}_t)$. Based on the Markov assumption, we have the following Bayesian inference
\begin{equation}  \label{eq:pf_recursive}
p(\mathbf{x}_{0:t} | \mathbf{z}_{1:t}) = \frac{p(\mathbf{z}_t | \mathbf{x}_{0:t}, \mathbf{z}_{1:t-1}) p(\mathbf{x}_{0:t} | \mathbf{z}_{1:t-1})} {p(\mathbf{z}_t | \mathbf{z}_{1:t-1})} \propto p(\mathbf{z}_t | \mathbf{x}_t)p (\mathbf{x}_t | \mathbf{x}_{t-1}) p(\mathbf{x}_{0:t-1} | \mathbf{z}_{1:t-1}),
\end{equation}
which gives a recursive formalism to compute $p(\mathbf{x}_{0:t} | \mathbf{z}_{1:t})$ from $p(\mathbf{x}_{0:t-1} | \mathbf{z}_{1:t-1})$.

If we substitute $p(\mathbf{x}_{0:t} | \mathbf{z}_{1:t})$ for $p(\mathbf{x}_{0:t})$ and substitute $p(\mathbf{z}_t | \mathbf{x}_t) p(\mathbf{x}_t | \mathbf{x}_{t-1})$ for $p(\mathbf{x}_t^i | \mathbf{x}_{0:t-1}^i)$ in the generic sequential importance sampling with resampling method, then we have a recursive estimation version of the method, namely the \textbf{particle filter}, as follows.
\\

\noindent \rule{\columnwidth}{2pt}
\begin{center}
\textbf{Particle Filter (PF)} 
\end{center}
\noindent \textbf{Initialization} ($t=0$, $i \in \{1, \cdots, N\}$): \\
$\bullet$ Generate particles $\mathbf{x}_0^i$ as 
\begin{align*} 
\mathbf{x}_0^i \sim q(\mathbf{x}_0).
\end{align*}
$\bullet$ Compute weights $w_0^i$ as
\begin{align*} 
w_0^i &= \frac{p(\mathbf{x}_0^i)}{q(\mathbf{x}_0^i)} 
\end{align*}
\quad and normalize the weights. 

\noindent \textbf{Iteration} ($t\geq 1$, $i \in \{1, \cdots, N\}$): \\
$\bullet$ Generate particles $\mathbf{x}_t^i$ as 
\begin{align*} 
\mathbf{x}_t^i \sim q(\mathbf{x}_t | \mathbf{x}_{t-1}^i).
\end{align*}
$\bullet$ Compute weights $w_t^i$ as
\begin{align*} 
w_t^i = w_{t-1}^i \frac{p(\mathbf{z}_t | \mathbf{x}_t^i) p(\mathbf{x}_t^i | \mathbf{x}_{t-1}^i)}{q(\mathbf{x}_t^i | \mathbf{x}_{t-1}^i)}
\end{align*}
\quad and normalize the weights. \\
$\bullet$ Compute $N_{eff}$ of $w_t^i$ via (\ref{eq:w_uniform}). If 
\begin{align*} 
N_{eff}<N_{thr}, 
\end{align*}
\quad then resample $\mathbf{x}_t^i$ and $w_t^i$. \\
\noindent \rule{\columnwidth}{2pt} 
\\

In the particle filter, samples are figuratively called \textit{particles} instead of \textit{samples}
\footnote{Each particle or sample $\mathbf{x}_t^i$ implicitly represents a hypothesis for the whole historical trajectory of the state, namely a hypothesis for $\mathbf{x}_{0:t}$.}. In the formalism of the particle filter, the system model $p(\mathbf{x}_t | \mathbf{x}_{t-1})$ and the measurement model $p(\mathbf{z}_t | \mathbf{x}_t)$ can be arbitrary and can handle arbitrary distributions. So the particle filter is a rather generic method for recursive estimation, at least theoretically.
A natural way of setting the proposal density $q(\mathbf{x}_t | \mathbf{x}_{t-1})$ is to set it as $p(\mathbf{x}_t | \mathbf{x}_{t-1})$ and hence the weight update formalism is simplified as
\begin{align*} 
w_t^i = w_{t-1}^i p(\mathbf{z}_t | \mathbf{x}_t^i).
\end{align*}
On the other hand, a proposal density $q(\mathbf{x}_t | \mathbf{x}_{t-1})$ other than $p(\mathbf{x}_t | \mathbf{x}_{t-1})$ can also be adopted according to concrete needs in practical applications.

\subsection{Handle correlation: road until split covariance intersection filter}  \label{sec:split_cif}

Recursive estimation methods described by so far are based on an implicit assumption: \textbf{data independence}. More specifically, in recursive estimation, whenever some pieces of data (including state estimates, system control input, and measurements) are \textit{fused}, they are assumed to be independent of each other or have no correlation with each other. For example, when current measurement $\mathbf{z}_t$ is used to update current \textit{a priori} (or predicted) estimate $\mathbf{\bar{x}}_t$, it is assumed that $\mathbf{z}_t$ is independent of $\mathbf{\bar{x}}_t$, namely $\mathbf{z}_t$ and $\mathbf{\bar{x}}_t$ have no correlation. On the other hand, \textbf{data correlation} may exist in practical applications.

\subsubsection*{Harm due to naive handling of data correlation}

The simplest way to handle data correlation in estimation is \textit{naive handling of data correlation}, namely to totally neglect data correlation. Naive handling of data correlation tends to cause harm to estimation: sometimes the harm caused might be negligible, and other times the harm caused can be severe and even disastrous. To understand harm due to naive handling of data correlation, we need to introduce the concept of \textbf{estimate consistency} \cite{Julier1997b, Julier2001, Li2013a}.

Given a generic estimate $\{\mathbf{x}, \mathbf{\Sigma}\}$ where $\mathbf{x}$ and $\mathbf{\Sigma}$ denote the estimated state vector and the estimated covariance matrix respectively. Let 
\begin{align*}  
\mathbf{x_m}, \quad \mathbf{\tilde{x}}, \quad \mathbf{\Sigma_m} 
\end{align*}
denote the ground truth of $\mathbf{x}$, that of $\mathbf{x}$ error with respect to $\mathbf{x_m}$, and that of the $\mathbf{x}$ covariance. \textbf{Consistency} means that \textit{the estimated covariance matrix is no smaller than its true covariance}, formalized as
\begin{align}  \label{eq:consistency}
\mathbf{\Sigma} \geq \mathbf{\Sigma_m} \equiv \mathbf{E} [\mathbf{\tilde{x}} \mathbf{\tilde{x}}^\mathrm{T}] = \mathbf{E} [(\mathbf{x} - \mathbf{x_m}) (\mathbf{x} - \mathbf{x_m})^\mathrm{T}]
\end{align}
namely
\begin{align*}
\mathbf{\Sigma} - \mathbf{\Sigma_m} \geq 0.
\end{align*}
where $\geq 0$ means \textit{positive semi-definite} and $>0$ means \textit{positive definite}. Simply speaking, \textbf{estimate consistency} means that the estimate is not over-confident, whereas \textbf{estimate inconsistency} means that the estimate is over-confident. 

The author has no intention to argue that defining estimate consistency as 
\begin{align*}
\mathbf{\Sigma} \geq \mathbf{\Sigma_m}
\end{align*}
would be the best choice
\footnote{The author has to admit that the definition of estimate consistency seems a bit strange. Why cannot we simply define estimate consistency as $\mathbf{\Sigma} = \mathbf{\Sigma_m}$? In practical applications, $\mathbf{\Sigma} = \mathbf{\Sigma_m}$ or even $\mathbf{\Sigma} \approx \mathbf{\Sigma_m}$ is a too exigent requirement that can hardly be satisfied, so defining estimate consistency as $\mathbf{\Sigma} = \mathbf{\Sigma_m}$ seems more comprehensible but is of little practical use.}
but would just like to point out that such definition is appropriate enough to explain harmful phenomena due to bad handling especially naive handling of data correlation. To facilitate understanding, consider two extreme cases, namely two estimates 
\begin{align*}
\{\mathbf{x}, \mathbf{\Sigma} = \infty\}, \quad \{\mathbf{x}, \mathbf{\Sigma} = 0\}. 
\end{align*}
The first estimate is consistent whereas the second estimate is inconsistent. The infinite covariance of the first estimate, i.e. 
\begin{align*}
\mathbf{\Sigma} = \infty
\end{align*}
implies that the first estimate does not have any self-confidence, which is undesirable. The zero covariance of the second estimate, i.e. 
\begin{align*}
\mathbf{\Sigma} = 0
\end{align*}
implies that the second estimate is absolutely over-confident, which is undesirable either. Although both estimates are undesirable, they are essentially different: the first estimate is somewhat acceptable whereas the second estimate is unacceptable at all. The reason is that the first estimate still has the chance to improve itself by fusion with more estimates. In contrast, the second estimate has no chance at all to improve itself because it is so over-confident that it does not allow for any change of itself --- A social analogue for this phenomenon is that a person who currently lacks self-confidence severely still has the potential to progress because this person is willing to learn from others' merits, whereas another person who is severely over-confident has no space for progress because this other person is unwilling to listen to others' advices.   

Naive handling of data correlation may cause estimate inconsistency. For example, let two estimates \{$\mathbf{x}_1$, $\mathbf{\Sigma}_1$\} and \{$\mathbf{x}_2$, $\mathbf{\Sigma}_2$\} be the same to a single source estimate $\{\mathbf{x}, \mathbf{\Sigma}\}$ and let the two estimates be fused by the Kalman filter (\ref{eq:WAfullDim}) with naive handling of data correlation
\begin{align*} 
\hat{\mathbf{\Sigma}}^{-1} &= \mathbf{\Sigma}_1^{-1} + \mathbf{\Sigma}_2^{-1} = 2 \mathbf{\Sigma}^{-1},  \\
\hat{\mathbf{x}} &= \hat{\mathbf{\Sigma}} (\mathbf{\Sigma}_1^{-1} \mathbf{x}_1 + \mathbf{\Sigma}_2^{-1} \mathbf{x}_2) = \mathbf{x}.
\end{align*}
So the fusion estimate is $\{\mathbf{x}, \mathbf{\Sigma}/2\}$. However, it is obvious that the ground truth of the fusion estimate should still be the same to the source estimate $\{\mathbf{x}, \mathbf{\Sigma}\}$, which implies that the fusion estimate $\{\mathbf{x}, \mathbf{\Sigma}/2\}$ is inconsistent.

Naive handling of data correlation may cause even severe estimate inconsistency in the form of \textit{circular reasoning} that involves agents of network relationship, such as in connected robots applications \cite{Fox2000, Howard2006, Li2013b}. To demonstrate harm due to naive handling of data correlation in the form of circular reasoning, we adopt an example of circular reasoning between two estimators $\mathbf{E_A}$ and $\mathbf{E_B}$ that provide estimates of the same entity. Besides, suppose their initial estimates are the same to a single source estimate $\{\mathbf{x}, \mathbf{\Sigma}\}$. They alternately send their own estimate to the other for fusion, and each round of fusion is performed by the Kalman filter with naive handling of data correlation. The estimates of $\mathbf{E_A}$ and $\mathbf{E_B}$ after each round of circular reasoning are listed as
\begin{align*}
\begin{matrix}
\mbox{Round} & 0 & 1 & 2 & 3 & 4 & 5 & \cdots & \infty  \\
\mathbf{E_A} & \{\mathbf{x}, \mathbf{\Sigma}\} & \{\mathbf{x}, \mathbf{\Sigma}\} & \{\mathbf{x}, \mathbf{\Sigma}/3\} & \{\mathbf{x}, \mathbf{\Sigma}/3\} & \{\mathbf{x}, \mathbf{\Sigma}/8\} & \{\mathbf{x}, \mathbf{\Sigma}/8\} & \cdots & \{\mathbf{x}, \mathbf{0}\}  \\
\mathbf{E_B} & \{\mathbf{x}, \mathbf{\Sigma}\} & \{\mathbf{x}, \mathbf{\Sigma}/2\} & \{\mathbf{x}, \mathbf{\Sigma}/2\} & \{\mathbf{x}, \mathbf{\Sigma}/5\} & \{\mathbf{x}, \mathbf{\Sigma}/5\} & \{\mathbf{x}, \mathbf{\Sigma}/13\} & \cdots & \{\mathbf{x}, \mathbf{0}\}
\end{matrix}
\end{align*}
The only independent data ever involved in above process of circular reasoning is the single source estimate $\{\mathbf{x}, \mathbf{\Sigma}\}$ at the very beginning, so that the ground truth of the fusion estimate should be $\{\mathbf{x}, \mathbf{\Sigma}\}$ as well. However, the estimated covariances of $\mathbf{E_A}$ and $\mathbf{E_B}$ both converge to zero, which implies that the estimates of $\mathbf{E_A}$ and $\mathbf{E_B}$ become severely inconsistent after circular reasoning with naive handling of data correlation. For agents with more complicated network relationship, circular reasoning with naive handling of data correlation tends to incur even severe estimate inconsistency. One piece of source data may be indirectly fused again and again during circular reasoning with naive handling of data correlation. Consequently, agents may have severe over-confidence in this single piece of source data --- This phenomenon is similar to how a rumour becomes ``true'' in society.

\subsubsection*{Handle known data correlation}

Given two estimates \{$\mathbf{x}_1$, $\mathbf{\Sigma}_1$\} and \{$\mathbf{x}_2$, $\mathbf{\Sigma}_2$\} with cross covariances $\mathbf{\Sigma}_{12}$ and $\mathbf{\Sigma}_{21}$, satisfying 
\begin{align*} 
\mathbf{\Sigma}_{21} = \mathbf{\Sigma}_{12}^\mathrm{T}. 
\end{align*}
Let \{$\hat{\mathbf{x}}$, $\hat{\mathbf{\Sigma}}$\} denote the fusion estimate of estimates \{$\mathbf{x}_1$, $\mathbf{\Sigma}_1$\} and \{$\mathbf{x}_2$, $\mathbf{\Sigma}_2$\}, then the fusion estimate \{$\hat{\mathbf{x}}$, $\hat{\mathbf{\Sigma}}$\} can be approximately obtained \cite{Bar-Shalom1986} as
\begin{subequations}  \label{eq:known_correlation_fuse_approx}
\begin{align}
\hat{\mathbf{\Sigma}} &= \mathbf{\Sigma}_1 - (\mathbf{\Sigma}_1 - \mathbf{\Sigma}_{12}) (\mathbf{\Sigma}_1 + \mathbf{\Sigma}_2 - \mathbf{\Sigma}_{12} - \mathbf{\Sigma}_{21})^{-1} (\mathbf{\Sigma}_1 - \mathbf{\Sigma}_{21}),  \\
\hat{\mathbf{x}} &= \mathbf{x}_1 + (\mathbf{\Sigma}_1 - \mathbf{\Sigma}_{12}) (\mathbf{\Sigma}_1 + \mathbf{\Sigma}_2 - \mathbf{\Sigma}_{12} - \mathbf{\Sigma}_{21})^{-1} (\mathbf{x}_2 - \mathbf{x}_1).
\end{align}
\end{subequations}

Fusion of source estimates with known data correlation can also be performed in the spirit of \textit{information} analysis \cite{Chang1997}. Given two estimates \{$\mathbf{x}_1$, $\mathbf{\Sigma}_1$\} and \{$\mathbf{x}_2$, $\mathbf{\Sigma}_2$\} that share common information represented by a source estimate \{$\mathbf{x}_0$, $\mathbf{\Sigma}_0$\}. Besides, the estimate \{$\mathbf{x}_1$, $\mathbf{\Sigma}_1$\} contains independent information represented by a source estimate \{$\mathbf{x}_a$, $\mathbf{\Sigma}_a$\} that fuses with the source estimate \{$\mathbf{x}_0$, $\mathbf{\Sigma}_0$\} to form \{$\mathbf{x}_1$, $\mathbf{\Sigma}_1$\}, namely
\begin{align*} 
\mathbf{\Sigma}_1^{-1} &= \mathbf{\Sigma}_0^{-1} + \mathbf{\Sigma}_a^{-1},  \\
\mathbf{x}_1 &= \mathbf{\Sigma}_1 (\mathbf{\Sigma}_0^{-1} \mathbf{x}_0 + \mathbf{\Sigma}_a^{-1} \mathbf{x}_a) \iff \mathbf{\Sigma}_1^{-1} \mathbf{x}_1 = \mathbf{\Sigma}_0^{-1} \mathbf{x}_0 + \mathbf{\Sigma}_a^{-1} \mathbf{x}_a.
\end{align*}
The estimate \{$\mathbf{x}_2$, $\mathbf{\Sigma}_2$\} contains independent information represented by a source estimate \{$\mathbf{x}_b$, $\mathbf{\Sigma}_b$\} that fuses with the source estimate \{$\mathbf{x}_0$, $\mathbf{\Sigma}_0$\} to form \{$\mathbf{x}_2$, $\mathbf{\Sigma}_2$\}, namely
\begin{align*} 
\mathbf{\Sigma}_2^{-1} &= \mathbf{\Sigma}_0^{-1} + \mathbf{\Sigma}_b^{-1},  \\
\mathbf{x}_2 &= \mathbf{\Sigma}_2 (\mathbf{\Sigma}_0^{-1} \mathbf{x}_0 + \mathbf{\Sigma}_b^{-1} \mathbf{x}_b) \iff \mathbf{\Sigma}_2^{-1} \mathbf{x}_2 = \mathbf{\Sigma}_0^{-1} \mathbf{x}_0 + \mathbf{\Sigma}_b^{-1} \mathbf{x}_b.
\end{align*}

The fusion estimate \{$\hat{\mathbf{x}}$, $\hat{\mathbf{\Sigma}}$\} is actually contributed by three independent estimates \{$\mathbf{x}_0$, $\mathbf{\Sigma}_0$\}, \{$\mathbf{x}_a$, $\mathbf{\Sigma}_a$\}, and \{$\mathbf{x}_b$, $\mathbf{\Sigma}_b$\}. 
\begin{align*}
\hat{\mathbf{\Sigma}}^{-1} &= \mathbf{\Sigma}_0^{-1} + \mathbf{\Sigma}_a^{-1} + \mathbf{\Sigma}_b^{-1} \\
  &= \mathbf{\Sigma}_0^{-1} + (\mathbf{\Sigma}_1^{-1} - \mathbf{\Sigma}_0^{-1}) + (\mathbf{\Sigma}_2^{-1} - \mathbf{\Sigma}_0^{-1}) \\
  &= \mathbf{\Sigma}_1^{-1} + \mathbf{\Sigma}_2^{-1} - \mathbf{\Sigma}_0^{-1},  \\
\hat{\mathbf{\Sigma}}^{-1} \hat{\mathbf{x}} &= \mathbf{\Sigma}_0^{-1} \mathbf{x}_0 + \mathbf{\Sigma}_a^{-1} \mathbf{x}_a + \mathbf{\Sigma}_b^{-1} \mathbf{x}_b  \\
  &= \mathbf{\Sigma}_0^{-1} \mathbf{x}_0 + (\mathbf{\Sigma}_1^{-1} \mathbf{x}_1 - \mathbf{\Sigma}_0^{-1} \mathbf{x}_0) + (\mathbf{\Sigma}_2^{-1} \mathbf{x}_2 - \mathbf{\Sigma}_0^{-1} \mathbf{x}_0)  \\
  &= \mathbf{\Sigma}_1^{-1} \mathbf{x}_1 + \mathbf{\Sigma}_2^{-1} \mathbf{x}_2 - \mathbf{\Sigma}_0^{-1} \mathbf{x}_0
\end{align*} 
namely
\begin{subequations}  \label{eq:imf}
\begin{align}
\hat{\mathbf{\Sigma}}^{-1} &= \mathbf{\Sigma}_1^{-1} + \mathbf{\Sigma}_2^{-1} - \mathbf{\Sigma}_0^{-1},  \\
\hat{\mathbf{x}} &= \hat{\mathbf{\Sigma}} (\mathbf{\Sigma}_1^{-1} \mathbf{x}_1 + \mathbf{\Sigma}_2^{-1} \mathbf{x}_2 - \mathbf{\Sigma}_0^{-1} \mathbf{x}_0).
\end{align}
\end{subequations}
The fusion method formalized in (\ref{eq:imf}) is the \textbf{information matrix filter} which provides a mechanism to fuse estimates with known correlation.

A typical example of known data correlation in practical applications is that caused by \textit{common process noise}. Both the fusion method formalized in (\ref{eq:known_correlation_fuse_approx}) and the information matrix filter formalized in (\ref{eq:imf}) stem from the concrete need to handle common process noise in practical applications. Common process noise will also be involved in practical applications for which the \textbf{federated Kalman filter} \cite{Carlson1990} is applied.

\begin{figure}[h!]
\begin{center}
\includegraphics[width=0.7\columnwidth]{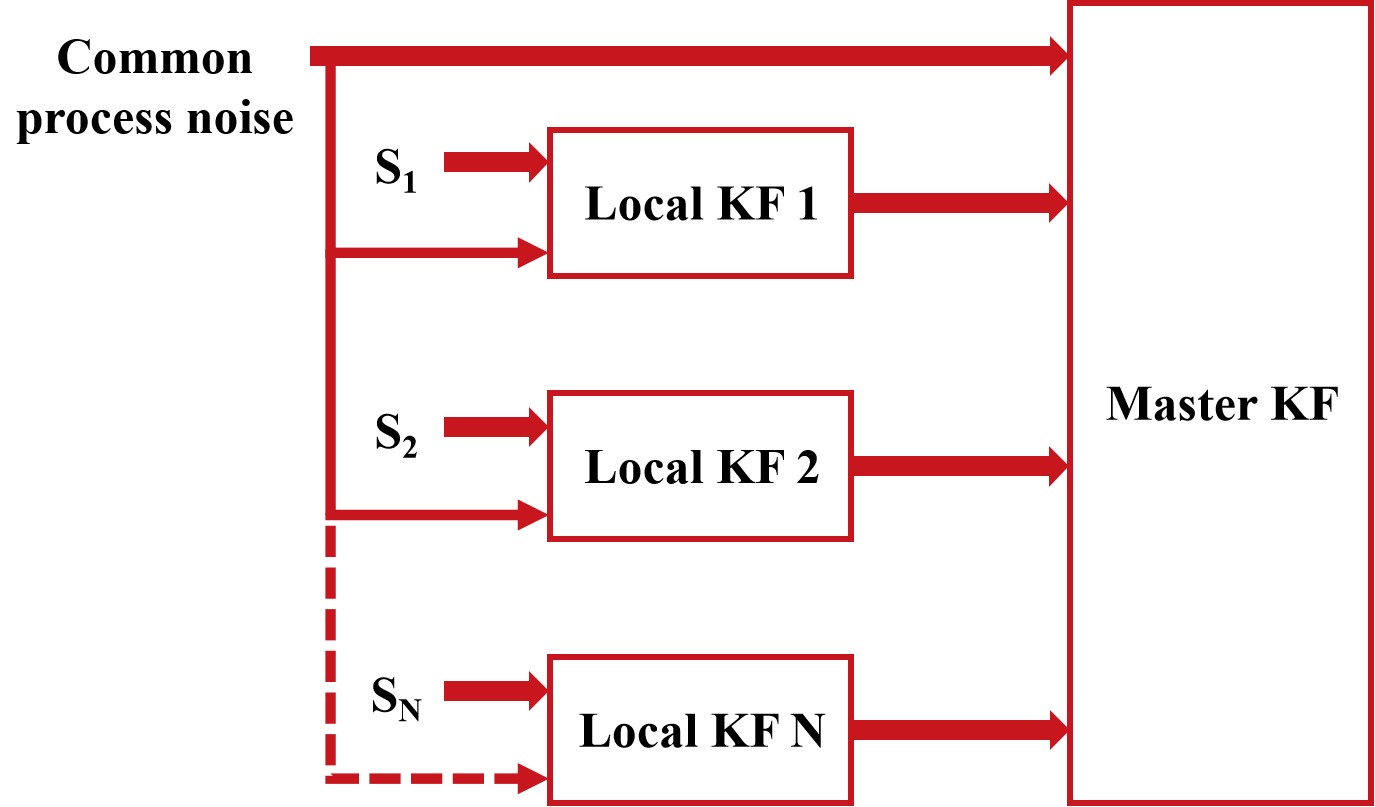}
\end{center}
\caption{Federated Kalman filter}
\label{fig:federated_kalman_filter}
\end{figure}

The spirit of the federated Kalman filter consists in a hierarchical fusion architecture, where the low-level layer consists of a number of \textit{local Kalman filters} that fuse various raw sensor data and the high-level layer is a \textit{master Kalman filter} that further fuses estimates from local Kalman filters. The hierarchical fusion mechanism of the federated Kalman filter is illustrated in Figure \ref{fig:federated_kalman_filter}. 

The local Kalman filters fuse their respective sensor data on one hand, and fuse a common source of (virtual) data namely common process noise on the other hand. To handle known data correlation caused by common process noise (denote its covariance as $\mathbf{\Sigma}_0$), the federated Kalman filter debuts the covariance expansion technique
\begin{equation}  \label{eq:federatedKF_cov_expansion}
\begin{bmatrix} \mathbf{\Sigma}_0 \\ \mathbf{\Sigma}_0 \\ \vdots \\ \mathbf{\Sigma}_0 \end{bmatrix} \quad \implies \quad \begin{bmatrix} \frac{1}{w_1} \mathbf{\Sigma}_0 \\ \frac{1}{w_2} \mathbf{\Sigma}_0 \\ \vdots \\ \frac{1}{w_N} \mathbf{\Sigma}_0 \end{bmatrix},
\end{equation}
where the expansion factors $w_i$ ($i \in \{1, \cdots, N\}$) satisfy the normalization condition
\begin{align*}
w_1 + w_2 + \cdots + w_N = 1.
\end{align*}
The covariance expansion technique specified in (\ref{eq:federatedKF_cov_expansion}) is theoretically rooted in the matrix inequality
\begin{equation}  \label{eq:cov_expansion_matrix_ineq}
\begin{bmatrix} \mathbf{\Sigma}_0 & \mathbf{\Sigma}_0 & \cdots & \mathbf{\Sigma}_0 \\ \mathbf{\Sigma}_0 & \mathbf{\Sigma}_0 & \cdots & \mathbf{\Sigma}_0 \\ \vdots & \vdots & \ddots & \vdots \\ \mathbf{\Sigma}_0 & \mathbf{\Sigma}_0 & \cdots & \mathbf{\Sigma}_0 \end{bmatrix} \leq \begin{bmatrix} \frac{1}{w_1} \mathbf{\Sigma}_0 & & & \\ & \frac{1}{w_2} \mathbf{\Sigma}_0 & & \\ & & \ddots & \\ & & & \frac{1}{w_N} \mathbf{\Sigma}_0 \end{bmatrix}.
\end{equation}

\begin{proof}
The matrix inequality (\ref{eq:cov_expansion_matrix_ineq}) is equivalent to
\begin{equation}  \label{eq:cov_expansion_matrix_ineq2}
\forall \mathbf{x}_i, \cdots, \mathbf{x}_N \qquad \sum_{i=1}^N \sum_{j=1}^N \mathbf{x}_i^\mathrm{T} \mathbf{\Sigma}_0 \mathbf{x}_j \leq \sum_{i=1}^N \frac{1}{w_i} \mathbf{x}_i^\mathrm{T} \mathbf{\Sigma}_0 \mathbf{x}_i.
\end{equation}
Decompose $\mathbf{\Sigma}_0$ as
\begin{align*}
\mathbf{\Sigma}_0 = \mathbf{S}^\mathrm{T} \mathbf{S}
\end{align*}
and denote
\begin{align*}
\mathbf{y}_i \equiv \mathbf{S} \mathbf{x}_i, \quad i \in \{1, \cdots, N\}.
\end{align*}

In the light of the Lagrange identity \cite{Mitrinovic1970} and note that
\begin{align*}
\sum_{i=1}^N w_i = 1,
\end{align*}
we have
\begin{align*}
&\sum_{i=1}^N \frac{1}{w_i} \mathbf{x}_i^\mathrm{T} \mathbf{\Sigma}_0 \mathbf{x}_i - \sum_{i=1}^N \sum_{j=1}^N \mathbf{x}_i^\mathrm{T} \mathbf{\Sigma}_0 \mathbf{x}_j = (\sum_{i=1}^N w_i) (\sum_{i=1}^N \frac{1}{w_i} \mathbf{y}_i^\mathrm{T} \mathbf{y}_i) - (\sum_{i=1}^N \mathbf{y}_i)^\mathrm{T} (\sum_{i=1}^N \mathbf{y}_i)  \\
=& \sum_{1 \leq i < j \leq m} (\sqrt{w_i} \frac{1}{\sqrt{w_j}} \mathbf{y}_j - \sqrt{w_j} \frac{1}{\sqrt{w_i}} \mathbf{y}_i)^\mathrm{T} (\sqrt{w_i} \frac{1}{\sqrt{w_j}} \mathbf{y}_j - \sqrt{w_j} \frac{1}{\sqrt{w_i}} \mathbf{y}_i) \geq 0.
\end{align*}
The proof is done.
\end{proof}

\subsubsection*{Handle unknown data correlation}

Data correlation is not always known in practical applications. We may encounter recursive estimation that needs to handle unknown data correlation, which is especially common in decentralized recursive estimation. Given two estimates \{$\mathbf{x}_1$, $\mathbf{\Sigma}_1$\} and \{$\mathbf{x}_2$, $\mathbf{\Sigma}_2$\} whose correlation is unknown. How to fuse these two estimates so that their fusion estimate \{$\mathbf{x}$, $\mathbf{\Sigma}$\} can be guaranteed consistent? This problem is first solved by the covariance intersection method (or filter) \cite{Julier1997b}, which is formalized as
\begin{subequations}  \label{eq:cif}
\begin{align}
\mathbf{\Sigma}^{-1} &= w \mathbf{\Sigma}_1^{-1} + (1-w) \mathbf{\Sigma}_2^{-1},  \\
\mathbf{x} &= \mathbf{\Sigma} (w \mathbf{\Sigma}_1^{-1} \mathbf{x}_1 + (1-w) \mathbf{\Sigma}_2^{-1} \mathbf{x}_2).
\end{align}
\end{subequations}
The fusion consistency of the covariance intersection filter is theoretically guaranteed for arbitrary choice of $w \in [0, 1]$. On the other hand, the covariance intersection filter has a drawback of yielding pessimistic estimates, because it treats all source data as being totally correlated and has no mechanism to exploit independent information contained potentially in data \cite{Li2013a}.

\subsubsection*{Handle both known independence and unknown correlation}

The Kalman filter has the merit of being optimal in fusing two independent (Gaussian-modelled) estimates
\footnote{This is proved in Section \ref{sec:opt_wgt_average}.}
and the covariance intersection filter has the merit of guaranteeing fusion consistency. A question arises naturally: \textit{can we have a method that possesses merits of both the Kalman filter and the covariance intersection filter}? The attempt to find an answer to this question leads to the birth of the split covariance intersection filter (Split CIF) \cite{Julier2001, Li2013a}. The split covariance intersection filter is presented briefly in \cite{Julier2001} only as a heuristic extension of the covariance intersection filter, whereas a complete theoretical foundation for the split covariance intersection filter is established in \cite{Li2013a}.

The split covariance intersection filter can be regarded as a generalization of both the Kalman filter and the covariance intersection filter. It enjoys formalism simplicity and implementation convenience. It overcomes the drawback of the Kalman filter when fusing correlated data and overcomes the drawback of the covariance intersection filter when fusing data that contain known independent information. The split covariance intersection filter is a useful tool for general data fusion and recursive estimation: it can reasonably handle both unknown correlated information and known independent information in source data. It has the potential to be applied in a variety of engineering activities. Representative application of the split covariance intersection filter can be in cooperative intelligent systems \cite{Li2024ITS, Li2013b, Pierre2018, ChenX2020, Li2022TITS} and in single intelligent system operation \cite{Li2013d, Li2022RAL, Li2023AprilTagNavigation} as well.

Represent a generic estimate in split form 
\begin{align*}
\{\mathbf{x}, \mathbf{\Sigma}_d+\mathbf{\Sigma}_i\}, 
\end{align*}
where the covariance component $\mathbf{\Sigma}_d$ represents the maximum degree to which the estimate is potentially correlated with others, and the covariance component $\mathbf{\Sigma}_i$ represents the degree of its independence. Both $\mathbf{\Sigma}_d$ and $\mathbf{\Sigma}_i$ are always positive semi-definite and their sum is always positive definite. To understand the split form, we may imagine that $\mathbf{x}$ consists of a correlated component $\mathbf{x}_d$ and an independent component $\mathbf{x}_i$ as 
\begin{align*}
\mathbf{x} = \mathbf{x}_d + \mathbf{x}_i, 
\end{align*}
where the estimated covariance matrices for $\mathbf{x}_d$ and $\mathbf{x}_i$ are regarded to be $\mathbf{\Sigma}_d$ and $\mathbf{\Sigma}_i$ respectively. Given two source estimates in split form 
\begin{align*}
\{\mathbf{x}_1, \mathbf{\Sigma}_{1d}+\mathbf{\Sigma}_{1i}\}, \quad \{\mathbf{x}_2, \mathbf{\Sigma}_{2d}+\mathbf{\Sigma}_{2i}\}, 
\end{align*}
the fusion estimate can be obtained via the \textbf{split covariance intersection filter} 
\begin{subequations}  \label{eq:scif}
\begin{align}
\mathbf{\Sigma}_1 &= \mathbf{\Sigma}_{1d}/w + \mathbf{\Sigma}_{1i},  \\
\mathbf{\Sigma}_2 &= \mathbf{\Sigma}_{2d}/(1-w) + \mathbf{\Sigma}_{2i},  \\
\mathbf{\Sigma} &= (\mathbf{\Sigma}_1^{-1} + \mathbf{\Sigma}_2^{-1})^{-1},  \\
\mathbf{x} &= \mathbf{\Sigma}(\mathbf{\Sigma}_1^{-1} \mathbf{x}_1 + \mathbf{\Sigma}_2^{-1} \mathbf{x}_2), \\
\mathbf{\Sigma}_i &= \mathbf{\Sigma}(\mathbf{\Sigma}_1^{-1} \mathbf{\Sigma}_{1i} \mathbf{\Sigma}_1^{-1} + \mathbf{\Sigma}_2^{-1} \mathbf{\Sigma}_{2i} \mathbf{\Sigma}_2^{-1}) \mathbf{\Sigma},  \\
\mathbf{\Sigma}_d &= \mathbf{\Sigma} - \mathbf{\Sigma}_i,
\end{align}
\end{subequations}
where $w \in [0,1]$. 

The weight $w$ is determined by minimizing the determinant of the new covariance, namely by solving the following $w$-optimization problem 
\begin{equation}  \label{eq:wopt}
w = \arg \min_{w \in [0,1]} \det(\mathbf{\Sigma}(w)),
\end{equation}
where 
\begin{align*}
\mathbf{\Sigma}(w) = (\mathbf{\Sigma}_{1}(w)^{-1} + \mathbf{\Sigma}_{2}(w)^{-1})^{-1}. 
\end{align*}
When $w$ takes the value $0$ or $1$, the new covariance $\mathbf{\Sigma}(w)$ denotes the limit value corresponding to $0$ or $1$ respectively, i.e.
\begin{align*}
\mathbf{\Sigma}(0) &= \lim_{w \to 0} \mathbf{\Sigma}(w),  \\
\mathbf{\Sigma}(1) &= \lim_{w \to 1} \mathbf{\Sigma}(w).
\end{align*}

The concept of estimate consistency defined by (\ref{eq:consistency}) is not enough to characterize the consistency property of an estimate in split form. To characterize consistency properties of estimates in split form, we resort to the concept of \textbf{split consistency} \cite{Li2013a}: Given a generic estimate 
\begin{align*}
\{\mathbf{x}, \mathbf{\Sigma}_d+\mathbf{\Sigma}_i\} 
\end{align*}
or imaginarily as 
\begin{align*}
\{\mathbf{x}_d + \mathbf{x}_i, \mathbf{\Sigma}_d+\mathbf{\Sigma}_i\}, 
\end{align*}
it is \textbf{split consistent} if it satisfies
\begin{subequations}  \label{eq:split_consistency}
\begin{align}
\mathbf{\Sigma}_d &\geq \mathbf{\Sigma_m}_d = \mathbf{E} [\mathbf{\tilde{x}}_d \mathbf{\tilde{x}}_d^\mathrm{T}],  \\
\mathbf{\Sigma} = \mathbf{\Sigma}_d + \mathbf{\Sigma}_i &\geq \mathbf{\Sigma_m}_d + \mathbf{\Sigma_m}_i = \mathbf{\Sigma_m} = \mathbf{E} [\mathbf{\tilde{x}} \mathbf{\tilde{x}}^\mathrm{T}].
\end{align}
\end{subequations}
The fusion consistency of the split covariance intersection filter is theoretically guaranteed for arbitrary choice of $w \in [0, 1]$. More specifically, given two source estimates 
\begin{align*}
\{\mathbf{x}_1, \mathbf{\Sigma}_{1d}+\mathbf{\Sigma}_{1i}\}, \quad \{\mathbf{x}_2, \mathbf{\Sigma}_{2d}+\mathbf{\Sigma}_{2i}\}
\end{align*}
that are split consistent, namely
\begin{align*}
\mathbf{\Sigma}_{1d} &\geq \mathbf{E} [\mathbf{\tilde{x}}_{1d} \mathbf{\tilde{x}}_{1d}^\mathrm{T}], \quad \mathbf{\Sigma}_{1d} + \mathbf{\Sigma}_{1i} \geq \mathbf{E} [\mathbf{\tilde{x}}_1 \mathbf{\tilde{x}}_1^\mathrm{T}],  \\
\mathbf{\Sigma}_{2d} &\geq \mathbf{E} [\mathbf{\tilde{x}}_{2d} \mathbf{\tilde{x}}_{2d}^\mathrm{T}], \quad \mathbf{\Sigma}_{2d} + \mathbf{\Sigma}_{2i} \geq \mathbf{E} [\mathbf{\tilde{x}}_2 \mathbf{\tilde{x}}_2^\mathrm{T}].
\end{align*}
For any choice of $w \in [0,1]$, the fusion estimate \{$\mathbf{x}$, $\mathbf{\Sigma}_d+\mathbf{\Sigma}_i$\} obtained via the split covariance intersection filter (\ref{eq:scif}) is guaranteed to be split consistent, namely
\begin{align*}
\mathbf{\Sigma}_d &\geq \mathbf{E} [\mathbf{\tilde{x}}_d \mathbf{\tilde{x}}_d^\mathrm{T}], \quad \mathbf{\Sigma}_d + \mathbf{\Sigma}_i \geq \mathbf{E} [\mathbf{\tilde{x}} \mathbf{\tilde{x}}^\mathrm{T}].
\end{align*}

\begin{proof}
Denote 
\begin{align*}
\bar{w} = 1-w
\end{align*}
and obtain
\begin{align*}
\mathbf{\Sigma}_d &= \mathbf{\Sigma} - \mathbf{\Sigma}_i = \mathbf{\Sigma} - \mathbf{\Sigma}(\mathbf{\Sigma}_{1}^{-1} \mathbf{\Sigma}_{1i} \mathbf{\Sigma}_{1}^{-1} + \mathbf{\Sigma}_{2}^{-1} \mathbf{\Sigma}_{2i} \mathbf{\Sigma}_{2}^{-1}) \mathbf{\Sigma}  \\
  &= \mathbf{\Sigma} (\mathbf{\Sigma}^{-1} - \mathbf{\Sigma}_{1}^{-1} \mathbf{\Sigma}_{1i} \mathbf{\Sigma}_{1}^{-1} - \mathbf{\Sigma}_{2}^{-1} \mathbf{\Sigma}_{2i} \mathbf{\Sigma}_{2}^{-1}) \mathbf{\Sigma}  \\
  &= \mathbf{\Sigma} (\mathbf{\Sigma}_1^{-1} \mathbf{\Sigma}_1 \mathbf{\Sigma}_1^{-1} + \mathbf{\Sigma}_2^{-1} \mathbf{\Sigma}_2 \mathbf{\Sigma}_2^{-1} - \mathbf{\Sigma}_1^{-1} \mathbf{\Sigma}_{1i} \mathbf{\Sigma}_1^{-1} - \mathbf{\Sigma}_2^{-1} \mathbf{\Sigma}_{2i} \mathbf{\Sigma}_2^{-1}) \mathbf{\Sigma}  \\
  &= \mathbf{\Sigma} (\mathbf{\Sigma}_1^{-1} \frac{\mathbf{\Sigma}_{1d}}{w} \mathbf{\Sigma}_1^{-1} + \mathbf{\Sigma}_2^{-1} \frac{\mathbf{\Sigma}_{2d}}{\bar{w}}  \mathbf{\Sigma}_2^{-1}) \mathbf{\Sigma}  \\
  &\geq \mathbf{\Sigma} (\mathbf{\Sigma}_1^{-1} \frac{\mathbf{E} [\mathbf{\tilde{x}}_{1d} \mathbf{\tilde{x}}_{1d}^\mathrm{T}]}{w} \mathbf{\Sigma}_1^{-1} + \mathbf{\Sigma}_2^{-1} \frac{\mathbf{E} [\mathbf{\tilde{x}}_{2d} \mathbf{\tilde{x}}_{2d}^\mathrm{T}]}{\bar{w}}  \mathbf{\Sigma}_2^{-1}) \mathbf{\Sigma}.
\end{align*}
Note that
\begin{align*}
\mathbf{\tilde{x}}_d = \mathbf{\Sigma}(\mathbf{\Sigma}_1^{-1} \mathbf{\tilde{x}}_{1d} + \mathbf{\Sigma}_2^{-1} \mathbf{\tilde{x}}_{2d}),
\end{align*}
we further have
\begin{align*}
\mathbf{\Sigma}_d - \mathbf{E} [\mathbf{\tilde{x}}_d \mathbf{\tilde{x}}_d^\mathrm{T}] 
&\geq \mathbf{\Sigma} \{ \mathbf{\Sigma}_1^{-1} \frac{\mathbf{E} [\mathbf{\tilde{x}}_{1d} \mathbf{\tilde{x}}_{1d}^\mathrm{T}]}{w} \mathbf{\Sigma}_1^{-1} + \mathbf{\Sigma}_2^{-1} \frac{\mathbf{E} [\mathbf{\tilde{x}}_{2d} \mathbf{\tilde{x}}_{2d}^\mathrm{T}]}{\bar{w}}  \mathbf{\Sigma}_2^{-1} \} \mathbf{\Sigma}  \\
  &\quad - \mathbf{\Sigma} \{ \mathbf{\Sigma}_1^{-1} \mathbf{E} [\mathbf{\tilde{x}}_{1d} \mathbf{\tilde{x}}_{1d}^\mathrm{T}] \mathbf{\Sigma}_1^{-1} + \mathbf{\Sigma}_2^{-1} \mathbf{E} [\mathbf{\tilde{x}}_{2d} \mathbf{\tilde{x}}_{2d}^\mathrm{T}] \mathbf{\Sigma}_2^{-1}   \\ 
  &\quad + \mathbf{\Sigma}_1^{-1} \mathbf{E} [\mathbf{\tilde{x}}_{1d} \mathbf{\tilde{x}}_{2d}^\mathrm{T}] \mathbf{\Sigma}_2^{-1} + \mathbf{\Sigma}_2^{-1} \mathbf{E} [\mathbf{\tilde{x}}_{2d} \mathbf{\tilde{x}}_{1d}^\mathrm{T}] \mathbf{\Sigma}_1^{-1} \} \mathbf{\Sigma}  \\
&= \mathbf{\Sigma} \{ \mathbf{\Sigma}_1^{-1} \frac{\bar{w}}{w} \mathbf{E} [\mathbf{\tilde{x}}_{1d} \mathbf{\tilde{x}}_{1d}^\mathrm{T}] \mathbf{\Sigma}_1^{-1} + \mathbf{\Sigma}_2^{-1} \frac{w}{\bar{w}} \mathbf{E} [\mathbf{\tilde{x}}_{2d} \mathbf{\tilde{x}}_{2d}^\mathrm{T}] \mathbf{\Sigma}_2^{-1}   \\
  &\quad - \mathbf{\Sigma}_1^{-1} \mathbf{E} [\mathbf{\tilde{x}}_{1d} \mathbf{\tilde{x}}_{2d}^\mathrm{T}] \mathbf{\Sigma}_2^{-1} - \mathbf{\Sigma}_2^{-1} \mathbf{E} [\mathbf{\tilde{x}}_{2d} \mathbf{\tilde{x}}_{1d}^\mathrm{T}] \mathbf{\Sigma}_1^{-1} \} \mathbf{\Sigma}  \\
&= \frac{\mathbf{\Sigma}}{w \bar{w}}  \mathbf{E}[(\bar{w} \mathbf{\Sigma}_1^{-1} \mathbf{\tilde{x}}_{1d} - w \mathbf{\Sigma}_2^{-1} \mathbf{\tilde{x}}_{2d}) (\bar{w} \mathbf{\Sigma}_1^{-1} \mathbf{\tilde{x}}_{1d} - w \mathbf{\Sigma}_2^{-1} \mathbf{\tilde{x}}_{2d})^\mathrm{T}] \mathbf{\Sigma}  \\
&\geq 0.
\end{align*}
So the inequality
\begin{align*}
\mathbf{\Sigma}_d \geq \mathbf{E} [\mathbf{\tilde{x}}_d \mathbf{\tilde{x}}_d^\mathrm{T}]
\end{align*}
is proved. 

From
\begin{align*}
\frac{\mathbf{\Sigma}_{1d}}{w} + \mathbf{\Sigma}_{1i} &= \mathbf{\Sigma}_{1i} + \frac{\mathbf{\Sigma}_{1d} - \mathbf{E} [\mathbf{\tilde{x}}_{1d} \mathbf{\tilde{x}}_{1d}^\mathrm{T}]}{w} + \frac{\mathbf{E} [\mathbf{\tilde{x}}_{1d} \mathbf{\tilde{x}}_{1d}^\mathrm{T}]}{w}  \\
  &\geq \mathbf{\Sigma}_{1i} + \mathbf{\Sigma}_{1d} - \mathbf{E} [\mathbf{\tilde{x}}_{1d} \mathbf{\tilde{x}}_{1d}^\mathrm{T}] + \frac{\mathbf{E} [\mathbf{\tilde{x}}_{1d} \mathbf{\tilde{x}}_{1d}^\mathrm{T}]}{w}  \\
  &\geq \mathbf{E} [\mathbf{\tilde{x}}_{1i} \mathbf{\tilde{x}}_{1i}^\mathrm{T}] + \frac{\mathbf{E} [\mathbf{\tilde{x}}_{1d} \mathbf{\tilde{x}}_{1d}^\mathrm{T}]}{w}
\end{align*}
and similarly 
\begin{align*}
\frac{\mathbf{\Sigma}_{2d}}{\bar{w}} + \mathbf{\Sigma}_{2i} \geq \mathbf{E} [\mathbf{\tilde{x}}_{2i} \mathbf{\tilde{x}}_{2i}^\mathrm{T}] + \frac{\mathbf{E} [\mathbf{\tilde{x}}_{2d} \mathbf{\tilde{x}}_{2d}^\mathrm{T}]}{\bar{w}}
\end{align*}
we have 
\begin{align*}
& \mathbf{\Sigma}_d + \mathbf{\Sigma}_i = \mathbf{\Sigma} (\mathbf{\Sigma}_1^{-1} \mathbf{\Sigma}_1 \mathbf{\Sigma}_1^{-1} + \mathbf{\Sigma}_2^{-1} \mathbf{\Sigma}_2 \mathbf{\Sigma}_2^{-1}) \mathbf{\Sigma}  \\
&= \mathbf{\Sigma} \{ \mathbf{\Sigma}_1^{-1} (\frac{\mathbf{\Sigma}_{1d}}{w} + \mathbf{\Sigma}_{1i}) \mathbf{\Sigma}_1^{-1} + \mathbf{\Sigma}_2^{-1} (\frac{\mathbf{\Sigma}_{2d}}{\bar{w}} + \mathbf{\Sigma}_{2i}) \mathbf{\Sigma}_2^{-1} \} \mathbf{\Sigma}  \\
&\geq \mathbf{\Sigma} \{ \mathbf{\Sigma}_1^{-1} (\mathbf{E} [\mathbf{\tilde{x}}_{1i} \mathbf{\tilde{x}}_{1i}^\mathrm{T}] + \frac{\mathbf{E} [\mathbf{\tilde{x}}_{1d} \mathbf{\tilde{x}}_{1d}^\mathrm{T}]}{w}) \mathbf{\Sigma}_1^{-1}  \\
&\quad + \mathbf{\Sigma}_2^{-1} (\mathbf{E} [\mathbf{\tilde{x}}_{2i} \mathbf{\tilde{x}}_{2i}^\mathrm{T}] + \frac{\mathbf{E} [\mathbf{\tilde{x}}_{2d} \mathbf{\tilde{x}}_{2d}^\mathrm{T}]}{\bar{w}}) \mathbf{\Sigma}_2^{-1} \} \mathbf{\Sigma}.
\end{align*}
Note that
\begin{align*}
\mathbf{\tilde{x}}_d = \mathbf{\Sigma}(\mathbf{\Sigma}_1^{-1} \mathbf{\tilde{x}}_{1d} + \mathbf{\Sigma}_2^{-1} \mathbf{\tilde{x}}_{2d}), \quad \mathbf{\tilde{x}}_i = \mathbf{\Sigma}(\mathbf{\Sigma}_1^{-1} \mathbf{\tilde{x}}_{1i} + \mathbf{\Sigma}_2^{-1} \mathbf{\tilde{x}}_{2i}),
\end{align*}
we further have
\begin{align*}
& \mathbf{\Sigma}_d + \mathbf{\Sigma}_i - \mathbf{E} [\mathbf{\tilde{x}} \mathbf{\tilde{x}}^\mathrm{T}] = \mathbf{\Sigma}_d + \mathbf{\Sigma}_i - \mathbf{E} [\mathbf{\tilde{x}}_d \mathbf{\tilde{x}}_d^\mathrm{T}] - \mathbf{E} [\mathbf{\tilde{x}}_i \mathbf{\tilde{x}}_i^\mathrm{T}]   \\
&\geq \mathbf{\Sigma} \{ \mathbf{\Sigma}_1^{-1} (\mathbf{E} [\mathbf{\tilde{x}}_{1i} \mathbf{\tilde{x}}_{1i}^\mathrm{T}] + \frac{\mathbf{E} [\mathbf{\tilde{x}}_{1d} \mathbf{\tilde{x}}_{1d}^\mathrm{T}]}{w}) \mathbf{\Sigma}_1^{-1}  \\
  &\quad + \mathbf{\Sigma}_2^{-1} (\mathbf{E} [\mathbf{\tilde{x}}_{2i} \mathbf{\tilde{x}}_{2i}^\mathrm{T}] + \frac{\mathbf{E} [\mathbf{\tilde{x}}_{2d} \mathbf{\tilde{x}}_{2d}^\mathrm{T}]}{\bar{w}}) \mathbf{\Sigma}_2^{-1}  \\
  &\quad - \mathbf{\Sigma}_1^{-1} \mathbf{E} [\mathbf{\tilde{x}}_{1d} \mathbf{\tilde{x}}_{1d}^\mathrm{T}] \mathbf{\Sigma}_1^{-1} - \mathbf{\Sigma}_2^{-1} \mathbf{E} [\mathbf{\tilde{x}}_{2d} \mathbf{\tilde{x}}_{2d}^\mathrm{T}] \mathbf{\Sigma}_2^{-1}   \\ 
  &\quad - \mathbf{\Sigma}_1^{-1} \mathbf{E} [\mathbf{\tilde{x}}_{1d} \mathbf{\tilde{x}}_{2d}^\mathrm{T}] \mathbf{\Sigma}_2^{-1} - \mathbf{\Sigma}_2^{-1} \mathbf{E} [\mathbf{\tilde{x}}_{2d} \mathbf{\tilde{x}}_{1d}^\mathrm{T}] \mathbf{\Sigma}_1^{-1}  \\
  &\quad - \mathbf{\Sigma}_1^{-1} \mathbf{E} [\mathbf{\tilde{x}}_{1i} \mathbf{\tilde{x}}_{1i}^\mathrm{T}] \mathbf{\Sigma}_1^{-1} - \mathbf{\Sigma}_2^{-1} \mathbf{E} [\mathbf{\tilde{x}}_{2i} \mathbf{\tilde{x}}_{2i}^\mathrm{T}] \mathbf{\Sigma}_2^{-1}  \} \mathbf{\Sigma}  \\
&= \frac{\mathbf{\Sigma}}{w \bar{w}}  \mathbf{E}[(\bar{w} \mathbf{\Sigma}_1^{-1} \mathbf{\tilde{x}}_{1d} - w \mathbf{\Sigma}_2^{-1} \mathbf{\tilde{x}}_{2d}) (\bar{w} \mathbf{\Sigma}_1^{-1} \mathbf{\tilde{x}}_{1d} - w \mathbf{\Sigma}_2^{-1} \mathbf{\tilde{x}}_{2d})^\mathrm{T}] \mathbf{\Sigma}  \\
&\geq 0.
\end{align*}
So the inequality
\begin{align*}
\mathbf{\Sigma}_d + \mathbf{\Sigma}_i \geq \mathbf{E} [\mathbf{\tilde{x}} \mathbf{\tilde{x}}^\mathrm{T}] 
\end{align*} 
is also proved.
\end{proof}

The consistency defined by (\ref{eq:consistency}) is naturally satisfied if the split consistency defined by (\ref{eq:split_consistency}) is satisfied. Besides, above proofs show that if the source estimates 
\begin{align*}
\{\mathbf{x}_1, \mathbf{\Sigma}_{1d}+\mathbf{\Sigma}_{1i}\}, \quad \{\mathbf{x}_2, \mathbf{\Sigma}_{2d}+\mathbf{\Sigma}_{2i}\}
\end{align*}
are assumed split consistent regardless of whether they are split consistent in reality, then the fusion estimate 
\begin{align*}
\{\mathbf{x}, \mathbf{\Sigma}_d+\mathbf{\Sigma}_i\} 
\end{align*}
obtained via the split covariance intersection filter (\ref{eq:scif}) will also be split consistent from this assumption perspective. This implies that the split covariance intersection filter itself will not add any extra confidence into the fusion estimate and hence cause estimate inconsistency. In one word, the fusion consistency of the split covariance intersection filter is always guaranteed.

\subsubsection*{Formalism for partial observation}

Suppose there exist a complete source estimate 
\begin{align*}
\{\mathbf{x}_1, \mathbf{\Sigma}_{1d}+\mathbf{\Sigma}_{1i}\} 
\end{align*}
and a partial source estimate 
\begin{align*}
\{\mathbf{z} = \mathbf{H} \mathbf{x}_2, \mathbf{\Sigma}_{zd}+\mathbf{\Sigma}_{zi}\}. 
\end{align*}
Augment $\mathbf{z}$ to a complete source estimate 
\begin{align*}
\begin{bmatrix} \mathbf{z} \\ \mathbf{z}_0 \end{bmatrix}
= \begin{bmatrix} \mathbf{H} \\ \mathbf{H}_0  \end{bmatrix} \mathbf{x}_2 
\end{align*}
where $\mathbf{z}_0$ is set arbitrary and its covariance is set as 
\begin{align*}
\mathbf{\Sigma}_{\mathbf{z}_0} = \infty.
\end{align*}
Besides, the augmented measurement matrix
\begin{align*}
\begin{bmatrix} \mathbf{H} \\ \mathbf{H}_0 \end{bmatrix}
\end{align*}
is assumed to be an invertible matrix, so we have
\begin{align*}
&\mathbf{x}_2 = \begin{bmatrix} \mathbf{H} \\ \mathbf{H}_0  \end{bmatrix}^{-1} \begin{bmatrix} \mathbf{z} \\ \mathbf{z}_0  \end{bmatrix}, \\
&\mathbf{\Sigma}_{2d}+\mathbf{\Sigma}_{2i} = \begin{bmatrix} \mathbf{H} \\  \mathbf{H}_0 \end{bmatrix}^{-1} \begin{bmatrix} \mathbf{\Sigma}_{zd}+\mathbf{\Sigma}_{zi} & \mathbf{0} \\  \mathbf{0} & \infty \end{bmatrix} \begin{bmatrix} \mathbf{H} \\  \mathbf{H}_0 \end{bmatrix}^{-\mathrm{T}}  \\
&= \begin{bmatrix} \mathbf{H} \\  \mathbf{H}_0 \end{bmatrix}^{-1} \begin{bmatrix} \mathbf{\Sigma}_{zd} & \mathbf{0} \\  \mathbf{0} & \infty \end{bmatrix} \begin{bmatrix} \mathbf{H} \\  \mathbf{H}_0 \end{bmatrix}^{-\mathrm{T}}   +  \begin{bmatrix} \mathbf{H} \\  \mathbf{H}_0 \end{bmatrix}^{-1} \begin{bmatrix} \mathbf{\Sigma}_{zi} & \mathbf{0} \\  \mathbf{0} & \infty \end{bmatrix} \begin{bmatrix} \mathbf{H} \\  \mathbf{H}_0 \end{bmatrix}^{-\mathrm{T}}.
\end{align*}

Fuse the two estimates in split form 
\begin{align*}
\{\mathbf{x}_1, \mathbf{\Sigma}_{1d}+\mathbf{\Sigma}_{1i}\}, \quad \{\mathbf{x}_2, \mathbf{\Sigma}_{2d}+\mathbf{\Sigma}_{2i}\} 
\end{align*}
via the split covariance intersection filter (\ref{eq:scif}) as
\begin{align*}
\mathbf{\Sigma}_1 &= \mathbf{\Sigma}_{1d}/w + \mathbf{\Sigma}_{1i},  \\  
\mathbf{\Sigma}_z &= \mathbf{\Sigma}_{zd}/(1-w) + \mathbf{\Sigma}_{zi},  \\
\mathbf{\Sigma}_2^{-1} &= \begin{bmatrix} \mathbf{H} \\  \mathbf{H}_0 \end{bmatrix}^\mathrm{T} \begin{bmatrix} \mathbf{\Sigma}_z^{-1} & \mathbf{0} \\  \mathbf{0} & \mathbf{0} \end{bmatrix} \begin{bmatrix} \mathbf{H} \\  \mathbf{H}_0 \end{bmatrix} = \mathbf{H}^\mathrm{T} \mathbf{\Sigma}_z^{-1} \mathbf{H},  \\
\mathbf{\Sigma} &= (\mathbf{\Sigma}_1^{-1} + \mathbf{\Sigma}_2^{-1})^{-1} = (\mathbf{\Sigma}_1^{-1} + \mathbf{H}^\mathrm{T} \mathbf{\Sigma}_z^{-1} \mathbf{H})^{-1} \\
  &= (\mathbf{I} + \mathbf{\Sigma}_1 \mathbf{H}^\mathrm{T} \mathbf{\Sigma}_z^{-1} \mathbf{H})^{-1} \mathbf{\Sigma}_1 = [\sum\limits_0^\infty (-\mathbf{\Sigma}_1 \mathbf{H}^\mathrm{T} \mathbf{\Sigma}_z^{-1} \mathbf{H})^i] \mathbf{\Sigma}_1 \\
  &= \{\mathbf{I} - \mathbf{\Sigma}_1 \mathbf{H}^\mathrm{T} \mathbf{\Sigma}_z^{-1} [\sum\limits_0^\infty (-\mathbf{H} \mathbf{\Sigma}_1 \mathbf{H}^\mathrm{T} \mathbf{\Sigma}_\mathbf{z}^{-1})^i] \mathbf{H}\} \mathbf{\Sigma}_1 \\
  &= \{\mathbf{I} - \mathbf{\Sigma}_1 \mathbf{H}^\mathrm{T} \mathbf{\Sigma}_z^{-1} (\mathbf{I} + \mathbf{H} \mathbf{\Sigma}_1 \mathbf{H}^\mathrm{T} \mathbf{\Sigma}_z^{-1})^{-1} \mathbf{H}\} \mathbf{\Sigma}_1 \\
  &= \{\mathbf{I} - \mathbf{\Sigma}_1 \mathbf{H}^\mathrm{T} (\mathbf{\Sigma}_z + \mathbf{H} \mathbf{\Sigma}_1 \mathbf{H}^\mathrm{T})^{-1} \mathbf{H}\} \mathbf{\Sigma}_1  \\
  &= (\mathbf{I} - \mathbf{K} \mathbf{H}) \mathbf{\Sigma}_1
\end{align*}
where
\footnote{It is worth noting that the infinite matrix series expansion in above derivation holds true only when the eigenvalues of relevant matrices are within the unit circle in the complex plane. But this does not influence the equality $\mathbf{\Sigma} = (\mathbf{I} - \mathbf{K} \mathbf{H}) \mathbf{\Sigma}_1$ which is equivalent to the equality of two finite-order polynomials. Since the equality holds true for infinite choices of matrix elements involved, the equality must always hold true. Thus above derivation result $\mathbf{\Sigma} = (\mathbf{I} - \mathbf{K} \mathbf{H}) \mathbf{\Sigma}_1$ always holds true.}
\begin{align*}
\mathbf{K} = \mathbf{\Sigma}_1 \mathbf{H}^\mathrm{T} (\mathbf{\Sigma}_z + \mathbf{H} \mathbf{\Sigma}_1 \mathbf{H}^\mathrm{T})^{-1}.
\end{align*}

Based on above derivation, we further have
\begin{align*}
\mathbf{x} &= \mathbf{\Sigma} (\mathbf{\Sigma}_1^{-1} \mathbf{x}_1 + \mathbf{\Sigma}_2^{-1} \mathbf{x}_2) = \mathbf{\Sigma} \mathbf{\Sigma}_1^{-1} \mathbf{x}_1 + \mathbf{\Sigma} \mathbf{\Sigma}_2^{-1} \mathbf{x}_2 \\
  &= \mathbf{\Sigma} \mathbf{\Sigma}_1^{-1} \mathbf{x}_1 + (\mathbf{I} - \mathbf{\Sigma} \mathbf{\Sigma}_1^{-1}) \mathbf{x}_2 = (\mathbf{I} - \mathbf{K} \mathbf{H}) \mathbf{x}_1 + \mathbf{K} \mathbf{H} \mathbf{x}_2 \\
  &= (\mathbf{I} - \mathbf{K} \mathbf{H}) \mathbf{x}_1 + \mathbf{K} \mathbf{H} \begin{bmatrix} \mathbf{H} \\  \mathbf{H}_0 \end{bmatrix}^{-1} \begin{bmatrix} \mathbf{z} \\  \mathbf{z}_0 \end{bmatrix} \\
  &= (\mathbf{I} - \mathbf{K} \mathbf{H}) \mathbf{x}_1 + \mathbf{K} \begin{bmatrix} \mathbf{I} & \mathbf{0} \end{bmatrix} \begin{bmatrix}
\mathbf{z} \\  \mathbf{z}_0 \end{bmatrix} \\
  &= (\mathbf{I} - \mathbf{K} \mathbf{H}) \mathbf{x}_1 + \mathbf{K} \mathbf{z}  \\
  &= \mathbf{x}_1 + \mathbf{K} (\mathbf{z} - \mathbf{H} \mathbf{x}_1)
\end{align*}
and
\begin{align*}
\mathbf{\Sigma}_i &= \mathbf{\Sigma}(\mathbf{\Sigma}_1^{-1} \mathbf{\Sigma}_{1i} \mathbf{\Sigma}_1^{-1} + \mathbf{\Sigma}_2^{-1} \mathbf{\Sigma}_{2i} \mathbf{\Sigma}_2^{-1}) \mathbf{\Sigma} \nonumber \\
  &= (\mathbf{I} - \mathbf{K} \mathbf{H}) \mathbf{\Sigma}_{1i} (\mathbf{I} - \mathbf{K} \mathbf{H})^\mathrm{T} + \mathbf{K} \mathbf{H} \begin{bmatrix} \mathbf{H} \\  \mathbf{H}_0 \end{bmatrix}^{-1} \begin{bmatrix} \mathbf{\Sigma}_{zi} & \mathbf{0} \\  \mathbf{0} & \infty \end{bmatrix} \begin{bmatrix} \mathbf{H} \\  \mathbf{H}_0 \end{bmatrix}^{-\mathrm{T}} \mathbf{H}^\mathrm{T} \mathbf{K}^\mathrm{T}  \\
  &= (\mathbf{I} - \mathbf{K} \mathbf{H}) \mathbf{\Sigma}_{1i} (\mathbf{I} - \mathbf{K} \mathbf{H})^\mathrm{T} + \mathbf{K} \begin{bmatrix} \mathbf{I} & \mathbf{0} \end{bmatrix} \begin{bmatrix} \mathbf{\Sigma}_{zi} & \mathbf{0} \\  \mathbf{0} & \infty \end{bmatrix} \begin{bmatrix} \mathbf{I} \\ \mathbf{0} \end{bmatrix} \mathbf{K}^\mathrm{T}  \\
&= (\mathbf{I} - \mathbf{K} \mathbf{H}) \mathbf{\Sigma}_{1i} (\mathbf{I} - \mathbf{K} \mathbf{H})^\mathrm{T} + \mathbf{K} \mathbf{\Sigma}_{zi} \mathbf{K}^\mathrm{T},  \\
\mathbf{\Sigma}_d &= \mathbf{\Sigma} - \mathbf{\Sigma}_i.
\end{align*}
Therefore the split covariance intersection filter for partial observation is formalized as
\begin{subequations}  \label{eq:scif_partial_observation}
\begin{align}
\mathbf{\Sigma}_1 &= \mathbf{\Sigma}_{1d}/w + \mathbf{\Sigma}_{1i},  \\
\mathbf{\Sigma}_z &= \mathbf{\Sigma}_{zd}/(1-w) + \mathbf{\Sigma}_{zi},  \\
\mathbf{K} &= \mathbf{\Sigma}_1 \mathbf{H}^\mathrm{T} (\mathbf{\Sigma}_z + \mathbf{H} \mathbf{\Sigma}_1 \mathbf{H}^\mathrm{T})^{-1},  \\
\mathbf{x} &= \mathbf{x}_1 + \mathbf{K} (\mathbf{z} - \mathbf{H} \mathbf{x}_1),  \\
\mathbf{\Sigma} &= (\mathbf{I} - \mathbf{K} \mathbf{H}) \mathbf{\Sigma}_1,  \\
\mathbf{\Sigma}_i &= (\mathbf{I} - \mathbf{K} \mathbf{H}) \mathbf{\Sigma}_{1i} (\mathbf{I} - \mathbf{K} \mathbf{H})^\mathrm{T} + \mathbf{K} \mathbf{\Sigma}_{zi} \mathbf{K}^\mathrm{T},  \\
\mathbf{\Sigma}_d &= \mathbf{\Sigma} - \mathbf{\Sigma}_i.
\end{align}
\end{subequations}

\subsubsection*{Efficient and accurate implementation}

Matlab code for implementing the split covariance intersection filter is given as follows
\footnote{In fact, the author already posed the following version of the Split CIF code on the open-source website \textit{GitHub} a few years ago to facilitate potential researchers' understanding of the author's theoretical works on the split covariance intersection filter.}.

\begin{framed} 
\noindent \textbf{SplitCIF.m} \\
\noindent \% Suppose \{X1, P1i+P1d\} is a complete estimate i.e. X1 = X\_true \\
\%$~~~~$ $~~~~$  \{X2, P2i+P2d\} can be a partial estimate i.e. X2 = H * X\_true \\
\% For each estimate X in the split form,  \\
\%$~~~~$ i part i.e. Pi represents the degree of X being independent \\
\%$~~~~$ d part i.e. Pd represents the degree of being potentially correlated  \\
\% \\
\% Special cases of the Split CIF: \\
\% Case 1: P1d = 0 matrix , P2d = 0 matrix \\
\%$~~~~$ The Split CIF is simply reduced to the Kalman filter (KF) \\
\%$~~~~$ So if you want to use the KF, then you can still use the Split CIF  \\
\%$~~~~$ by simplying setting both P1d and P2d to zero matrices \\
\% Case 2: P1i = 0 matrix , P2i = 0 matrix \\
\%$~~~~$ The Split CIF is simply reduced to the Covariance Intersection (CI) \\
\%$~~~~$ So if you want to use the CI, then you can still use the Split CIF  \\
\%$~~~~$ by simplying setting both P1i and P2i to zero matrices \\
\% Note: The Split CIF is a generalization of both the KF and the CI \\
function [X, Pi, Pd] = SplitCIF(X1, P1i, P1d, X2, P2i, P2d, H) \\
$~~~~$ I = eye(length(X1)); \\
$~~~~$ numeric\_eps = 0.00000000001; \\
$~~~~$ function val = DetP(w) \\
$~~~~$ $~~~~$ if (w$<$numeric\_eps) \\
$~~~~$ $~~~~$ $~~~~$ w = numeric\_eps; \\
$~~~~$ $~~~~$ elseif (w$>$1-numeric\_eps) \\
$~~~~$ $~~~~$ $~~~~$ w = 1-numeric\_eps; \\
$~~~~$ $~~~~$ end \\
$~~~~$ $~~~~$ P1 = P1d/w + P1i; \\
$~~~~$ $~~~~$ P2 = P2d/(1-w) + P2i; \\
$~~~~$ $~~~~$ K = P1*H'*inv(H*P1*H'+P2); \\
$~~~~$ $~~~~$ P = (I-K*H)*P1; \\
$~~~~$ $~~~~$ val = det(P); \\
$~~~~$ end \\
 \\
$~~~~$ \% The w-optimization problem is theoretically guaranteed to be convex \\
$~~~~$ \% So some convex optimization technique can be used  \\
$~~~~$ \% The golden section technique is adopted to search for the optimal w \\
$~~~~$ function w\_opt = GetW\_GoldenSection(P1i, P1d, P2i, P2d, H) \\
$~~~~$ $~~~~$ if (trace(abs(P2d))$<$numeric\_eps) \\
$~~~~$ $~~~~$ $~~~~$ w\_opt = 1; \\
$~~~~$ $~~~~$ $~~~~$ return \\
$~~~~$ $~~~~$ $~~~~$ fprintf('No continuation$\backslash$n'); \\
$~~~~$ $~~~~$ elseif (trace(abs(P1d))$<$numeric\_eps) \\
$~~~~$ $~~~~$ $~~~~$ w\_opt = 0; \\
$~~~~$ $~~~~$ $~~~~$ return \\
$~~~~$ $~~~~$ $~~~~$ fprintf('No continuation$\backslash$n'); \\
$~~~~$ $~~~~$ end$~~~~$ $~~~~$  \\
$~~~~$ $~~~~$ wL = 0; fwL = DetP(wL); \\
$~~~~$ $~~~~$ wR = 1; fwR = DetP(wR); \\
$~~~~$ $~~~~$ sL = 0.382; fsL = DetP(sL); \\
$~~~~$ $~~~~$ sR = 0.618; fsR = DetP(sR); \\
$~~~~$ $~~~~$ \% Already very accurate for many applications \\
$~~~~$ $~~~~$ err\_tol = 0.00001;   \\
$~~~~$ $~~~~$ while(wR-wL$>$err\_tol) \\
$~~~~$ $~~~~$ $~~~~$ if (fsL $<$ fsR) \\
$~~~~$ $~~~~$ $~~~~$ $~~~~$ wR = sR; fwR = fsR; \\
$~~~~$ $~~~~$ $~~~~$ $~~~~$ sR = sL; fsR = fsL; \\
$~~~~$ $~~~~$ $~~~~$ $~~~~$ sL = wL + 0.618*(sL-wL); fsL = DetP(sL); \\
$~~~~$ $~~~~$ $~~~~$ else \\
$~~~~$ $~~~~$ $~~~~$ $~~~~$ wL = sL; fwL = fsL; \\
$~~~~$ $~~~~$ $~~~~$ $~~~~$ sL = sR; fsL = fsR; \\
$~~~~$ $~~~~$ $~~~~$ $~~~~$ sR = wR - 0.618*(wR-sR); fsR = DetP(sR); \\
$~~~~$ $~~~~$ $~~~~$ end \\
$~~~~$ $~~~~$ end \\
$~~~~$ $~~~~$ fmin = min([fwL, fsL, fsR, fwR]); \\
$~~~~$ $~~~~$ if (fwL == fmin) \\
$~~~~$ $~~~~$ $~~~~$ w\_opt = wL; \\
$~~~~$ $~~~~$ elseif (fwR == fmin) \\
$~~~~$ $~~~~$ $~~~~$ w\_opt = wR; \\
$~~~~$ $~~~~$ else \\
$~~~~$ $~~~~$ $~~~~$ w\_opt = (wL+wR)/2; \\
$~~~~$ $~~~~$ end \\
$~~~~$ end \\
 \\
$~~~~$ w\_opt = GetW\_GoldenSection(P1i, P1d, P2i, P2d, H); \\
$~~~~$ if (w\_opt$<$numeric\_eps) \\
$~~~~$ $~~~~$ w\_opt = numeric\_eps; \\
$~~~~$ elseif (w\_opt$>$1-numeric\_eps) \\
$~~~~$ $~~~~$ w\_opt = 1-numeric\_eps; \\
$~~~~$ end \\
$~~~~$ P1 = P1d/w\_opt + P1i; \\
$~~~~$ P2 = P2d/(1-w\_opt) + P2i; \\
$~~~~$ K = P1*H'*inv(H*P1*H'+P2); \\
$~~~~$ X = X1 + K*(X2-H*X1); \\
$~~~~$ P = (I-K*H)*P1; \\
$~~~~$ Pi = (I-K*H)*P1i*(I-K*H)' + K*P2i*K'; \\
$~~~~$ Pd = P-Pi; \\
end
\end{framed}

Above implementation of the split covariance intersection filter relies on convexity of the $w$-optimization problem (\ref{eq:wopt}) which is theoretically guaranteed. The proof details for convexity of (\ref{eq:wopt}) are saved here. Readers may refer to the author's another book \cite{Li2022FARET_2, Li2022FARET_1} for the proof details.

\section{Data association}  \label{sec:data_association}

Recall the linear measurement model (\ref{eq:SDE_linear_measurement_discrete})
\begin{align*}
\mathbf{z}_t = \mathbf{H} \mathbf{x}_t
\end{align*}
or the generic measurement model (\ref{eq:generic_msr_model_discrete})
\begin{align*}
\mathbf{z}_t = h(\mathbf{x}_t).
\end{align*}
As clarified in Section \ref{sec:measurement_model}, the measurement model is a mathematical model that describes the causal relationship between the state and the measurement, as if the measurement model ``predicts'' what the measurement would be given a known state. 

By so far, we have been adopting an implicit assumption for the measurement model: we know the causal relationship or the ``predicting'' relationship is actually between \textit{which state} (or \textit{which state part}) and \textit{which measurement} (or \textit{which measurement part}). Simply speaking, we know \textit{which state is associated with which measurement} or inversely \textit{which measurement is associated with which state}. 
However, above implicit assumption cannot be taken for granted in practical applications. How to know which state and which measurement are mutually associated is a problem that has no \textit{a priori} answer but needs to be solved on-line along with recursive estimation. The problem is coined as \textbf{data association}.

\subsection{Multiple-target tracking}  \label{sec:multiple_target_tracking}

Data association may be easy in some practical applications. For example, if there is a single entity whose state needs to be estimated and there is hardly any possibility to have false positives (i.e. wrong measurements associated with the entity state) \cite{Li2025VTT}, then data association would be a trivial process.

On the other hand, data association in some practical applications may also be much more complicated than the trivial kind. The probably most complicated kind of data association is in the context of \textbf{multiple-target tracking} \cite{Bar-Shalom1975, Blackman1986} which is historically rooted in radar applications. It is worth noting that methods intended for such complicated kind of data association are not limited to radar applications, but can be extended to various other kinds of applications such as intelligent vehicle navigation in highly dynamic environments \cite{WangZ2023} and intelligent video surveillance \cite{YuanY2017}.

\subsubsection*{Joint probabilistic data association}

A representative data association method for multiple-target tracking is the \textbf{joint probabilistic data association} method \cite{Bar-Shalom1975}. Its key idea is to enable a track to be updated by a probability weighted sum of all measurements that are regarded as being potentially associated with the track, formalized as
\begin{equation}  \label{eq:joint_prob_data_association}
\hat{\mathbf{x}}_{\chi,t} = \sum_{k=1}^{m_{\chi}} E[\mathbf{x}_t | \chi = \chi_{\mathbf{z}_{t,k}}, \mathbf{z}_{t,k}] p(\chi = \chi_{\mathbf{z}_{t,k}} | \mathbf{z}_{t,k}).
\end{equation}
The measurements can be sifted by a threshold or gate conditioned on the track to determine which ones are potentially associated with the track. It is worth noting that on one hand a track may be updated by more than one measurement and on the other hand a measurement may contribute to the update of more than one track.

\subsubsection*{Multiple hypothesis tracking}

Another representative data association method for multiple-target tracking is the \textbf{multiple hypothesis tracking} method \cite{Blackman1986, Blackman2004}, which relies on deferred inference rather than immediate inference of data association hypotheses. More specifically, whenever there is measurement-to-track ambiguity, alternative data association hypotheses are taken into account. The estimator propagates all potential data association hypotheses into the future with an expectation that their uncertainty would be resolved by further measurements, instead of immediately choosing the one optimal in terms of certain criterion among them or immediately combining them heuristically into one ``virtual'' hypothesis as in the joint probabilistic data association method.

\subsubsection*{Probability hypothesis density filter}

One more representative data association method for multiple-target tracking is the \textbf{probability hypothesis density filter} \cite{Mahler2003}. It treats the states of multiple targets holistically as a \textit{random finite set} called the \textit{multiple-target state}
\begin{equation}  \label{eq:multi_target_state_RFS}
\mathbf{X}_t \equiv \{\mathbf{x}_{1,t}, \mathbf{x}_{2,t}, \cdots, \mathbf{x}_{n_t,t}\}
\end{equation}
and treats the measurements holistically as another \textit{random finite set} called the \textit{multiple-target measurement}
\begin{equation}  \label{eq:multi_target_msr_RFS}
\mathbf{Z}_t \equiv \{\mathbf{z}_{1,t}, \mathbf{z}_{2,t}, \cdots, \mathbf{z}_{m_t,t}\}.
\end{equation}

The multiple-target state $\mathbf{X}$ at time $t$ is given by the union of the spontaneous births, the surviving targets, and the spawned targets
\begin{equation}  \label{eq:multi_target_state_RFS2}
\mathbf{X}_t = \Gamma_t \cup \{ \mathop{\cup}\limits_{\zeta \in \mathbf{X}_{t-1}} SU_t (\zeta) \} \cup \{ \mathop{\cup}\limits_{\zeta \in \mathbf{X}_{t-1}} SP_t (\zeta) \},
\end{equation}
where $\Gamma_t$ denotes the random finite set of spontaneous births at time $t$, $SP_t (\zeta)$ denotes the random finite set of targets spawned at time $t$ from a target with previous state $\zeta$, and $SU_t (\zeta)$ denotes the random finite set that can take on either a state predicted from $\zeta$ when the target survives or $\emptyset$ when the target dies.

The multiple-target measurement $\mathbf{Z}$ at time $t$ is given by the union of false measurements (or clutter) and target generated measurements
\begin{equation}  \label{eq:multi_target_msr_RFS2}
\mathbf{Z}_t = K_t \cup \{ \mathop{\cup}\limits_{x \in \mathbf{X}_{t}} G_t (x) \},
\end{equation}
where $K_t$ denotes the random finite set of false measurements (false positive) and $G_t (x)$ denotes the random finite set that can take on either a measurement generated from $x$ when the target is detected or $\emptyset$ when the target is not detected (false negative).

Recall the generic system model formalism (\ref{eq:generic_sys_model_discrete})
\begin{align*}
\mathbf{x}_t = g(\mathbf{x}_{t-1}, \mathbf{u}_t)
\end{align*}
and ``paraphrase'' it in the conditional probability density function form as $p_g (\mathbf{x}_t | \mathbf{x}_{t-1})$. Recall the generic measurement model formalism (\ref{eq:generic_msr_model_discrete})
\begin{align*}
\mathbf{z}_t = h(\mathbf{x}_t)
\end{align*}
and ``paraphrase'' it also in the conditional probability density function form as $p_h (\mathbf{z}_t | \mathbf{x}_t)$. It is worth noting that the subscripts $g$ and $h$ may be omitted such as in the formalism (\ref{eq:pf_recursive}) associated with the particle filter. However, the subscripts $g$ and $h$ are intentionally added here to explicitly distinguish the roles of $p_g (\mathbf{x}_t | \mathbf{x}_{t-1})$ and $p_h (\mathbf{z}_t | \mathbf{x}_t)$ as the system model and the measurement model respectively.

The probability hypothesis density filter provides a generic way of performing recursive estimation while performing implicit data association simultaneously as
\begin{subequations}  \label{eq:PHD_filter_recursive}
\begin{align}
\bar{p}_t (\mathbf{x}_t) &= \int p_{su,t}(\zeta) p_g(\mathbf{x}_t | \zeta) \hat{p}_{t-1}(\zeta) \mathrm{d} \zeta + \int p_{sp,t}(\mathbf{x}_t | \zeta) \hat{p}_{t-1}(\zeta) \mathrm{d} \zeta + \gamma_t(\mathbf{x}_t),   \\
\hat{p}_t (\mathbf{x}_t) &= [1 - p_{d,t}(\mathbf{x}_t)] \bar{p}_t (\mathbf{x}_t) + \sum_{\mathbf{z} \in \mathbf{Z}_t} \frac{p_{d,t}(\mathbf{x}_t) p_h (\mathbf{z}_t | \mathbf{x}_t) \bar{p}_t (\mathbf{x}_t)}{\kappa_t(\mathbf{z}_t) + \int p_{d,t}(\zeta) p_h (\mathbf{z}_t | \zeta) \bar{p}_t (\zeta) \mathrm{d} \zeta},
\end{align}
\end{subequations}
where $\bar{p}_t (\mathbf{x}_t)$ denotes the \textit{intensity} associated with the \textit{multiple-target predicted density}
\begin{equation}  \label{eq:multiple-target_predicted_density}
p(\mathbf{X}_t | \mathbf{Z}_{1:t-1}) = \int p_g(\mathbf{X}_t | \mathbf{X}) p(\mathbf{X} | \mathbf{Z}_{1:t-1}) \mu (\mathrm{d} \mathbf{X})
\end{equation}
and $\hat{p}_t (\mathbf{x}_t)$ denotes the \textit{intensity} associated with the \textit{multiple-target posterior density}
\begin{equation}  \label{eq:multiple-target_posterior_density}
p(\mathbf{X}_t | \mathbf{Z}_{1:t}) = \frac{p_h(\mathbf{Z}_t | \mathbf{X}_t) p(\mathbf{X}_t | \mathbf{Z}_{1:t-1})}{\int p_h(\mathbf{Z}_t | \mathbf{X}) p(\mathbf{X} | \mathbf{Z}_{1:t-1}) \mu (\mathrm{d} \mathbf{X})}.
\end{equation}
Besides, $p_{su,t}(\zeta)$ denotes the probability of a target still existing at time $t$ given its previous state $\zeta$, $p_{sp,t}(\cdot | \zeta)$ denotes the intensity of the random finite set $SP_t (\zeta)$ spawned at time $t$ by a target with previous state $\zeta$, $\gamma_t(\cdot)$ denotes the intensity of the random finite set $\Gamma_t$ at time t, $p_{d,t}(\mathbf{x}_t)$ denotes the probability of detecting a given state $\mathbf{x}_t$ at time t, and $\kappa_t(\cdot)$ denotes the intensity of the random finite set $K_t$ at time t. It is also worth noting that the integral variable $\zeta$ in integrals of (\ref{eq:PHD_filter_recursive}) takes states in the state space shared commonly by the multiple targets and $\mu (\cdot)$ in (\ref{eq:multiple-target_predicted_density}) and (\ref{eq:multiple-target_posterior_density}) denotes certain appropriate measure.

\subsection{Gaussian mixture probability hypothesis density filter}  \label{sec:GM_PHD_filter}

The probability hypothesis density filter provides a generic way of recursive estimation that can perform complicated data association implicitly and has the merit of completely avoiding combinatorial computations intended for handling \textit{a priori} unknown association between the set of measurements and the set of targets.

On the other hand, rather general implementation of (\ref{eq:PHD_filter_recursive}) either via numeric computation or via the \textit{Monte Carlo} method suffers from the \textit{curse of dimensionality} problem
\footnote{In as early as 1950s, \textit{Bellman} pointed out a problem that is nowadays commonly coined as the \textit{curse of dimensionality}, namely a problem of excessive dimensionality causing unaffordable processing price \cite{Bellman1954}. The curse of dimensionality occurs in many domains such as numerical computations, statistics, combinatorics, data mining, and machine learning.}.
Certain assumption is usually needed to realize more efficient implementation of (\ref{eq:PHD_filter_recursive}). The original version of the probability hypothesis density filter \cite{Mahler2003} resorts to the \textit{Poisson distribution assumption} which enables efficient characterization of an entire distribution just by a single variable and hence enables efficient implementation of (\ref{eq:PHD_filter_recursive}) as well --- Note that a Poisson point process that is interpreted as a random measure is uniquely determined by its intensity measure --- However, the Poisson distribution assumption, which always couples the expectation and the covariance of a distribution, is too strong or too restrictive and consequently has rather limited applicability in practice.

The \textbf{Gaussian mixture probability hypothesis density filter} \cite{VoB2006, VoB2006conf} is an improved version of the probability hypothesis density filter. The former inherits the latter's advantage on one hand and overcomes the latter's limitation on the other hand. The Gaussian mixture probability hypothesis density filter resorts to the \textit{Linear-Gaussian assumption} for the system model $p_g (\mathbf{x}_t | \mathbf{x}_{t-1})$ and the measurement model $p_h (\mathbf{z}_t | \mathbf{x}_t)$ as
\begin{subequations}  \label{eq:GM_PHD_sys_msr_linear_Gauss}
\begin{align}
p_g (\mathbf{x}_t | \mathbf{x}_{t-1}) &= N(\mathbf{x}_t \mbox{ } | \mathbf{A}_t \mathbf{x}_{t-1}, \mathbf{\Sigma}_{\epsilon,t}),  \\
p_h (\mathbf{z}_t | \mathbf{x}_t) &= N(\mathbf{z}_t \mbox{ } | \mathbf{H}_t \mathbf{x}_t, \mathbf{\Sigma}_{z,t}),
\end{align}
\end{subequations}
takes the survival and detection probabilities as state independent ones
\begin{equation}  \label{eq:GM_PHD_constant_su_d_prob}
p_{su,t}(\mathbf{x}_t) = p_{su,t}, \quad p_{d,t}(\mathbf{x}_t) = p_{d,t},
\end{equation}
and takes the intensity of the false measurement random finite set $K_t$, i.e. $\kappa_t(\mathbf{z}_t)$ as
\begin{equation}  \label{eq:GM_PHD_false_msr_kappa}
\kappa_t(\mathbf{z}_t) \propto \mbox{uniform}(\mathbf{z}_t).
\end{equation}

The Gaussian mixture probability hypothesis density filter also adopts \textit{Gaussian mixture modelling} for the intensity associated with the multiple-target posterior density, i.e. $\hat{p}_t (\mathbf{x}_t)$ as
\begin{equation}  \label{eq:GM_PHD_hat_p}
\hat{p}_t (\mathbf{x}_t) = \sum_{k=1}^{J_t} w_{k,t} N(\mathbf{x}_t \mbox{ } | \hat{\mathbf{x}}_{k,t}, \hat{\mathbf{\Sigma}}_{k,t}),
\end{equation}
for the intensity of the spawned random finite set $SP_t (\zeta)$, i.e. $p_{sp,t}(\mathbf{x}_t | \zeta)$ as
\begin{equation}  \label{eq:GM_PHD_spawn_p}
p_{sp,t}(\mathbf{x}_t | \zeta) = \sum_{k=1}^{J_{sp,t}} w_{sp,k,t} N(\mathbf{x}_t \mbox{ } | \mathbf{A}_{sp,k,t} \zeta + \mathbf{b}_{sp,k,t}, \mathbf{\Sigma}_{sp,k,t}),
\end{equation}
for the intensity of the spontaneous birth random finite set $\Gamma_t$, i.e. $\gamma_t(\mathbf{x}_t)$ as
\begin{equation}  \label{eq:GM_PHD_birth_gamma}
\gamma_t(\mathbf{x}_t) = \sum_{k=1}^{J_{\gamma,t}} w_{\gamma,k,t} N(\mathbf{x}_t \mbox{ } | \mathbf{x}_{\gamma,k,t}, \mathbf{\Sigma}_{\gamma,k,t}).
\end{equation}
Summary of the Gaussian mixture probability hypothesis density filter is as follows.

\noindent \textbf{Prediction}:

1. Compute the predicted intensity corresponding to surviving targets as
\begin{equation}  \label{eq:GM_PHD_predict_survive}
p_{su,t} \sum_{k=1}^{J_{t-1}} w_{k,t-1} N(\mathbf{x}_t \mbox{ } | \bar{\mathbf{x}}_{su,k,t}, \bar{\mathbf{\Sigma}}_{su,k,t})
\end{equation}
with
\begin{subequations}  \label{eq:GM_PHD_predict_survive_param}
\begin{align}
\bar{\mathbf{x}}_{su,k,t} &= \mathbf{A}_t \hat{\mathbf{x}}_{k,t-1},   \\
\bar{\mathbf{\Sigma}}_{su,k,t} &= \mathbf{A}_t \hat{\mathbf{\Sigma}}_{k,t-1} \mathbf{A}_t^\mathrm{T} + \mathbf{\Sigma}_{\epsilon,t},
\end{align}
\end{subequations}
where $k \in \{1, \cdots, J_{t-1}\}$.

2. Compute the predicted intensity corresponding to spawned targets as
\begin{equation}  \label{eq:GM_PHD_predict_spawn}
\sum_{k=1}^{J_{t-1}} \sum_{j=1}^{J_{sp,t}} w_{k,t-1} w_{sp,j,t} N(\mathbf{x}_t \mbox{ } | \bar{\mathbf{x}}_{sp,k,j,t}, \bar{\mathbf{\Sigma}}_{sp,k,j,t})
\end{equation}
with
\begin{subequations}  \label{eq:GM_PHD_predict_spawn_param}
\begin{align}
\bar{\mathbf{x}}_{sp,k,j,t} &= \mathbf{A}_{sp,j,t} \hat{\mathbf{x}}_{k,t-1} + \mathbf{b}_{sp,j,t},   \\ 
\bar{\mathbf{\Sigma}}_{sp,k,j,t} &= \mathbf{A}_{sp,j,t} \hat{\mathbf{\Sigma}}_{k,t-1} \mathbf{A}_{sp,j,t}^\mathrm{T} + \mathbf{\Sigma}_{sp,j,t},
\end{align}
\end{subequations}
where $k \in \{1, \cdots, J_{t-1}\}$ and $j \in \{1, \cdots, J_{sp,t}\}$.

3. Note that the sum of multiple Gaussian mixtures is still a Gaussian mixture. Merge (\ref{eq:GM_PHD_birth_gamma}), (\ref{eq:GM_PHD_predict_survive}), and (\ref{eq:GM_PHD_predict_spawn}) to synthesize the predicted (or \textit{a priori}) intensity as
\begin{align}  \label{eq:GM_PHD_predict_intensity}
\bar{p}_t (\mathbf{x}_t) =& \sum_{k=1}^{J_{\gamma,t}} w_{\gamma,k,t} N(\mathbf{x}_t \mbox{ } | \mathbf{x}_{\gamma,k,t}, \mathbf{\Sigma}_{\gamma,k,t}) + p_{su,t} \sum_{k=1}^{J_{t-1}} w_{k,t-1} N(\mathbf{x}_t \mbox{ } | \bar{\mathbf{x}}_{su,k,t}, \bar{\mathbf{\Sigma}}_{su,k,t})  \nonumber \\ 
  &+ \sum_{k=1}^{J_{t-1}} \sum_{j=1}^{J_{sp,t}} w_{k,t-1} w_{sp,j,t} N(\mathbf{x}_t \mbox{ } | \bar{\mathbf{x}}_{sp,k,j,t}, \bar{\mathbf{\Sigma}}_{sp,k,j,t}) 
  \equiv \sum_{k=1}^{\bar{J}_t} \bar{w}_{k,t} N(\mathbf{x}_t \mbox{ } | \bar{\mathbf{x}}_{k,t}, \bar{\mathbf{\Sigma}}_{k,t}).
\end{align}

\noindent \textbf{Update}:

4. Compute the updated intensity corresponding to missed targets as
\begin{equation}  \label{eq:GM_PHD_update_miss}
(1 - p_{d,t}) \sum_{k=1}^{\bar{J}_t} \bar{w}_{k,t} N(\mathbf{x}_t \mbox{ } | \bar{\mathbf{x}}_{k,t}, \bar{\mathbf{\Sigma}}_{k,t}).
\end{equation}

5. Compute the updated intensity corresponding to detected targets
\begin{equation}  \label{eq:GM_PHD_update_detect}
\sum_{\mathbf{z} \in \mathbf{Z}_t} \sum_{k=1}^{\bar{J}_t} w_{\mathbf{z},k,t} N(\mathbf{x}_t \mbox{ } | \hat{\mathbf{x}}_{\mathbf{z},k,t}, \hat{\mathbf{\Sigma}}_{k,t})
\end{equation}
with
\begin{subequations}  \label{eq:GM_PHD_update_detect_param}
\begin{align}
w_{\mathbf{z},k,t} &= \frac{p_{d,t} \bar{w}_{k,t} N(\mathbf{z} \mbox{ } | \mathbf{H}_t \bar{\mathbf{x}}_{k,t}, \mathbf{H}_t \bar{\mathbf{\Sigma}}_{k,t} \mathbf{H}_t^\mathrm{T} + \mathbf{\Sigma}_{z,t})}{\kappa_t(\mathbf{z}) + p_{d,t} \sum_{j=1}^{\bar{J}_t} \bar{w}_{j,t} N(\mathbf{z} \mbox{ } | \mathbf{H}_t \bar{\mathbf{x}}_{j,t}, \mathbf{H}_t \bar{\mathbf{\Sigma}}_{j,t} \mathbf{H}_t^\mathrm{T} + \mathbf{\Sigma}_{z,t})},  \\
\mathbf{K}_{k,t} &= \bar{\mathbf{\Sigma}}_{k,t} \mathbf{H}_t^\mathrm{T} (\mathbf{H}_t \bar{\mathbf{\Sigma}}_{k,t} \mathbf{H}_t^\mathrm{T} + \mathbf{\Sigma}_{z,t})^{-1},  \\
\hat{\mathbf{x}}_{\mathbf{z},k,t} &= \bar{\mathbf{x}}_{k,t} + \mathbf{K}_{k,t} (\mathbf{z} - \mathbf{H}_t \bar{\mathbf{x}}_{k,t}),  \\
\hat{\mathbf{\Sigma}}_{k,t} &= (\mathbf{I} - \mathbf{K}_{k,t} \mathbf{H}_t) \bar{\mathbf{\Sigma}}_{k,t}.
\end{align}
\end{subequations}

6. Note again that the sum of multiple Gaussian mixtures is still a Gaussian mixture. Merge (\ref{eq:GM_PHD_update_miss}) and (\ref{eq:GM_PHD_update_detect}) to synthesize the updated (or \textit{a posteriori}) intensity as
\begin{align}  \label{eq:GM_PHD_update_intensity}
\hat{p}_t (\mathbf{x}_t) &= (1 - p_{d,t}) \sum_{k=1}^{\bar{J}_t} \bar{w}_{k,t} N(\mathbf{x}_t \mbox{ } | \bar{\mathbf{x}}_{k,t}, \bar{\mathbf{\Sigma}}_{k,t}) + \sum_{\mathbf{z} \in \mathbf{Z}_t} \sum_{k=1}^{\bar{J}_t} w_{\mathbf{z},k,t} N(\mathbf{x}_t \mbox{ } | \hat{\mathbf{x}}_{\mathbf{z},k,t}, \hat{\mathbf{\Sigma}}_{k,t})  \nonumber \\
  &\equiv \sum_{k=1}^{J_t} w_{k,t} N(\mathbf{x}_t \mbox{ } | \hat{\mathbf{x}}_{k,t}, \hat{\mathbf{\Sigma}}_{k,t}).
\end{align}

7. Prune for the Gaussian mixture probability hypothesis density filter to reduce the number of Gaussian components propagated to next control period. First, discard Gaussian components with weights below certain threshold or keep only a certain number of Gaussian components with highest weights. Second, merge close Gaussian components into a single Gaussian component, formalized generically as
\begin{subequations}  \label{eq:GM_PHD_prune_merge}
\begin{align}
w_{\mathbf{L},t} &= \sum_{k \in \mathbf{L}} w_{k,t},  \\
\hat{\mathbf{x}}_{\mathbf{L},t} &= \frac{1}{w_{\mathbf{L},t}} \sum_{k \in \mathbf{L}} w_{k,t} \hat{\mathbf{x}}_{k,t},  \\
\hat{\mathbf{\Sigma}}_{\mathbf{L},t} &= \frac{1}{w_{\mathbf{L},t}} \sum_{k \in \mathbf{L}} w_{k,t} [\hat{\mathbf{\Sigma}}_{k,t} + (\hat{\mathbf{x}}_{k,t} - \hat{\mathbf{x}}_{\mathbf{L},t}) (\hat{\mathbf{x}}_{k,t} - \hat{\mathbf{x}}_{\mathbf{L},t})^\mathrm{T}],
\end{align}
\end{subequations}
where $\mathbf{L}$ denotes a generic cluster of Gaussian components that are merged into a single one --- Theoretically more sound clustering methods such as the \textit{mean shift} method \cite{Fukunaga1975, Cheng1995, Comaniciu2002} may be used for clustering of Gaussian components. However, a simple method based on \textit{Mahalanobis distance} \cite{Mahalanobis1936} thresholding would be already satisfactory in practice --- For simplicity yet without causing confusion, the author abuses the notation
\begin{align*}
\hat{p}_t (\mathbf{x}_t) = \sum_{k=1}^{J_t} w_{k,t} N(\mathbf{x}_t \mbox{ } | \hat{\mathbf{x}}_{k,t}, \hat{\mathbf{\Sigma}}_{k,t})
\end{align*}
to also denote the Gaussian mixture after pruning.

8. Extract the multiple-target state estimate $\hat{\mathbf{X}}_t$ as follows: Initialize
\begin{align*}
\hat{\mathbf{X}}_t = \emptyset.
\end{align*}
Then traverse the $J_t$ Gaussian components one by one. For a generic Gaussian component $N(\mathbf{x}_t \mbox{ } | \hat{\mathbf{x}}_{k,t}, \hat{\mathbf{\Sigma}}_{k,t})$ with its weight $w_{k,t}$, if
\begin{align*}
w_{k,t} > 0.5,
\end{align*}
then add $\mbox{round}(w_{k,t})$ copies of $\hat{\mathbf{x}}_{k,t}$ to $\hat{\mathbf{X}}_t$.

\section{Analog estimation}  \label{sec:analog_estimation}

Nowadays, digital devices including digital sensors (which provide measurements) and digital processors (which perform estimation) are ubiquitous, whereas analog devices are rarely used. So estimation methods encountered nowadays are naturally realized in discrete way. In other words, we by default estimate states at discrete time indices or control periods, namely states with discrete subscripts formalized such as
\begin{align*}
\cdots \quad \mathbf{x}_{t-2}, \quad \mathbf{x}_{t-1}, \quad \mathbf{x}_t, \quad \mathbf{x}_{t+1}, \quad \mathbf{x}_{t+2} \quad \cdots 
\end{align*}
So when people talk about estimation nowadays, they tend to mean \textit{digital estimation} or \textit{discrete-time estimation} by default
\footnote{It is like when people talk about image processing nowadays, they tend to mean digital image processing by default \cite{Li2025GFCV_SJTU}, though the terms ``image processing'' could fairly refer to processing analog images by analog electronic circuits or even to chemical processing of photo films captured by cameras in old days.}.

On the other hand, analog devices including analog sensors and analog processors may still be used for sake of state estimation in special contexts. Then we need effective estimation methods that are realized in continuous way
\footnote{In fact, dozens of years ago when digital devices were not so ubiquitous, estimation methods usually belonged to the analog category.}, 
or in other words, we need to perform \textit{analog estimation} or \textit{continuous-time estimation} in such contexts.

\subsection{Kalman–Bucy filter}

A representative method that performs analog estimation is the Kalman–Bucy filter \cite{Bucy2005}. It is a continuous-time version of the Kalman filter. Given a generic control system that adopts linear state-space modelling described by (\ref{eq:state_differential_equation_linear})
\begin{align*}
\frac{\mathrm{d}}{\mathrm{d} t} \mathbf{x} = \mathbf{A} \mathbf{x} + \mathbf{B} \mathbf{u}
\end{align*}
with its state denoted as $\mathbf{x}$ and its control input to the target process denoted as $\mathbf{u}$ and that adopts linear measurement modelling described by (\ref{eq:SDE_linear_measurement})
\begin{align*}
\mathbf{z} = \mathbf{H} \mathbf{x}.
\end{align*}
Let the covariance $\mathbf{\Sigma_e}$ denotes system modelling uncertainty
\footnote{This may be caused by control input uncertainty denoted by the covariance $\mathbf{\Sigma_u}$. Recall that the covariance $\mathbf{\Sigma_u}$ has already appeared in (\ref{eq:KFprediction}).}
and the covariance $\mathbf{\Sigma_z}$ denotes measurement uncertainty.

The Kalman–Bucy filter is formalized as
\begin{subequations}  \label{eq:Kalman–Bucy_filter}
\begin{align}
\frac{\mathrm{d}}{\mathrm{d} t} \hat{\mathbf{x}} &= \mathbf{A} \hat{\mathbf{x}} + \mathbf{B} \mathbf{u} + \hat{\mathbf{\Sigma}} \mathbf{H}^{\mathrm{T}} \mathbf{\Sigma}_\mathbf{z}^{-1} (\mathbf{z} - \mathbf{H} \hat{\mathbf{x}}),  \\
\frac{\mathrm{d}}{\mathrm{d} t} \hat{\mathbf{\Sigma}} &= \mathbf{A} \hat{\mathbf{\Sigma}} + \hat{\mathbf{\Sigma}} \mathbf{A}^\mathrm{T} - \hat{\mathbf{\Sigma}} \mathbf{H}^{\mathrm{T}} \mathbf{\Sigma}_\mathbf{z}^{-1} \mathbf{H} \hat{\mathbf{\Sigma}} + \mathbf{\Sigma_e},
\end{align}
\end{subequations}
where $\hat{\mathbf{x}}$ denotes the state estimate and $\hat{\mathbf{\Sigma}}$ denotes the state estimate covariance. The second equation of (\ref{eq:Kalman–Bucy_filter}) is a differential Riccati equation or matrix differential Riccati equation.

\begin{figure}[h!]
\begin{center}
\includegraphics[width=0.6\columnwidth]{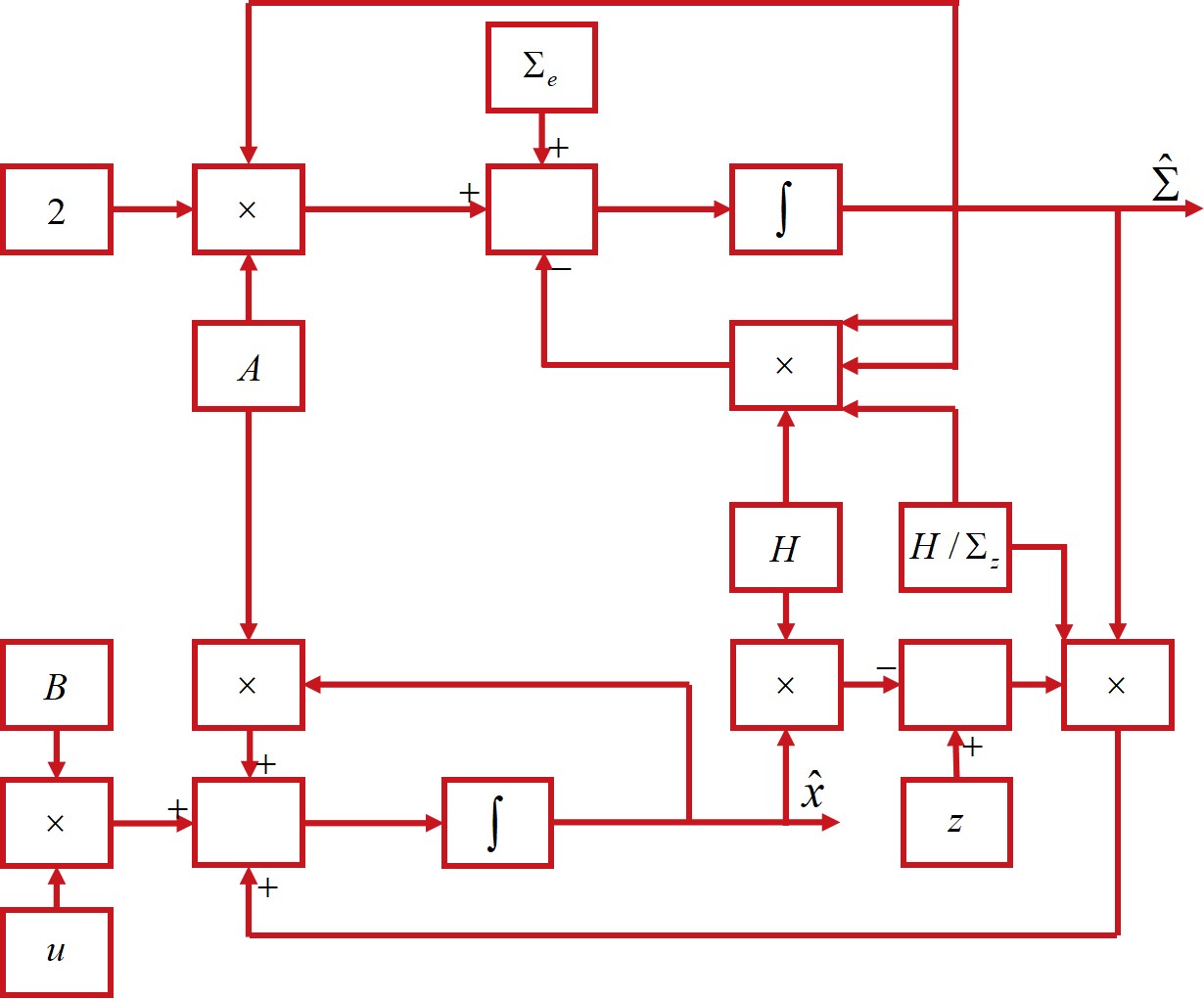}
\end{center}
\caption{Analog electronic circuits realizing the Kalman–Bucy filter example}
\label{fig:analog_kalman_bucy}
\end{figure}

From pure mathematics perspective, the differential Riccati equation can be solved by methods such as presented in \cite{Nguyen2010}. On the other hand, in practical applications, the differential equations described in the Kalman–Bucy filter (\ref{eq:Kalman–Bucy_filter}) do not need to be intentionally solved, but can be naturally realized via analog devices or numeric computation based analog simulators. To facilitate understanding, consider as application example the special case where all relevant variables are scalar variables. Then the Kalman–Bucy filter is reduced to
\begin{align*}
\frac{\mathrm{d}}{\mathrm{d} t} \hat{x} &= A \hat{x} + B u + \hat{\Sigma} \frac{H}{\Sigma_z} (z - H \hat{x}),  \\
\frac{\mathrm{d}}{\mathrm{d} t} \hat{\Sigma} &= 2 A \hat{\Sigma} - \hat{\Sigma}^2 \frac{H^2}{\Sigma_z} + \Sigma_e,
\end{align*}
which can be realized by the analog electronic circuits illustrated in Figure \ref{fig:analog_kalman_bucy}. The ``$\int$'' block represents the analog integrator, the ``$\times$'' block represents the analog multiplier, and the blank block with peripheral positive and negative signs represents the analog adder. For the analog adder block, a signal passed with the positive sign ``$+$'' is added to the block output, whereas a signal passed with the negative sign ``$-$'' is subtracted from the block output. Readers may refer to \cite{Agarwal2005, Asadi2023} for basic knowledge of analog electronic circuits.

The final result to which the state estimate covariance $\hat{\mathbf{\Sigma}}$ will converge is the positive definite solution of the Riccati equation
\begin{align*}
\mathbf{A} \hat{\mathbf{\Sigma}} + \hat{\mathbf{\Sigma}} \mathbf{A}^\mathrm{T} - \hat{\mathbf{\Sigma}} \mathbf{H}^{\mathrm{T}} \mathbf{\Sigma}_\mathbf{z}^{-1} \mathbf{H} \hat{\mathbf{\Sigma}} + \mathbf{\Sigma_e} = \mathbf{0}.
\end{align*}
Transform above Riccati equation into
\begin{align*}
(\mathbf{A} - \hat{\mathbf{\Sigma}} \mathbf{H}^{\mathrm{T}} \mathbf{\Sigma}_\mathbf{z}^{-1} \mathbf{H}) \hat{\mathbf{\Sigma}} + \hat{\mathbf{\Sigma}} (\mathbf{A} - \hat{\mathbf{\Sigma}} \mathbf{H}^{\mathrm{T}} \mathbf{\Sigma}_\mathbf{z}^{-1} \mathbf{H})^\mathrm{T} = - \hat{\mathbf{\Sigma}} \mathbf{H}^{\mathrm{T}} \mathbf{\Sigma}_\mathbf{z}^{-1} \mathbf{H} \hat{\mathbf{\Sigma}} - \mathbf{\Sigma_e} < 0.
\end{align*}
According to the \textit{Lyapunov criterion III} or the \textit{Lyapunov criterion III-B} presented in Section 1.4.1 in Chapter 1, 
\footnote{Namely Chapter 1 of the author's works \cite{Li2026ACTPA_SJTU_2, Li2026ACTPA_SJTU_1}. Note that this article is Chapter 3 of the works.}
the matrix 
\begin{align*}
\mathbf{A}_h \equiv \mathbf{A} - \hat{\mathbf{\Sigma}} \mathbf{H}^{\mathrm{T}} \mathbf{\Sigma}_\mathbf{z}^{-1} \mathbf{H}
\end{align*}
is stable.

Define the continuous-time estimation error as
\begin{equation}  \label{eq:analog_estimation_error}
\mathbf{e} (t) = \hat{\mathbf{x}} (t) - \mathbf{x} (t)
\end{equation}
and we have
\begin{align}  \label{eq:analog_kalman_bucy_error_DE}
\frac{\mathrm{d}}{\mathrm{d} t} \mathbf{e} &= \frac{\mathrm{d}}{\mathrm{d} t} \hat{\mathbf{x}} - \frac{\mathrm{d}}{\mathrm{d} t} \mathbf{x} = \mathbf{A} \hat{\mathbf{x}} + \mathbf{B} \mathbf{u} + \hat{\mathbf{\Sigma}} \mathbf{H}^{\mathrm{T}} \mathbf{\Sigma}_\mathbf{z}^{-1} (\mathbf{z} - \mathbf{H} \hat{\mathbf{x}}) - (\mathbf{A} \mathbf{x} + \mathbf{B} \mathbf{u})  \nonumber \\
  &= \mathbf{A} \hat{\mathbf{x}} + \hat{\mathbf{\Sigma}} \mathbf{H}^{\mathrm{T}} \mathbf{\Sigma}_\mathbf{z}^{-1} (\mathbf{H} \mathbf{x} - \mathbf{H} \hat{\mathbf{x}}) - \mathbf{A} \mathbf{x}  \nonumber \\
  &= (\mathbf{A} - \hat{\mathbf{\Sigma}} \mathbf{H}^{\mathrm{T}} \mathbf{\Sigma}_\mathbf{z}^{-1} \mathbf{H}) (\hat{\mathbf{x}} - \mathbf{x}) = \mathbf{A}_h \mathbf{e}.
\end{align}
Since the matrix $\mathbf{A}_h$ is stable, the estimation error $\mathbf{e}$ will converge to zero. In other words, the state estimate $\hat{\mathbf{x}}$ will converge to the true state $\mathbf{x}$.

\subsection{Luenberger filter}

Another representative method that performs analog estimation is the Luenberger filter \cite{Luenberger1964}. Given a generic control system that adopts linear state-space modelling described by (\ref{eq:state_differential_equation_linear})
\begin{align*}
\frac{\mathrm{d}}{\mathrm{d} t} \mathbf{x} = \mathbf{A} \mathbf{x} + \mathbf{B} \mathbf{u}
\end{align*}
with its state denoted as $\mathbf{x}$ and its control input to the target process denoted as $\mathbf{u}$ and that adopts linear measurement modelling described by (\ref{eq:SDE_linear_measurement})
\begin{align*}
\mathbf{z} = \mathbf{H} \mathbf{x}.
\end{align*}

The Luenberger filter is formalized as
\begin{equation}  \label{eq:Luenberger_filter}
\frac{\mathrm{d}}{\mathrm{d} t} \hat{\mathbf{x}} = \mathbf{A} \hat{\mathbf{x}} + \mathbf{B} \mathbf{u} + \mathbf{L} (\mathbf{z} - \mathbf{H} \hat{\mathbf{x}}).
\end{equation}
Substitute the constant matrix $\mathbf{L}$ for the time-variant matrix $\hat{\mathbf{\Sigma}} \mathbf{H}^{\mathrm{T}} \mathbf{\Sigma}_\mathbf{z}^{-1}$ in the first equation of (\ref{eq:Kalman–Bucy_filter}) and obtain the formalism of (\ref{eq:Luenberger_filter}). So (\ref{eq:Luenberger_filter}) may be regarded as a simplified version of (\ref{eq:Kalman–Bucy_filter}).

Consider the continuous-time estimation error defined in (\ref{eq:analog_estimation_error}) and perform analysis similar to that in (\ref{eq:analog_kalman_bucy_error_DE}) as
\begin{align}  \label{eq:analog_Luenberger_error_DE}
\frac{\mathrm{d}}{\mathrm{d} t} \mathbf{e} &= \frac{\mathrm{d}}{\mathrm{d} t} \hat{\mathbf{x}} - \frac{\mathrm{d}}{\mathrm{d} t} \mathbf{x} = \mathbf{A} \hat{\mathbf{x}} + \mathbf{B} \mathbf{u} + \mathbf{L} (\mathbf{z} - \mathbf{H} \hat{\mathbf{x}}) - (\mathbf{A} \mathbf{x} + \mathbf{B} \mathbf{u})  \nonumber \\
  &= \mathbf{A} \hat{\mathbf{x}} + \mathbf{L} (\mathbf{H} \mathbf{x} - \mathbf{H} \hat{\mathbf{x}}) - \mathbf{A} \mathbf{x} = (\mathbf{A} - \mathbf{L} \mathbf{H}) \mathbf{e} \equiv \mathbf{A}_h \mathbf{e},
\end{align}
where
\begin{align*}
\mathbf{A}_h \equiv \mathbf{A} - \mathbf{L} \mathbf{H}.
\end{align*}
As implied by (\ref{eq:analog_Luenberger_error_DE}), the sufficient and necessary condition for the Luenberger filter to converge is that the matrix $\mathbf{A}_h$ is stable, or in other words, the matrix $\mathbf{A}_h$ has eigenvalues all with negative real part.

Suppose the control system is observable, then according to the \textit{control system observability criterion} presented in Section \ref{sec:observability}, the observability matrix
\begin{align*}
\mathbf{O}_{\mathbf{H}, \mathbf{A}} = \begin{bmatrix} \mathbf{H} \\ \mathbf{H} \mathbf{A} \\ \mathbf{H} \mathbf{A}^2 \\ \vdots \\ \mathbf{H} \mathbf{A}^{n-1} \end{bmatrix}
\end{align*}
is of full rank in terms of row vectors.

Consider the dual control system
\begin{equation}  \label{eq:state_differential_equation_linear_dual}
\frac{\mathrm{d}}{\mathrm{d} t} \mathbf{x}^* = \mathbf{A}^\mathrm{T} \mathbf{x}^* + \mathbf{H}^\mathrm{T} \mathbf{u}^*.
\end{equation}
The controllability matrix of the dual control system described by (\ref{eq:state_differential_equation_linear_dual}) is
\begin{align*}
\mathbf{C}_{\mathbf{A}^\mathrm{T}, \mathbf{H}^\mathrm{T}} = \begin{bmatrix} \mathbf{H}^\mathrm{T} & \mathbf{A}^\mathrm{T} \mathbf{H}^\mathrm{T} & (\mathbf{A}^\mathrm{T})^2 \mathbf{H}^\mathrm{T} & \cdots & (\mathbf{A}^\mathrm{T})^{n-1} \mathbf{H}^\mathrm{T} \end{bmatrix}
\end{align*}
which is right the transpose of the observability matrix $\mathbf{O}_{\mathbf{H}, \mathbf{A}}$. Since $\mathbf{O}_{\mathbf{H}, \mathbf{A}}$ is of full rank in terms of row vectors, the controllability matrix $\mathbf{C}_{\mathbf{A}^\mathrm{T}, \mathbf{H}^\mathrm{T}}$ is of full rank in terms of column vectors or of full rank for short. This implies that the dual control system is controllable and we can design the gain matrix $\mathbf{L}$ such that
\begin{align*}
\mathbf{A}_h^\mathrm{T} = \mathbf{A}^\mathrm{T} - \mathbf{H}^\mathrm{T} \mathbf{L}^\mathrm{T}
\end{align*}
is stable and hence
\begin{align*}
\mathbf{A}_h = (\mathbf{A}_h^\mathrm{T})^\mathrm{T}
\end{align*}
is stable as well.

\subsection{Integrated full-state feedback control with estimator}  \label{sec:integrated_FS_feedback+estimator}

\begin{figure}[h!]
\begin{center}
\includegraphics[width=0.8\columnwidth]{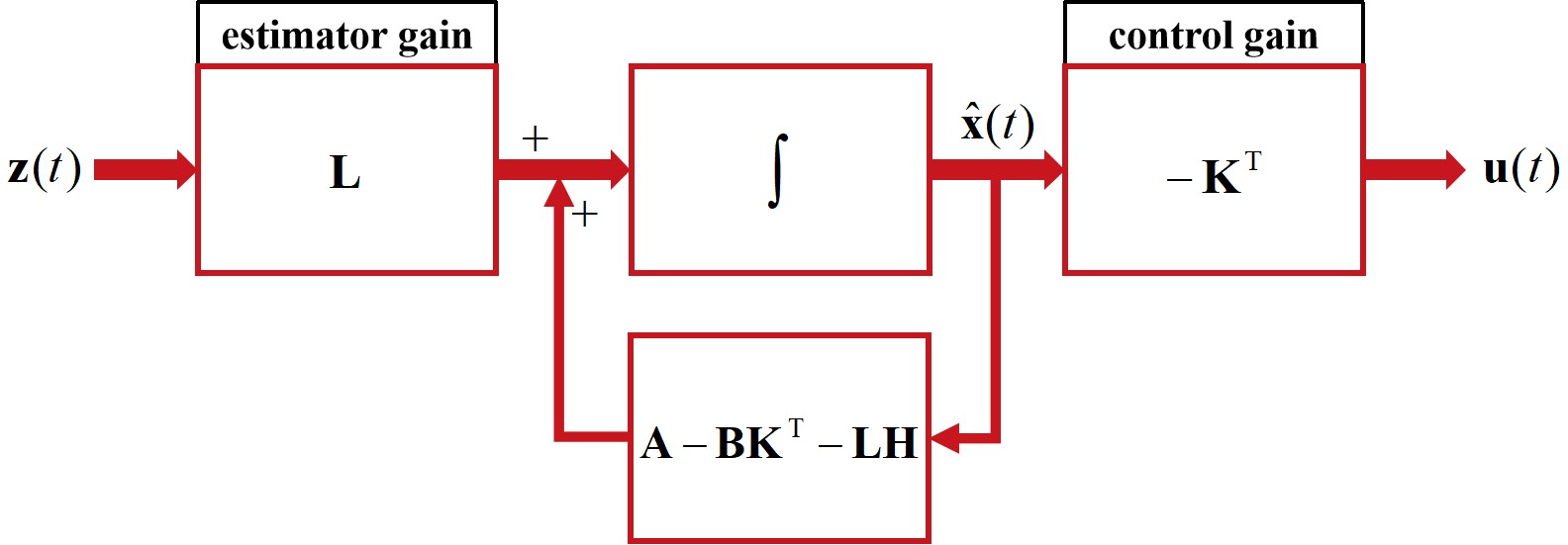}
\end{center}
\caption{Integrated full-state feedback control with estimator}
\label{fig:architecture_integrated_control_estimator}
\end{figure}

Given a generic control system that adopts linear state-space modelling described by (\ref{eq:state_differential_equation_linear})
\begin{align*}
\frac{\mathrm{d}}{\mathrm{d} t} \mathbf{x} = \mathbf{A} \mathbf{x} + \mathbf{B} \mathbf{u}
\end{align*}
with its state denoted as $\mathbf{x}$ and its control input to the target process denoted as $\mathbf{u}$ and that adopts linear measurement modelling described by (\ref{eq:SDE_linear_measurement})
\begin{align*}
\mathbf{z} = \mathbf{H} \mathbf{x}.
\end{align*}
Suppose the Luenberger filter is adopted for continuous-time state estimation. Then we can formalize the integrated full-state feedback control with estimator as
\begin{subequations}  \label{eq:integrated_FS_feedback+estimator}
\begin{align}
\frac{\mathrm{d}}{\mathrm{d} t} \hat{\mathbf{x}} &= \mathbf{A} \hat{\mathbf{x}} + \mathbf{B} \mathbf{u} + \mathbf{L} (\mathbf{z} - \mathbf{H} \hat{\mathbf{x}}) = (\mathbf{A} - \mathbf{B} \mathbf{K}^\mathrm{T} - \mathbf{L} \mathbf{H}) \hat{\mathbf{x}} + \mathbf{L} \mathbf{z},  \\
\mathbf{u} &= -\mathbf{K}^\mathrm{T} \hat{\mathbf{x}}.
\end{align}
\end{subequations}
The architecture of the integrated full-state feedback control with estimator formalized in (\ref{eq:integrated_FS_feedback+estimator}) is illustrated in Figure \ref{fig:architecture_integrated_control_estimator}.

Consider the continuous-time estimation error defined in (\ref{eq:analog_estimation_error}) and obtain (\ref{eq:analog_Luenberger_error_DE})
\begin{align*}
\frac{\mathrm{d}}{\mathrm{d} t} \mathbf{e} &= \frac{\mathrm{d}}{\mathrm{d} t} (\hat{\mathbf{x}} - \mathbf{x}) = (\mathbf{A} - \mathbf{L} \mathbf{H}) (\hat{\mathbf{x}} - \mathbf{x}) = \mathbf{A}_h \mathbf{e}.
\end{align*}
Express the state estimate $\hat{\mathbf{x}}$ in terms of the state $\mathbf{x}$ and the estimation error $\mathbf{e}$ according to (\ref{eq:analog_estimation_error}) as
\begin{equation}  \label{eq:analog_estimation_xhat=x-error}
\hat{\mathbf{x}} (t) = \mathbf{x} (t) + \mathbf{e} (t).
\end{equation}
Substitute (\ref{eq:analog_estimation_xhat=x-error}) into (\ref{eq:state_differential_equation_linear}) and obtain
\begin{align}  \label{eq:state_differential_equation_linear_x+e}
\frac{\mathrm{d}}{\mathrm{d} t} \mathbf{x} = \mathbf{A} \mathbf{x} - \mathbf{B} \mathbf{K}^\mathrm{T} \hat{\mathbf{x}} = \mathbf{A} \mathbf{x} - \mathbf{B} \mathbf{K}^\mathrm{T} (\mathbf{x} + \mathbf{e}) = (\mathbf{A} - \mathbf{B} \mathbf{K}^\mathrm{T}) \mathbf{x} - \mathbf{B} \mathbf{K}^\mathrm{T} \mathbf{e}.
\end{align}
Augment the state $\mathbf{x}$ with the estimation error $\mathbf{e}$ to form the augmented state
\begin{align*}
\begin{bmatrix} \mathbf{x} \\ \mathbf{e} \end{bmatrix}.
\end{align*}
Combine (\ref{eq:analog_Luenberger_error_DE}) and (\ref{eq:state_differential_equation_linear_x+e}) to formalize dynamics of the augmented state as
\begin{equation}  \label{eq:integrated_dynamics_linear_x+e}
\frac{\mathrm{d}}{\mathrm{d} t} \begin{bmatrix} \mathbf{x} \\ \mathbf{e} \end{bmatrix} = \begin{bmatrix} \mathbf{A} - \mathbf{B} \mathbf{K}^\mathrm{T} & - \mathbf{B} \mathbf{K}^\mathrm{T} \\ 0 & \mathbf{A} - \mathbf{L} \mathbf{H} \end{bmatrix} \begin{bmatrix} \mathbf{x} \\ \mathbf{e} \end{bmatrix}.
\end{equation}

The characteristic polynomial associated with the state transition matrix specified in (\ref{eq:integrated_dynamics_linear_x+e}) is
\begin{align*}
\det (s \mathbf{I} - \begin{bmatrix} \mathbf{A} - \mathbf{B} \mathbf{K}^\mathrm{T} & - \mathbf{B} \mathbf{K}^\mathrm{T} \\ 0 & \mathbf{A} - \mathbf{L} \mathbf{H} \end{bmatrix}) = \det (s \mathbf{I} - (\mathbf{A} - \mathbf{B} \mathbf{K}^\mathrm{T})) \cdot \det (s \mathbf{I} - (\mathbf{A} - \mathbf{L} \mathbf{H})).
\end{align*}
So if and only if both the matrices
\begin{align*}
\mathbf{A}_c \equiv \mathbf{A} - \mathbf{B} \mathbf{K}^\mathrm{T}
\end{align*}
and
\begin{align*}
\mathbf{A}_h \equiv \mathbf{A} - \mathbf{L} \mathbf{H}
\end{align*}
are stable, namely if and only if they both have eigenvalues all with negative real part, the overall system of the state $\mathbf{x}$ and the estimation error $\mathbf{e}$ is stable.

\subsubsection*{Application: single inverted pendulum control with estimator}

Apply the method of integrated full-state feedback control with estimator formalized in (\ref{eq:integrated_FS_feedback+estimator}) to handle the single inverted pendulum control problem. The single inverted pendulum control system adopts linear state-space modelling described by
\begin{align*}  
\frac{\mathrm{d}}{\mathrm{d} t} \mathbf{x} \equiv \frac{\mathrm{d}}{\mathrm{d} t} \begin{bmatrix} \theta \\ \frac{\mathrm{d} \theta}{\mathrm{d} t} \\ x \\ \frac{\mathrm{d} x}{\mathrm{d} t} \end{bmatrix} = \begin{bmatrix} 0 & 1 & 0 & 0 \\ \frac{g}{L} & 0 & 0 & 0 \\ 0 & 0 & 0 & 1  \\ 0 & 0 & 0 & 0 \end{bmatrix} \begin{bmatrix} \theta \\ \frac{\mathrm{d} \theta}{\mathrm{d} t} \\ x \\ \frac{\mathrm{d} x}{\mathrm{d} t} \end{bmatrix} + \begin{bmatrix} 0 \\ -\frac{1}{L} \\ 0 \\ 1 \end{bmatrix} a \equiv \mathbf{A} \mathbf{x} + \mathbf{B} a.
\end{align*}
For concrete configuration of parameters, let 
\begin{align*}  
L = 1, \quad g = 10.
\end{align*}
In fact, the concrete value of $m$ does not influence the model. Then the state transition matrix $\mathbf{A}$ and the control input matrix $\mathbf{B}$ are
\begin{align*}
\mathbf{A} = \begin{bmatrix} 0 & 1 & 0 & 0 \\ 10 & 0 & 0 & 0 \\ 0 & 0 & 0 & 1  \\ 0 & 0 & 0 & 0 \end{bmatrix}, \quad \mathbf{B} = \begin{bmatrix} 0 \\ -1 \\ 0 \\ 1 \end{bmatrix}.
\end{align*}
Suppose we can measure the inverted pendulum angle and the cart position, then the measurement model is
\begin{align*}
\mathbf{z} \equiv \begin{bmatrix} z_{\theta} \\ z_x \end{bmatrix} = \begin{bmatrix} 1 & 0 & 0 & 0 \\ 0 & 0 & 1 & 0 \end{bmatrix} \begin{bmatrix} \theta \\ \frac{\mathrm{d} \theta}{\mathrm{d} t} \\ x \\ \frac{\mathrm{d} x}{\mathrm{d} t} \end{bmatrix} \equiv \mathbf{H} \mathbf{x}.
\end{align*}

Set the expected closed-loop characteristic polynomial as
\begin{align*}
C_\mathrm{E}(s) = (s + 4)^4 = s^4 + 16 s^3 + 96 s^2 + 256 s + 256
\end{align*}
for both controller design and estimator design --- There is no special reason to set the expected characteristic polynomial so, but simply for demonstration purpose --- Then compute the gain matrix $\mathbf{K}$ for control and the gain matrix $\mathbf{L}$ for estimation as
\begin{align*}
\mathbf{K} = \begin{bmatrix}  -131.60 \\ -41.60 \\ -25.60 \\ -25.60 \end{bmatrix}, \qquad \mathbf{L} = \begin{bmatrix} 8 & 0 \\ 26 & 0 \\ 0 & 8 \\ 0 & 16 \end{bmatrix}
\end{align*}
respectively. Matlab simulation code for complete demonstration of single inverted pendulum control with estimator is given as follows. The visualization code \textbf{DisplaySIP.m} and the single inverted pendulum dynamics code \textbf{DynamicsSIP.m} in following code are already given in Section 2.2.3 in Chapter 2. The gain matrix designing code \textbf{DesignGainMatrix.m} is given in Section 2.3.2 in Chapter 2.
\footnote{Namely Chapter 2 of the author's works \cite{Li2026ACTPA_SJTU_2, Li2026ACTPA_SJTU_1}. Note that this article is Chapter 3 of the works.}

\begin{framed} 
\noindent \textbf{SingleInvertedPendulumWithEstimator.m} \\
\noindent \%\% Single inverted pendulum parameters \\
m1 = 1; L1 = 1; g = 10; \\
\%\% Simulation preliminary configuration \\
dt = 0.001; \% Numerical computation step \\
tSpan = 0:dt:4; \% Simulation time span \\
x = 0.2; \% Cart position  \\
dx = 0; \% Cart velocity \\
y = 0.2; \% Inverted pendulum angle theta  \\
dy = 0;  \% Inverted pendulum angular velocity \\
stt = [y; dy; x; dx]; \% Single inverted pendulum state \\
\% State estimate with random error corrupted initialization \\
sttE = stt + random('norm',0,0.2,4,1);  \\
sttAll = zeros(length(stt), length(tSpan)); k = 0; \% Record states \\
sttEAll = zeros(length(sttE), length(tSpan)); k = 0; \% Record estimates \\
xExpected = 0; yExpected = 0; \% Expected equilibrium status \\
SimConfig = [m1, L1, g, dt]; \\
 \\
\%\% Design the gain matrix via the general method \\
A = [0, 1, 0, 0; g/L1, 0, 0, 0; 0, 0, 0, 1; 0, 0, 0, 0]; \\
B = [0; -1/L1; 0; 1]; \\
H = [1, 0, 0, 0; 0, 0, 1, 0]; \\
lambdaE = [-4;-4;-4;-4]; \% Expected eigenvalues \\
\% Design the gain matrix for control \\
sttK = DesignGainMatrix(A, B, lambdaE); \\
fprintf('Gain matrix K: '); disp(sttK'); \\
\% Design the gain matrix for estimation \\
sttL = DesignGainMatrix(A', H', lambdaE); \\
fprintf('Gain matrix L: '); disp(sttL'); \\
 \\
\%\% Simulation of single inverted pendulum control \\
for t = tSpan \\
$~~~~$ \%\% Integrated control with estimator \\
$~~~~$ \% Noisy measurements of cart position and inverted pendulum angle \\
$~~~~$ z = H*stt + random('norm',0,0.01,2,1); \\
$~~~~$ \% State estimation \\
$~~~~$ sttE = sttE + ((A-B*sttK'-sttL*H)*sttE+sttL*z)*dt; \\
$~~~~$ \% Full-state feedback control of the cart acceleration \\
$~~~~$ acc = -sttK'*sttE;  \\
 \\
$~~~~$ \%\% Single inverted pendulum dynamics \\
$~~~~$ stt = DynamicsSIP(SimConfig, stt, acc); \\
$~~~~$ sttC = num2cell(stt); [y, dy, x, dx] = sttC\{:\}; \\
$~~~~$ if (abs(y)$>$=pi/2) fprintf('Control failure!$\backslash$n'); break; end \\
$~~~~$ k = k+1; sttAll(:,k) = stt; sttEAll(:,k) = sttE; \\
$~~~~$ \%\% Single inverted pendulum visualization \\
$~~~~$ if (rem(k,20) == 0) \\
$~~~~$ $~~~~$ DisplaySIP(x, y, L1); pause(dt); \\
$~~~~$ end \\
end
\end{framed}

Readers can try the simulation code \textbf{SingleInvertedPendulumWithEstimator.m} and will find that the objective of inverted pendulum control is successfully achieved. It is worth noting that from the perspective of practical applications, the simulation code \textbf{SingleInvertedPendulumWithEstimator.m} is the most realistic one among all demonstration code presented in this book by so far. It is most realistic in the sense that the control law involved in it, namely the control input of cart acceleration, is no longer generated ideally with the ground truth of the state. Instead, the control law is generated with the state estimate. The realism consists right in the fact that in practical applications, the ground truth of the state can never be obtained, whereas it is the estimate of the state that can actually be obtained and used in feedback control. Pay attention to the code part extracted below, where the variable ``sttE'' in code denotes the estimate of the single inverted pendulum state, instead of the ground truth of the single inverted pendulum state
\footnote{The variable ``stt'' in the simulation code \textbf{SingleInvertedPendulumWithEstimator.m} denotes the ground truth of the single inverted pendulum state.}.

\begin{framed} 
\noindent 
$~~~~$ \% Full-state feedback control of the cart acceleration \\
$~~~~$ acc = -sttK'*sttE;
\end{framed}

\begin{figure}[h!]
\begin{center}
\includegraphics[width=0.95\columnwidth]{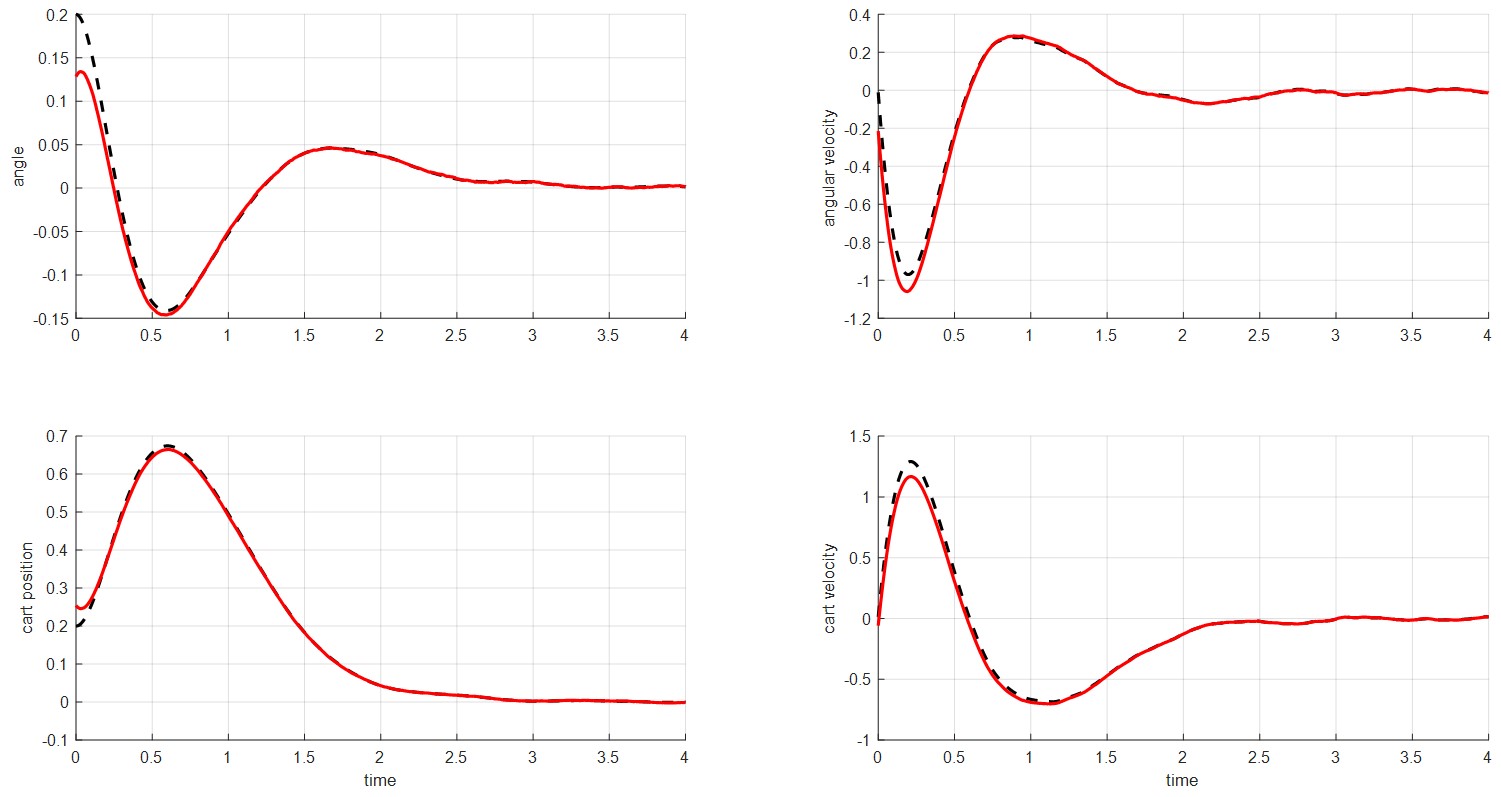}
\end{center}
\caption{Estimates of the single inverted pendulum state: (top-left) inverted pendulum angle; (top-right) inverted pendulum angular velocity; (bottom-left) cart position; (bottom-right) cart velocity}
\label{fig:SIP_integrated_control_estimator}
\end{figure}

Estimates and ground truths of the single inverted pendulum state, which are recorded during one trial of the simulation code \textbf{SingleInvertedPendulumWithEstimator.m}, are demonstrated in Figure \ref{fig:SIP_integrated_control_estimator}, where the top-left sub-figure, the top-right sub-figure, the bottom-left sub-figure, and the bottom-right sub-figure demonstrate estimates and ground truths of the inverted pendulum angle, the inverted pendulum angular velocity, the cart position, and the cart velocity respectively. In each sub-figure, the estimate is shown by the solid line whereas the ground truth is shown by the dashed line. As we can see, estimates of the inverted pendulum angle, the inverted pendulum angular velocity, the cart position, and the cart velocity all converge asymptotically on one hand to their ground truth counterparts which all converge to zero namely the equilibrium state on the other hand. In other words, the overall system of both the single inverted pendulum state and the state estimation error are successfully stabilized. For the method of integrated full-state feedback control with estimator, results demonstrated in Figure \ref{fig:SIP_integrated_control_estimator} provide quantitative verification of its effectiveness.

It is worth noting that by trying the simulation code \textbf{SingleInvertedPendulumWithEstimator.m}, readers may encounter simulation results different from those demonstrated in Figure \ref{fig:SIP_integrated_control_estimator}. The potential difference is caused by random errors added to relevant variables in the simulation code. Such random errors are added to simulate an effect of uncertainty such as single inverted pendulum state initialization uncertainty and measurement uncertainty.

\section{Note}

The author would like to highlight two important points again: \textit{First, for a control system that involves usage of the state, what can be used is not the ground truth of the state but the estimate of the state, which is obtained via certain estimation method.} This is a point that has just been mentioned for clarifying the realism of the simulation code \textbf{SingleInvertedPendulumWithEstimator.m} in Section \ref{sec:integrated_FS_feedback+estimator}. \textit{Second, state estimation is by no means trivial and we had better not take it for granted that the state can always be estimated easily or even feasibly in practical applications.} This is a point that is mentioned even at the beginning of this article. 
\footnote{Namely Chapter 3 of the author's works \cite{Li2026ACTPA_SJTU_2, Li2026ACTPA_SJTU_1}.}
It is true that in many practical applications, measurements of or related to the state can be conveniently obtained, and most part of difficulty involved in state estimation consists in estimation itself (or more specifically, using what estimation method to reveal the state). It is also true that in many practical applications, to even know how to obtain desirable or at least usable measurements of or related to the state would be much more difficult than estimation itself. Resolving difficulty in such latter case depends on field knowledge which can be of a wide range of variety and can by no means be covered in this book.

On the other hand, the author would also like to point out that despite ubiquitous involvement of state estimation in practical applications, we had better not exaggerate ubiquitous involvement of state estimation in study, research, and development concerning control theory. In many kinds of circumstances, \textit{we can fairly put state estimation aside, assume that the state is known, and focus on the control method itself.} Therefore, for presentation of control theory and relevant demonstrations throughout this book, we assume by default that the state is known and refrain from always entangling the state estimation part in the presented control methods.


\newpage
\addcontentsline{toc}{chapter}{Bibliography}

\fancyhf{} 

\bibliographystyle{unsrt}
\bibliography{LI_Hao_Refs_ACTPA}

\fancyhead[LE,RO]{\thepage}
\fancyhead[RE]{\textit{ \nouppercase{\leftmark}} }
\fancyhead[LO]{\textit{ \nouppercase{\rightmark}} }

\end{document}